\documentclass[onecolumn,prd,nofootinbib]{revtex4}
\usepackage{graphicx}
\usepackage{amsmath}
\usepackage{amsfonts}
\usepackage{bm}
\usepackage{mathrsfs}
\usepackage[leftcaption]{sidecap}
\usepackage{floatflt}
\usepackage{wrapfig}

\newcommand{\be}{\begin{equation}}
\newcommand{\ee}{\end{equation}}
\newcommand{\bea}{\begin{eqnarray}}
\newcommand{\eea}{\end{eqnarray}}
\newcommand{\ba}{\begin{eqnarray}}
\newcommand{\ea}{\end{eqnarray}}
\newcommand{\fl}{\hspace*{-\mathindent}}
 
\def \D {\mathrm{D}}
\def \d {\mbox{d}}

\def\c {\mbox{curl}\,}
\def\ep {\varepsilon}
\def\p  {\partial}
\def\a {a}
\def\b {b}
\newcommand{\e}{\rho}

\def\la {\langle}
\def\ra {\rangle}
\renewcommand{\div}{\mbox{div}\,}
\newcommand{\curl}{\mbox{curl}\,}

\newcommand{\hu}{\text{km\,s}^{-1}\text{Mpc}^{-1}}

\usepackage{bm}

\renewcommand{\>}{\rangle}

\newcommand{\del}{\nabla}
\newcommand{\li}{&}
\newcommand{\sfrac}{\frac}
\newcommand{\aperp}{a_{\perp }}

\newcommand{\apar}{a_{\parallel}}
\newcommand{\kapr}{\kappa(r)r^2}
\newcommand{\Hperp}{H_{\perp}}
\newcommand{\Hperpo}{H_{\perp_0}}
\newcommand{\Hpar}{H_{\parallel}}
\newcommand{\omegmo}{\Omega_{{m}}}
\newcommand{\omegko}{\Omega_{{\kappa}}}

\newcommand{\gmgeta}{\eta}
\newcommand{\gmgpsi}{\varsigma}
\newcommand{\gmgchi}{\chi}

\newcommand{\gmgk}{\varphi}
\newcommand{\gmgalpha}{v}

\newcommand{\gmggamma}{w}
\newcommand{\gmgomega}{\Delta}
\renewcommand{\H}{\mathcal{H}}
\newcommand{\T}{{\text{\tiny\it (T)}}}
\newcommand{\TF}{{\text{\tiny\it (TF)}}}
\newcommand{\GI}{{\text{\tiny\it (GI)}}}

\newcommand{\dec}{{*}}
\newcommand{\eq}{{\text{eq}}}
\newcommand{\sdel}{\tilde\nabla}
\renewcommand{\:}[2]{{\textstyle\frac{#1}{#2}}}
\renewcommand{\;}[2]{{\frac{#1}{#2}}}

\newcommand{\inflcdm}{\,\,\xrightarrow{{\scriptstyle\text{\,flat $\Lambda$CDM}\,}}\,\,}
\newcommand{\inlcdm}{\,\,\xrightarrow{{\scriptstyle\text{\,$\Lambda$CDM}\,}}\,\,}
\newcommand{\inflrw}{\,\,\xrightarrow{{\scriptstyle\text{\,FLRW}\,}}\,\,}
\newcommand{\withcp}{$\,\,\xrightarrow{{\scriptstyle+\text{\,CP}\,}}\,\,$}
\newcommand{\withoutcp}{{$\,\,\xrightarrow{{\scriptstyle\neg{\text{\,CP}\,}}}\,\,$}}

\begin{document}

\title{
Establishing homogeneity of the universe in the shadow of dark energy
}

\author{Chris Clarkson}

\address{ Astrophysics, Cosmology \& Gravity Centre, and, Department of Mathematics and Applied Mathematics, University of Cape Town, Rondebosch 7701, South Africa. \em\texttt{chris.clarkson@uct.ac.za}}

\date{16 April 2012}

\begin{abstract}

Assuming the universe is spatially homogeneous on the largest scales lays the foundation for almost all cosmology. This idea is based on the Copernican principle, that we are not at a particularly special place in the universe. Surprisingly, this philosophical assumption has yet to be rigorously demonstrated independently of the standard paradigm. This issue has been brought to light by cosmological models which can potentially explain apparent acceleration by spatial inhomogeneity rather than dark energy. These models replace the temporal fine tuning associated with $\Lambda$ with a spatial fine tuning, and so violate the Copernican assumption. While is seems unlikely that such models can really give a realistic solution to the dark energy problem, they do reveal how poorly constrained radial inhomogeneity actually is. So the bigger issue remains: How do we robustly test the Copernican principle independently of dark energy or theory of gravity?

\end{abstract}

\maketitle
\tableofcontents
\clearpage

\section{Introduction}

The standard model of cosmology is fabulous in its simplicity: based on linear perturbations about a spatially homogeneous and isotropic background model, it can easily account for just about all observations which probe a vast range of scales in space and time with a small handful of parameters. 
The bigger picture which emerges is of a model with an exponential expansion rate for much of the evolution of the universe, caused by the inflaton at the beginning and dark energy at the end. We are anthropically selected to live in the small era between these phases where structure forms and interesting things happen. 
Yet the physical matter which drives these accelerating periods is not understood at all in a fundamental sense. 
\begin{wrapfigure}[25]{r}[0pt]{0.6\textwidth}
\begin{center}
\includegraphics[width=0.6\textwidth]{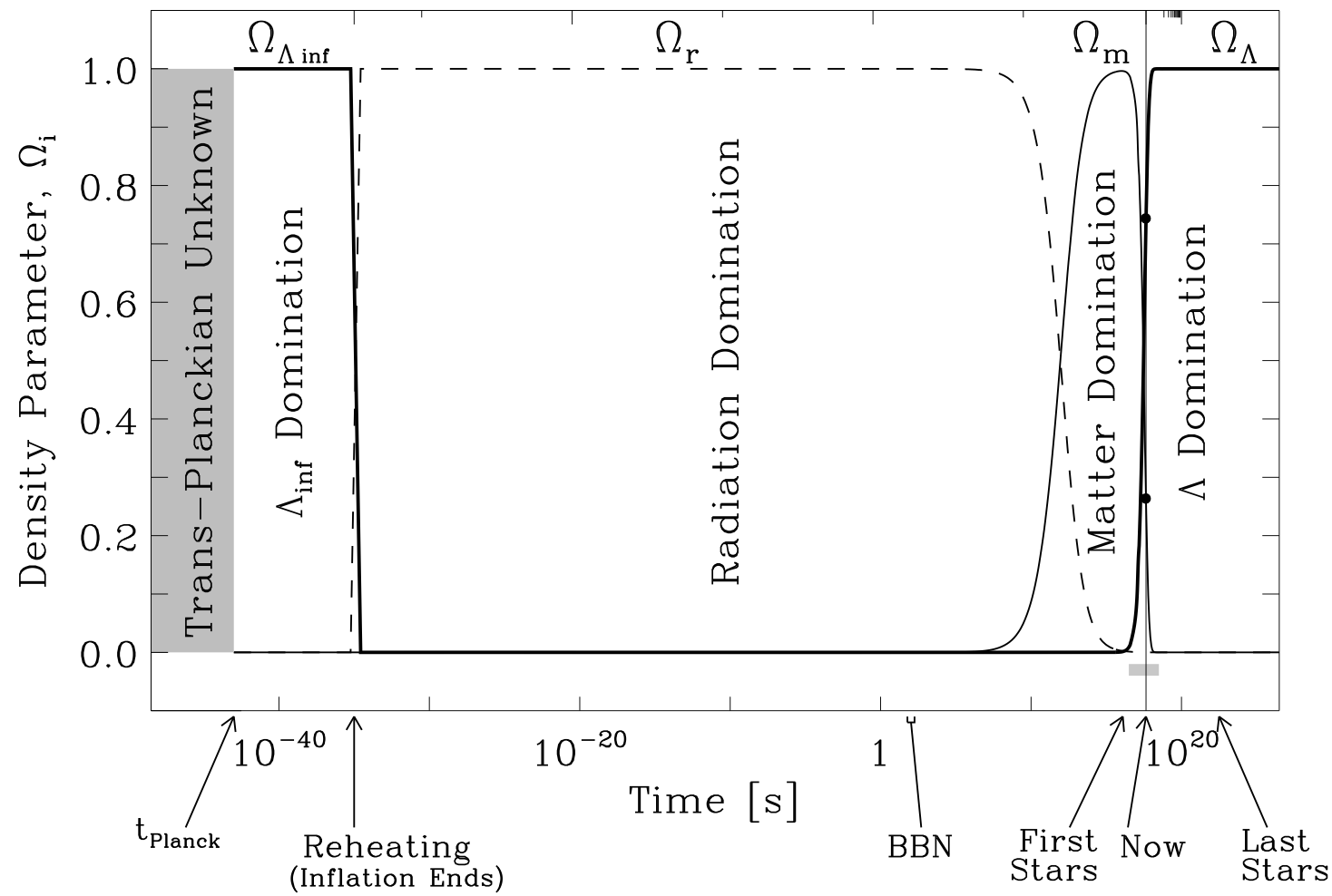}
\caption{The coincidence problem: why is $\Lambda$ as large as it can be? Any larger and the de Sitter phase would start before structure forms. (From~\cite{Egan:2010fc}.)}
\label{coincidence}
\end{center}
\end{wrapfigure}
Until they are, the standard model is unfortunately phenomenological in this critical aspect.  
Because of this, the anthropic fine tuning seems perverse: the cosmological constant, at odds with its `expected' value by some 120 orders of magnitude, has an energy density $\rho_\Lambda$ today about the same as that of matter $\rho_m$, despite the fact that the ratio of these grows with the volume of space: $\rho_\Lambda/\rho_m\sim a^3\sim1$. We are living through a phase transition (Fig.~\ref{coincidence}). 
Why?

The problem of understanding the physical origin and value of the cosmological constant is leading us to reconsider some of the foundational aspects of cosmological model building more carefully. 
In particular, it is an important fact that, at the moment, the spatial homogeneity of the universe when smoothed on equality scales exists by assumption, and is not yet an 
observationally proven fact established outside the paradigm of the standard model~-- which includes dark energy.  
Given this uncertainty, so-called void models can explain the observed distance modulus utilising a spatially varying energy density, 
Hubble rate and curvature on Gpc scales, without any unusual physical fields at late times
~\cite{MT1,MT2,Humphreys:1996fd,Mustapha:1997xb,MHE,zehavi,PascualSanchez:1999zr,Celerier:1999hp,Hellaby:2000ef,Tomita:2000jj,Tomita:2001gh,IKN,Mof1,Mof2,AAG,VFW,Alnes:2006pf,Chung:2006xh,Celerier:2006gy,garfinkle,BMN,EM,alnes,Romano:2007zz,celerier3,conley,Lu:2007gr,McClure:2007hy,sarkar,mattsson,ABNV,bolejko,Enqvist:2007vb,Caldwell:2007yu,CBL,UCE,GarciaBellido:2008nz,zibin,YKN,ishak,CFL,GarciaBellido:2008gd,huntsarkar,Araujo:2008rm,Jia:2008ti,ZMS,gbh1,CFZ,BW,CCF,CBKH,Tomita:2009wz,Araujo:2009zh,Celerier:2009sv,Sollerman:2009yu,tomita09,Kainulainen:2009sx,FLSC,mortsell,Quartin:2009xr,Mof3,Hellaby:2009vz,Romano:2009mr,Garfinkle:2009uf,Romano:2009qx,Kolb:2009hn,Romano:2009ej,Sussman:2010jf,Dunsby:2010ts,Lan:2010ky,RC,Saito:2010tr,Goto:2010gc,Yoo:2010qy,Sussman:2010gh,Clarkson:2010ej,BNV,Moss:2010jx,vanderWalt:2010zd,Yoo:2010ad,CM,Sussman:2010ew,Araujo:2010ir,Zhang:2010fa,Marra:2010pg,Foreman:2010uj,Yoo:2010hi,Yoo:2010qn,Alonso:2010zv,Araujo:2010ag,Chatterjee:2010dr,Duffy:2010bu,Nadathur:2010zm,Goto:2011ru,Marra:2011ct,Sussman:2011na,Bolejko:2011ys,Ellis:2011hk,Riess:2011yx,Romano:2011mx,Romano:2011tz,Belloso:2011ms,Moss:2011ze,Marra:2011zp,Celerier:2011zh,Bull:2011wi,Zibin:2011ma,Wang:2011kj,Winfield:2011tj,Yagi:2011bt,Zumalacarregui:2012pq,Roukema:2012ff}. 
The indication is that  models which are homogeneous at early times are incompatible with observations, as are adiabatic models with the simplest type of inhomogeneity present at early times~\cite{Bull:2011wi,Zibin:2011ma}. Isocurvature degrees of freedom and freedom in the initial power spectrum have not been explored in detail, however, and remain possible routes to constructing viable alternatives to $\Lambda$CDM~\cite{Yoo:2010ad,Clarkson:2010ej,Nadathur:2010zm}. 
They are therefore  a very significant departure from the standard model if they can be made to work, and would require a dramatic reworking of standard inflation (though there are inflationary models which can produce large spherical structures~-- see e.g.,~\cite{Linde:1994gy,Afshordi:2010wn}). 
Irrespective of all this, they have actually been neglected as a serious contender for dark energy because of the anti-Copernican fine tuning that exists: 
we have to be within tens of Mpc of the centre of spherical symmetry of the background~\cite{Alnes:2006pf,alnes,mortsell,Foreman:2010uj}, which 
implies a spatial coincidence of, roughly, (40 Mpc/15 Gpc)$^3\sim 10^{-8}$. This is just plain weird. However, it is not hard to imagine a bigger picture where there exist structures as large~-- or larger than~-- our Hubble scale, one of which we are just glimpsing a part of~\cite{Uzan:2009mx}. Perhaps there could be selection effects favouring stable solar systems in regions of lower dark matter density (or something), which would normalise the spatial coincidence. Who knows? 

While it is still not clear whether these models can really be made a viable alternative to dark energy, these models have brought into focus important questions: 
\emph{Is the universe spatially homogeneous and isotropic when viewed on the largest scales?}  
Perhaps we really have shown this already? If so, to what level of confidence?

The applicability of the Friedmann-Lema\^\i tre-Robertson-Walker (FLRW) metric is the underlying axiom from which we infer dark energy exists~-- whether in the form of $\Lambda$, exotic matter or as a modification of GR.  
It is necessary, therefore, to try to demonstrate that the FLRW paradigm is correct from a purely observational point of view~-- without assuming it a priori, and preferably independently of the field equations. 
There are different issues to consider:
\begin{description}
\item[The Copernican Principle:]  We are not at a special location in the universe.
\item[The Cosmological Principle:] Smoothed on large enough scales the universe is spatially homogeneous and isotropic.
\end{description}
A textbook formulation of the FLRW metric from these principles starts from the first then uses the high isotropy of the Cosmic Microwave Background (CMB), and approximate isotropy of local observables to conclude the second.
 This is correct in an exact sense, as we discuss below: e.g., if all observers measure the distance-redshift relation to be exactly isotropic, 
 then the spacetime is exactly FLRW. The realistic case is much more subtle. 
A statement such as `if most observers find their observables consistent with a small level of 
anisotropy, then the metric of the universe is roughly FLRW when smoothed over a suitably large scale' requires quite a few assumptions about spatial gradients which 
may or may not be realistic, and has only been theoretically argued in the case of the CMB. An additional  problem here is what we 
mean by smoothing in a spacetime: smoothing observables is not the same as spatial smoothing, and smoothing a geometry is ill-defined and is not the same geometry one arrives at from a smoothed energy-momentum tensor (see~\cite{Clarkson:2011zq,Buchert:2011sx} for recent reviews). 
We shall not consider this important problem further here as it is beyond the scope of this work, and assume that the conventional definition in terms of a smooth spacetime is sufficient. 

\begin{wrapfigure}[26]{r}[0pt]{0.5\textwidth}
\begin{center}
\vspace{-5mm}
\includegraphics[height=0.49\textwidth]{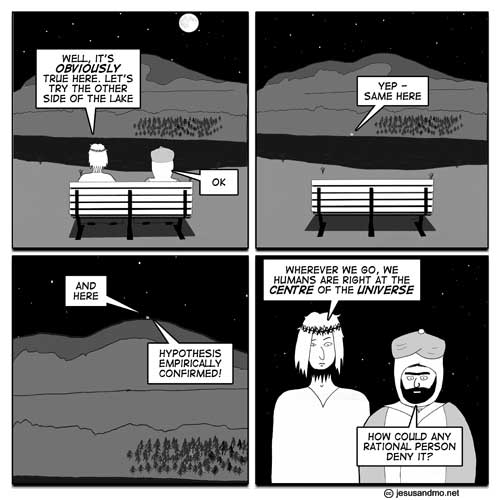}\hfill
\caption{A test for homogeneity. (From jesusandmo.net.)}
\label{cp1}
\end{center}
\end{wrapfigure}
These important subtleties aside, going from the Copernican principle (CP) to the FLRW geometry seems reasonable in a wide variety of circumstances, as we discuss in detail. Consequently, establishing spatial homogeneity of the universe~-- and with it the existence of dark energy~-- can be answered if we can observationally test the Copernican assumption. However obvious it seems,\footnote{The wisdom of the crowd can give the accurate weight of a cow~\cite{crowd}; it would be engaging to see what this method would give us here.} it is nevertheless a philosophical assumption at the heart of cosmology which should be demonstrated scientifically where possible~\cite{Ellis:2006fy}. 

The Copernican principle is hard  to test on large (Gpc) scales  simply because we view the universe effectively from one spacetime event (Fig.~\ref{cp1}), although it can be tested locally~\cite{Labini:2010qx}. Compounding this is the fact that it's hard to disentangle temporal evolution from spatial variation~-- especially so if we do not have a separately testable model for the matter present (dark energy!). A nice way to illustrate the difficulty is to consider an alternative way of making a large-scale cosmological model. Instead of postulating a model at an early time (i.e., an FLRW model with perturbations), evolving it forwards and comparing it with observations, we can take observations directly as `initial data' on our past lightcone and integrate into the interior to reconstruct our past history~\cite{7b,Araujo:2008rm,Hellaby:2008pp,Araujo:2009zh,Araujo:2010ag,vanderWalt:2010zd}. 
Would this \emph{necessarily} yield an FLRW model? What observables do we need? Under what assumptions of dark energy and theory of gravity: is a model based on general relativity which is free of dark energy a possible solution? While such a scheme is impractical in the near future, it is conceptually useful to consider cosmology as an inverse problem, and should be well-posed at least while local structure is in the linear regime.

With these ideas in mind, practical tests of the Copernican assumption can be developed. The are several basic ideas. One is to try to directly observe the universe as others see it. An alien civilisation at $z\sim1$ who had the foresight to send us a data file with their cosmological observations would be nice, but failing that placing limits on anisotropy around distant clusters can achieve the same ends with less slime. Another is to combine observables such that we can see if the data on our past lightcone would conflict with an FLRW model in the interior. This helps formulate the Copernican principle as a null hypothesis which can in principle be refuted. A third is to see if the thermal history is the same in widely separated regions which can be used to probe homogeneity at early times~\cite{1986MNRAS.218..605B}. 

This review is organised as follows. First we consider what isotropic observations tell us, and review models which violate the Copernican principle. Then we discuss general results which help us go from exact isotropy of observables to exact homogeneity of space. Finally we summarise the consistency tests available to test the FLRW assumption.

\section{Models without homogeneity as an alternative to dark energy}

\subsection{From isotropic observables to isotropy of space}

Without assuming the Copernican Principle, we have the following isotropy result~\cite{7,7a,7b}:
\begin{description}
\item[{\bf \footnotesize [ENMSW]}~{\em Matter lightcone-isotropy \em\withoutcp\em spatial isotropy}:]

If one observer comoving with the matter measures isotropic area distances, number counts, bulk peculiar velocities, and lensing, in an expanding dust Universe with $\Lambda$, then the spacetime is isotropic about the observer's worldline.
\end{description}

\begin{wrapfigure}[24]{l}[0pt]{0.6\textwidth}
\begin{center}
\includegraphics[width=0.55\textwidth]{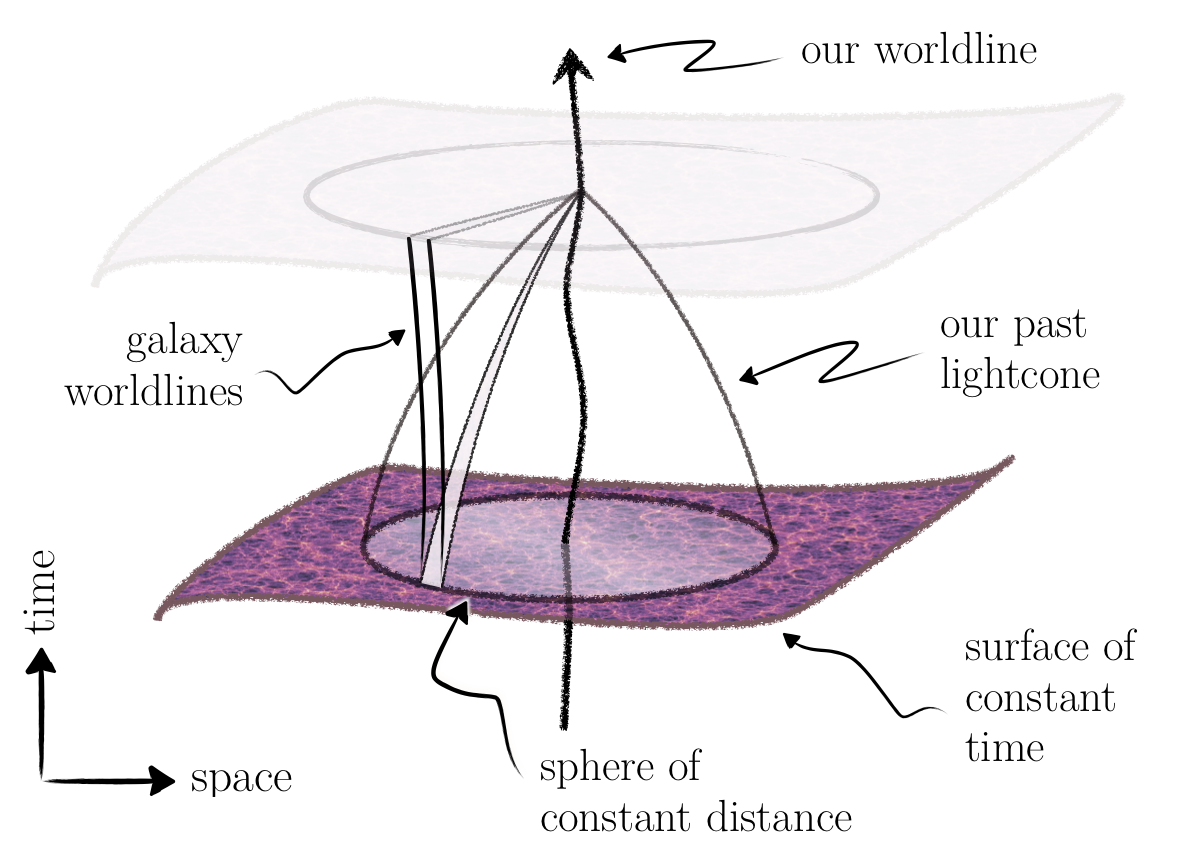}
\caption{The Copernican principle is hard to test because we are fixed to one event in spacetime. We make observations on our past nullcone which slices through spatial surfaces.}
\label{cp}
\end{center}
\end{wrapfigure}

This is an impressive selection of observables required.
 Note that isotropy of (bulk) peculiar velocities seen by the observer is equivalent to vanishing proper motions (tranverse velocities) on the observer's sky. Isotropy of lensing means that there is no distortion of images, only magnification.

The proof of this result requires a non-perturbative approach~-- there is no background to perturb around. Since the data is given on the past lightcone of the observer, we need the fully general metric, adapted to the past lightcones of the observer worldline ${\cal C}$. We define observational coordinates $x^\a=(w,y,\theta,\phi)$, where $x^a=(\theta,\phi)$ are the celestial coordinates, $w=\,$const are the past light cones on ${\cal C}$ ($y=0$), normalised so that $w$ measures proper time along ${\cal C}$, and $y$ measures distance down the light rays $(w,\theta,\phi)=\,$const. A convenient choice for $y$ is $y=z$ (redshift) on the lightcone of here-and-now, $w=w_0$, and then keep $y$ comoving with matter off the initial lightcone, so that $u^y=0$. (This is rather idealised of course, as redshift may not be monotonic, and caustics will form, and so on.) Then the matter 4-velocity and the photon wave-vector are
\begin{equation}\label{}
u^\a=(1+z) (1,0, V^a)\,,~~ k_\a=\partial_{\a}w\,, ~~ 
1+z=u_\a k^\a,
\end{equation}
where $V^a=\mathrm{d} x^a/\mathrm{d} w$ are the transverse velocity components on the observer's sky.
The metric is
\begin{eqnarray}
\mathrm{d} s^2 &=&\! -A^2\mathrm{d} w^2+ 2B\mathrm{d} w \mathrm{d} y+ 2C_a\mathrm{d} x^a \mathrm{d} w+D^2(\!\mathrm{d}\Omega^2+ L_{ab}\mathrm{d} x^a \mathrm{d} x^b )~~ \label{metric} \\
A^2 &=& (1+z)^{-2}+2C_aV^a+ g_{ab}V^a V^b\,,~~~ B={\mathrm{d} v \over \mathrm{d} y},
\end{eqnarray}
where the expression for $A^2$ follows from $u_au^a=-1$; $D$ is the area distance, and $L_{ab}$ determines the lensing distortion of images via the shear of lightrays,
\begin{equation}\label{}
\hat\sigma_{ab}= {D^2 \over 2B} {\p L_{ab} \over \p y}.
\end{equation}
The number of galaxies in a solid angle $\mathrm{d}\Omega$ and a null distance increment $\mathrm{d} y$ is
\begin{equation}\label{}
\mathrm{d} N = Sn(1+z)D^2B \mathrm{d}\Omega \mathrm{d} y\,,
\end{equation}
where $S$ is the selection function and $n$ is the number density.

Before specializing to isotropic observations, we identify how the observations in general and in principle determine the geometry of the past light cone $w=w_0$ of here-and-now, where $y=z$:
\begin{itemize}
\item
Area distances directly determine the metric function $D(w_0,z,x^a)$.
\item
The number counts (given a knowledge of $S$) determine $Bn$ and thus, assuming a knowledge of the bias, they determine $
B(w_0,z,x^a) \rho_m(w_0,z,x^a)$, where $\rho_m=\rho_b+\rho_c$ is the total matter density.
 \item
Transverse (proper) motions in principle directly determine $V^a(w_0,z,x^b)$.
\item
Image distortion determines $L_{ab}(w_0,z,x^c)$. (The differential lensing matrix $\hat\sigma_{ab}$ is determined by $L_{ab}, D,B$.)
\end{itemize}
Then~\cite{7,7a}:
\begin{description}
\item[{\bf \footnotesize [ENMSW]}~{\em  Lightcone observations $\bm\Longrightarrow$ spacetime metric}:]

Observations $(D, N, V^a,L_{ab})$ on the past lightcone $w=w_0$ determine in principle $(g_{ab},u^\a, B\rho_m)$ on the lightcone. This is exactly the information needed for Einstein's equations to determine $B,C_a$ on $w=w_0$, so that the metric and matter are fully determined on the lightcone. Finally, the past Cauchy development of this data determines $g_{ab},u^a,\rho_m$ in the interior of the past lightcone.
\end{description}
If we assume that observations are isotropic, then
\begin{equation}\label{misob}
{\p D \over \p x^a}= {\p N\over \p x^a}=V^a=L_{ab}=0\,.
\end{equation}
Momentum conservation  and the $yy$ field equation then give the following equations on $w=w_0$~\cite{7,7b}:
\begin{eqnarray}
C_a &=& (1+z)^{-1}\int_0^z(1+z)B_{,a} dz \\
B &=&{\mathrm{d} v \over \mathrm{d} z} =  2D'\left[2-\int_0^z(1+z)^2D\mu_m dz\right]^{-1},
\end{eqnarray}
where a prime denotes $\p/\p z$. These imply that $B_{,a}=0=C_a$, so that $\rho_{m,a}=0$ -- and hence the metric and matter are isotropic on $w=w_0$. This can only happen if the interior of $w=w_0$ is isotropic. If observations remain isotropic along ${\cal C}$, then the spacetime is isotropic.

\subsection{Cosmology with spherical symmetry}

Isotropic observations imply spherical symmetry in the presence of dust matter, leading to the Lema\^{i}tre-Tolman-Bondi (LTB) model, or the `LLTB' model if we include $\Lambda$. 
An interesting  explanation for the dark energy problem in cosmology is one where  the underlying geometry of the universe is significantly inhomogeneous on Hubble scales. Spacetimes used in this context are usually LTB models~-- so-called `void models', first introduced in~\cite{MT1}. These models can look like dark energy because we have direct access only to data on our lightcone and so we cannot disentangle temporal evolution in the scale factor from radial variations. The main focus has been on aiming to see if these models can fit the data without $\Lambda$, thus circumventing the dark energy problem. However, they can equally be used with $\Lambda$ to place constraints on radial inhomogeneity, though very little work has been done on this~\cite{Marra:2010pg}. We shall briefly review the LTB dust models, as they illustrate the kind of observations required to constrain homogeneity. 

An inhomogeneous void  may be modelled as a spherically symmetric dust LTB model with metric
\ba
\label{LTBmetric2}
\d s^2 = -\d t^2 + \frac{\apar^2(t,r)}{1-\kapr}\d r^2 + \aperp^2(t,r)r^2\d\Omega^2\,,
\ea
where the radial ($\apar$) and angular ($\aperp$) scale factors are related by
$\apar = \p(\aperp r)/\p r$. The curvature $\kappa=\kappa(r)$ is not constant but is instead a free function. The FLRW limit is $\kappa\to$\,const., and $a_\perp=a_\|$.
The two scale factors define two Hubble rates:
\ba\label{H}
\Hperp= \Hperp(t,r) \equiv \frac{\dot a_\perp}{\aperp},~~~~\Hpar=\Hpar(t,r) \equiv \frac{\dot a_{\|}}{\apar},
\ea
The analogue of the Friedmann equation in this space-time is
then given by
\be
H_{\perp}^2 = \frac{M}{a_{\perp}^3}-\frac{\kappa}{a_{\perp}^2},
\ee
where $M=M(r)$ is another free function of $r$, and the locally measured energy density is 
\be
8 \pi G \rho(t,r) = \frac{(M r^3)_{,r}}{a_{\parallel}a_{\perp}^2 r^2},
\ee
which obeys the conservation equation
\be
\dot{\rho}+ (2 H_{\perp}+H_{\parallel}) \rho =0.
\ee
The acceleration equations in the transverse and radial directions are
\be
\frac{\ddot{a}_{\perp}}{a_{\perp}} = -\frac{M}{2 a_{\perp}^3}
~~~~~\text{and}~~~~
\frac{\ddot{a}_{\parallel}}{a_{\parallel}} = -4 \pi \rho +\frac{M}{a_{\perp}^3}.
\ee
\begin{wrapfigure}[20]{r}[0pt]{0.41\textwidth}
\begin{center}
\includegraphics[width=0.4\textwidth]{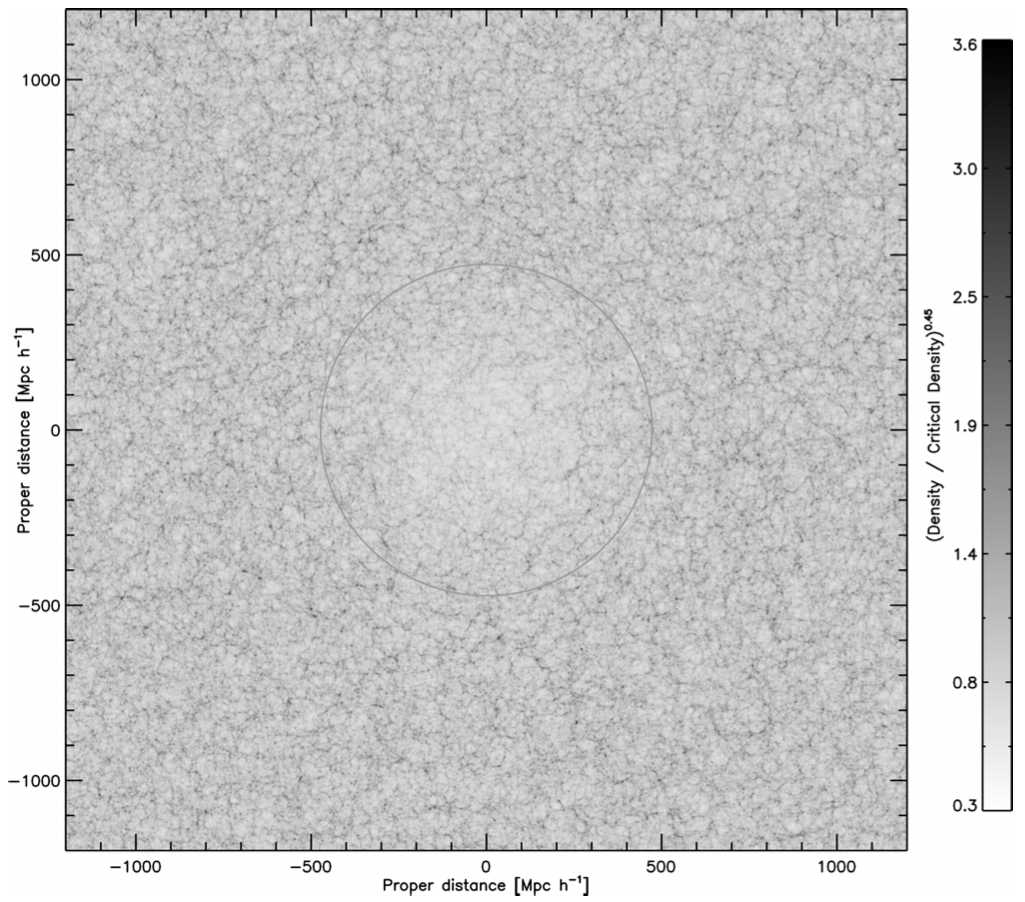}
\caption{A void model produced by a Newtonian N-body simulation. (From~\cite{Alonso:2010zv}.)}
\label{nbody}
\end{center}
\end{wrapfigure}
We introduce dimensionless density parameters for the CDM and curvature, by analogy with the FLRW models:
\be
 \omegko(r)=-\frac{\kappa}{\Hperpo^2},
~~~~~
  \omegmo(r)=\frac{M}{\Hperpo^2 } ,
\ee
using which, the Friedmann equation takes on its familiar form:
\be
\frac{\Hperp^2}{\Hperpo^2}=\omegmo\aperp^{-3} + \omegko \aperp^{-2},
\ee
so $\omegmo(r)+ \omegko(r)=1$. Integrating the Friedmann equation from the time of the big bang $t_B = t_B(r)$ to some later time $t$ yields the age of the universe at a given $(t,r)$:
\be
\label{solA}
\tau(t,r)= t - t_B = \frac{1}{\Hperpo(r)}\int^{\aperp (t,r)}_{0}\frac{\d  x}{\sqrt{\omegmo (r)x^{-1} + \omegko (r)}}\,.
\ee 
We now have two free functional degrees of freedom: $\omegmo(r)$ and $t_B(r)$, which can be specified as we wish (if the bang time function is not constant this represents a decaying mode if one tries to approximate the solution as perturbed FLRW~\cite{1977A&A....59...53S}). A coordinate choice which fixes $a_\perp(t_0,r)=1$ then fixes $H_{\perp_0}(r)$ from Eq.~(\ref{solA}). A value for $H_0=H_{\perp_0}(r=0)$ is used at this point.

The LTB model is actually also a Newtonian solution~-- that is, Newtonian spherically symmetric dust solutions have the same equations of motion~\cite{Szekeres:1995gy}. This was demonstrated explicitly in an N-body simulation of a void~\cite{Alonso:2010zv} (see Fig.~\ref{nbody}).

\subsection{Background observables}

In LTB models, there are several approaches to finding observables such as distances as a function of redshift~\cite{Mustapha:1997xb}. We refer to~\cite{Enqvist:2007vb} for details of the approach we use here. On the past light cone a central observer may write the $t,r$ coordinates as functions of $z$. These functions are determined by the system of differential equations
\ba
\label{dtdz}
\frac{\d t}{\d z} = -\frac{1}{(1+z)\Hpar}\,, ~~~~~
\label{drdz}
\frac{\d r}{\d z} = \frac{\sqrt{1-\kappa r^2}}{(1+z)\apar\Hpar},
\ea
where $H_\|(t,r)=H_\|(t(z),r(z))=H_\|(z)$, etc. The area distance is given by
\be
d_A(z)=a_\perp(t(z),r(z)) r(z)
\ee
and the luminosity distance is, as usual $d_L(z)=(1+z)^2 d_A(z)$. The volume element as a function of redshift is given by
\be
\frac{\d V}{\d z}=\frac{4\pi d_A(z)^2}{(1+z)H_\|(z)}\,.
\ee
This then implies the number counts as a function of redshift, provided the bias and mass functions are known; if not, there is an important degeneracy between source evolution when trying to use number counts to measure the volume element~\cite{MHE,Hellaby:2000ef}.

\begin{figure}[htbp]
\begin{center}
\begin{tabular}{l|c|r}
\includegraphics[width=0.32\textwidth]{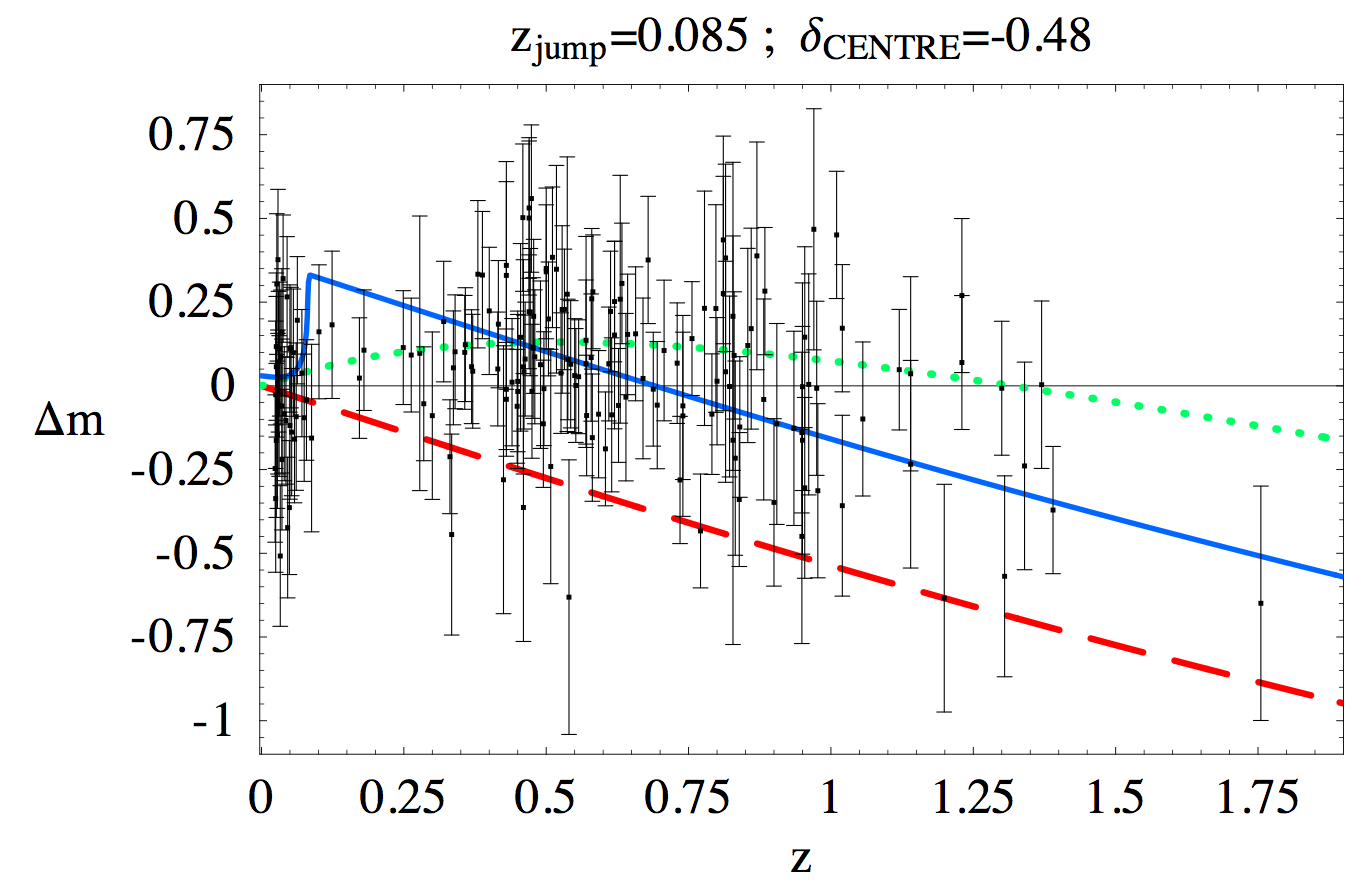}
&
\includegraphics[width=0.32\textwidth]{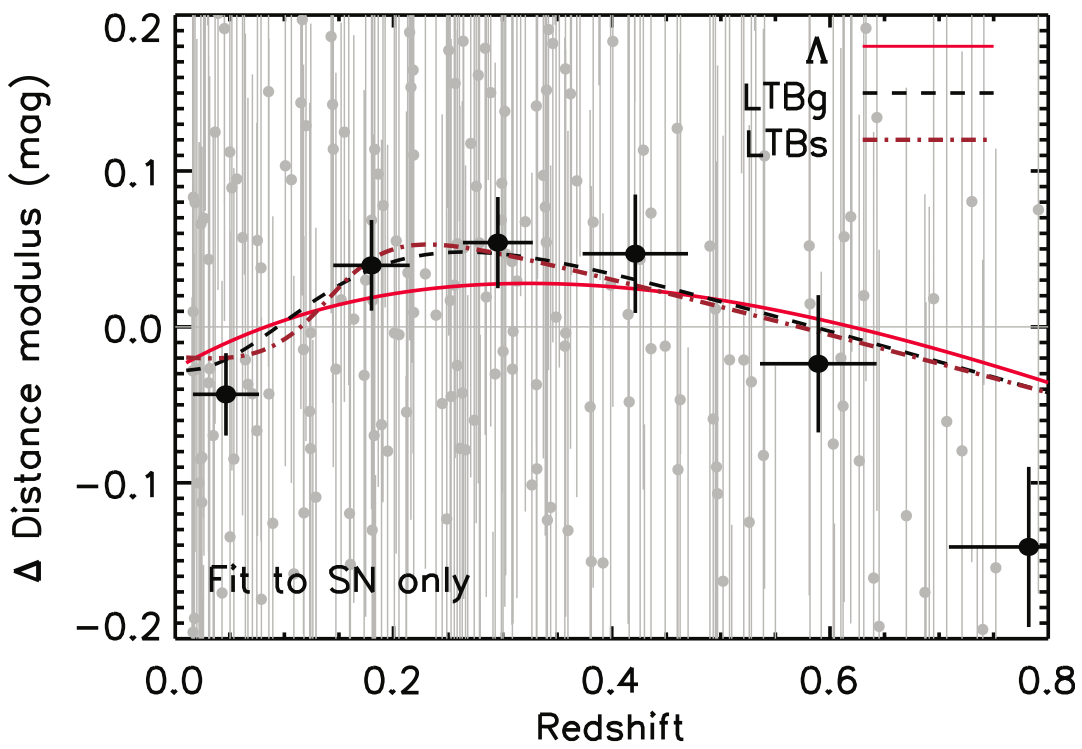}
&
\includegraphics[width=0.32\textwidth]{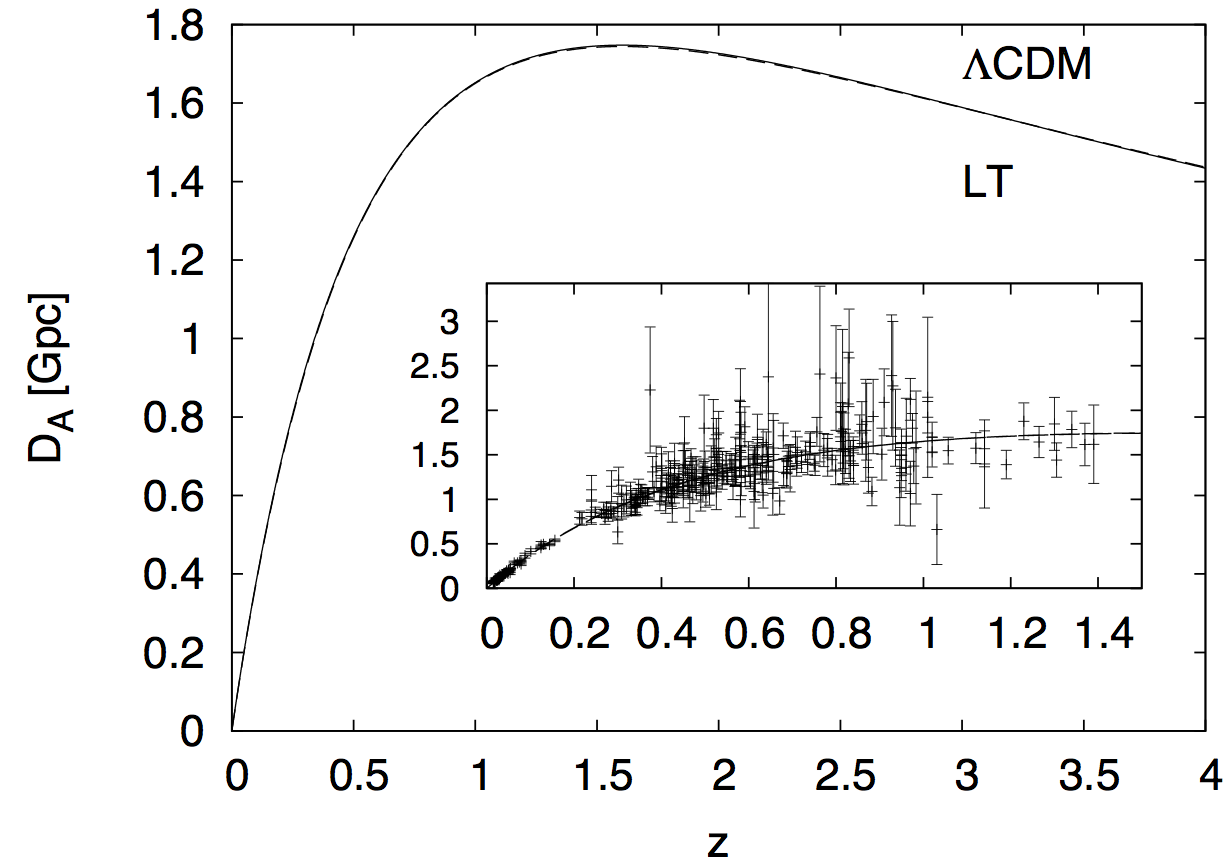}\\[-0mm]
\includegraphics[width=0.32\textwidth]{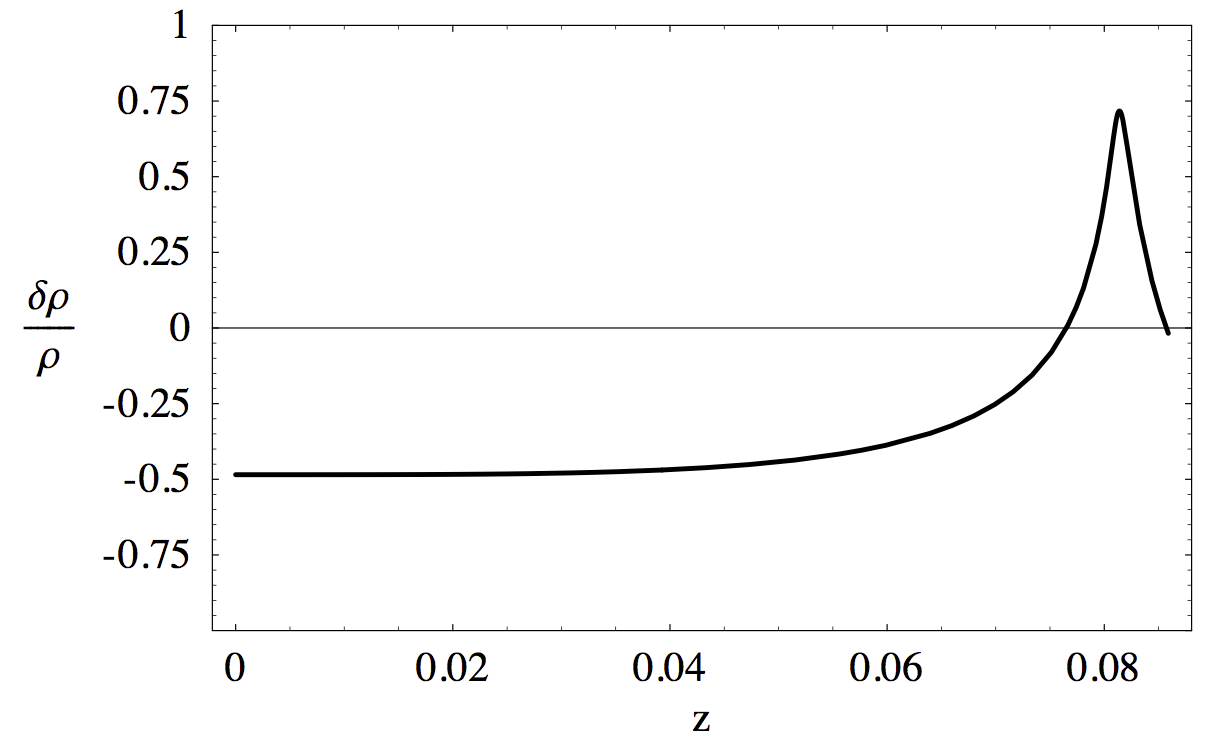}
&
\includegraphics[width=0.35\textwidth]{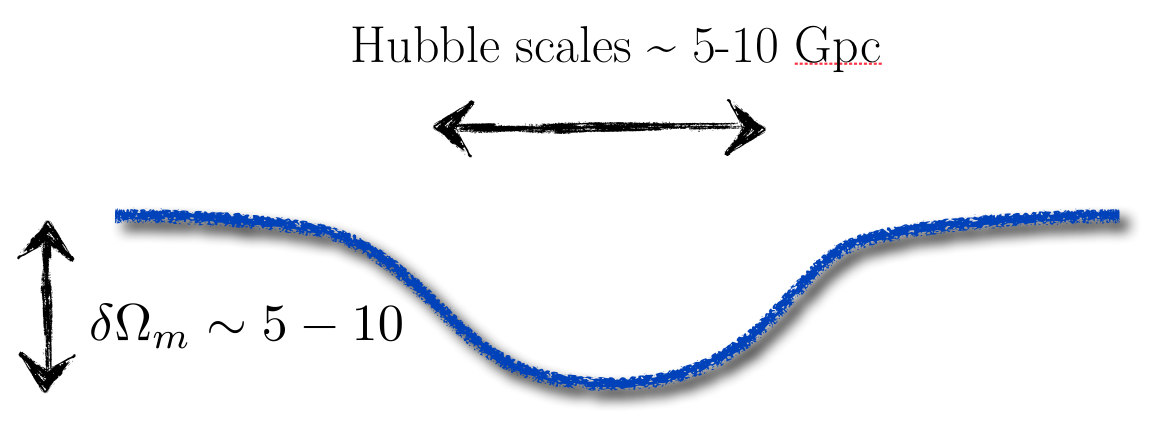}
&
\includegraphics[width=0.32\textwidth]{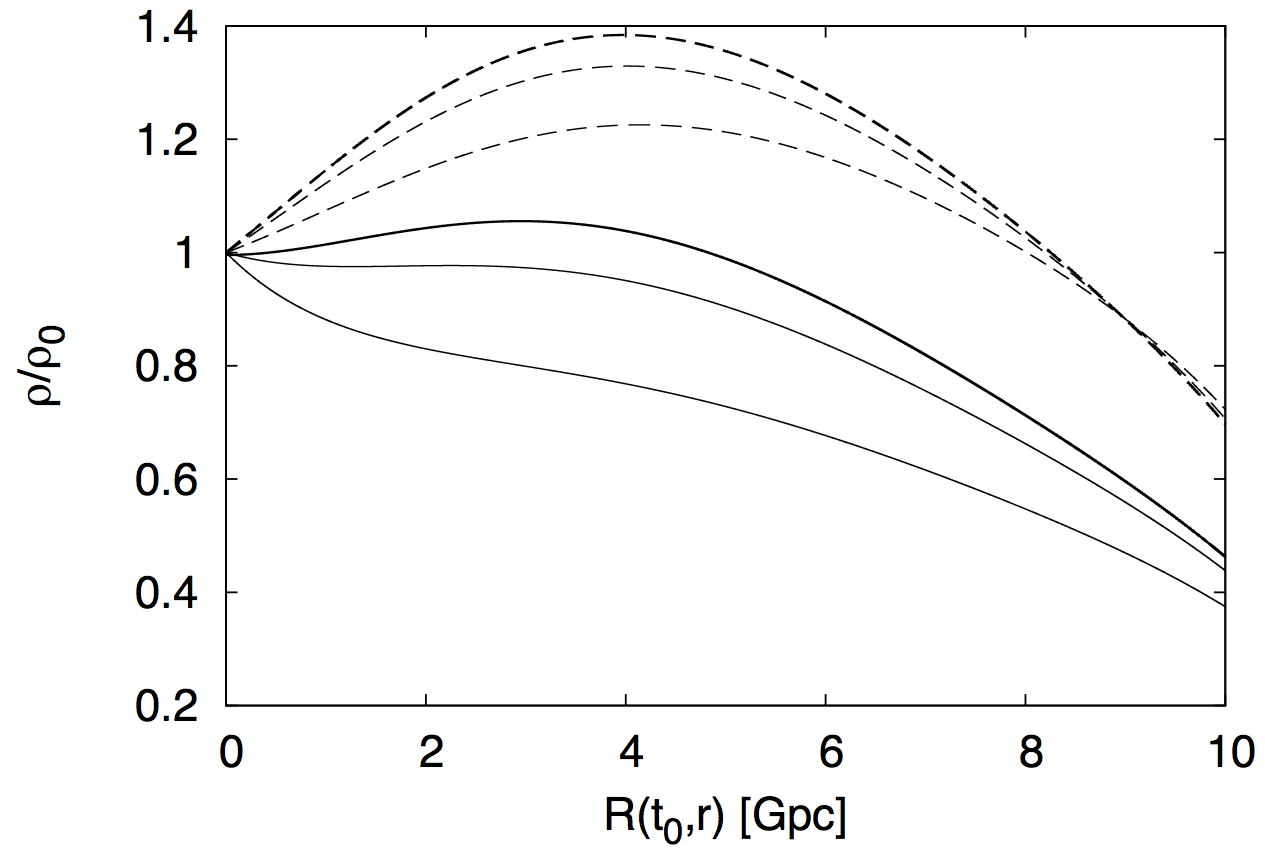}
\end{tabular}
\caption{LTB models have no problem fitting distance data. Left is an attempt to fit the early SNIa data of~\cite{Riess:2006fw}, using a very small void with an over-dense shell around it embedded in an EdS model, from~\cite{ABNV}. The hope was that we could be located in the sort of voids we observe all over the place, giving a jump in the distance modulus which can then fit the SNIa. The gap in the data at intermediate redshift was filled by the SDSS SNIa~\cite{Sollerman:2009yu} which ruled these out, leaving the possibility of giant voids several Gpc across with a Gaussian density profile as an alternative to $\Lambda$CDM (centre, top), with approximate dimensions shown (below). If the bang time function is non-zero then the data do not constrain the density to be a void profile (right); \cite{Celerier:2009sv} show that a central over-density can fit the SNIa data of~\cite{Kowalski:2008ez}.}
\label{snia}
\end{center}
\end{figure}

With one free function we can design models that give any distance modulus we like~(see e.g., \cite{MT2,Mustapha:1997xb,MHE,Celerier:1999hp,Chung:2006xh,AAG,garfinkle,BMN,bolejko,Enqvist:2007vb,GarciaBellido:2008nz,YKN,CFL,FLSC}). In Fig.~\ref{snia} we show a selection of different models which have been considered recently. 
Generically, those give rise to `void models': with the bang time function set to zero
and we choose $\omegmo(r)$ to reproduce exactly a $\Lambda$CDM $D(z)$, then the LTB model is a void with steep radial density profile, but a Gaussian profile fits the SNIa data just as well~\cite{FLSC}.

\subsection{{The small-scale CMB \& $H_0$}}

The physics of decoupling and line-of-sight effects contribute differently to the CMB, and have different dependency on the cosmological model. In sophisticated inhomogeneous models both pre- and post-decoupling effects will play a role, but Hubble-scale void models allow an important simplification for calculating the moderate to high $\ell$ part of the CMB.

The comoving scale of the voids which closely mimic the $\Lambda$CDM distance modulus are typically $O(\mbox{Gpc})$. The physical size of the sound horizon, which sets the largest scale seen in the pre-decoupling part of the power spectrum, is around $150\,$Mpc redshifted to today.
This implies that in any causally connected patch of the Universe prior to decoupling, the density gradient is very small. Furthermore, the comoving radius of decoupling is larger than $10\,$Gpc, on which scale the gradient of the void profile is small in the simplest models (or can be by assumption). For example, at decoupling the total fractional difference in energy density  between the centre of the void and the asymptotic region is around 10\%~\cite{Clarkson:2010ej}; hence, across a causal patch we expect a maximum 1\% change in the energy density in the radial direction, and much less at the radius of the CMB that we observe for a Gaussian profile. This suggests that before decoupling on small scales we can model the universe in disconnected FLRW shells at different radii, with the shell of interest located at the distance where we see the CMB. This may be calculated using standard FLRW codes, but with the line-of-sight parts corrected for~\cite{ZMS,CFZ}. The calculation for the high-$\ell$ spectrum was first presented in~\cite{ZMS}, and further developed in~\cite{CFZ,RC,Yoo:2010qy,Clarkson:2010ej}.

For line-of-sight effects, we need to use the full void model. These come in two forms. The simplest effect is via the background dynamics, which affects the area distance to the CMB, somewhat similar to a simple dark energy model. This is the important effect for the small-scale CMB. The more complicated effect is on the largest scales through the Integrated Sachs-Wolfe effect (see~\cite{Tomita:2009wz} for the general formulas in LTB).  This requires the solution of the perturbation equations presented below, and has not been addressed.

The CMB parameters (an asterisk denotes decoupling)
\ba
l_a=\pi \frac{d_A(z_{\dec})}{a_{\dec}r_s(a_{\dec})},\label{shift}~~~~~
l_{\eq}=\frac{k_{\eq}d_A(z_{\dec})}{a_{\dec}}= \frac{T_\dec}{T_{\eq}}\frac{d_A(z_\dec)}{H_{\eq}^{-1}}, \label{shift2}~~~~~
R_*=\left.\frac{3\rho_b}{4\rho_\gamma}\right|_{\dec}\,, \label{shift3}
\ea
are sufficient to characterise the key features of the first three peaks of the CMB~\cite{Hu:2001bc,Wang:2007mza} (see also~\cite{Vonlanthen:2010cd}). Within a standard thermal history, the physics that determines the first three peaks also fixes the details of the damping tail~\cite{Hu2008}. With the exception of $d_A(z_{\dec})$, all quantities are local to the surface of the CMB that we observe. 

\begin{figure}
~~~~~~~\includegraphics[width=0.4\columnwidth]{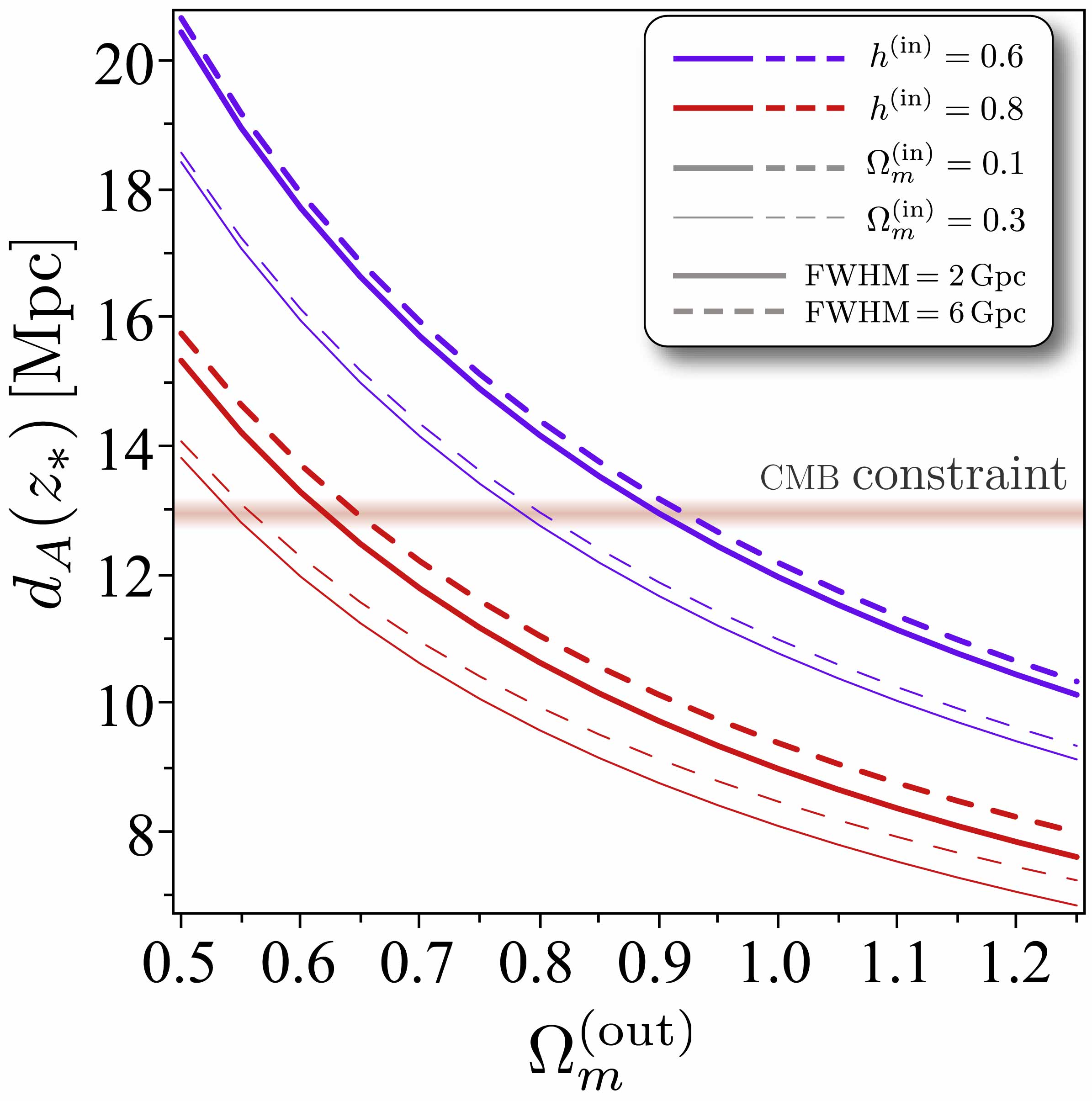}\hfill
\includegraphics[width=0.4\columnwidth]{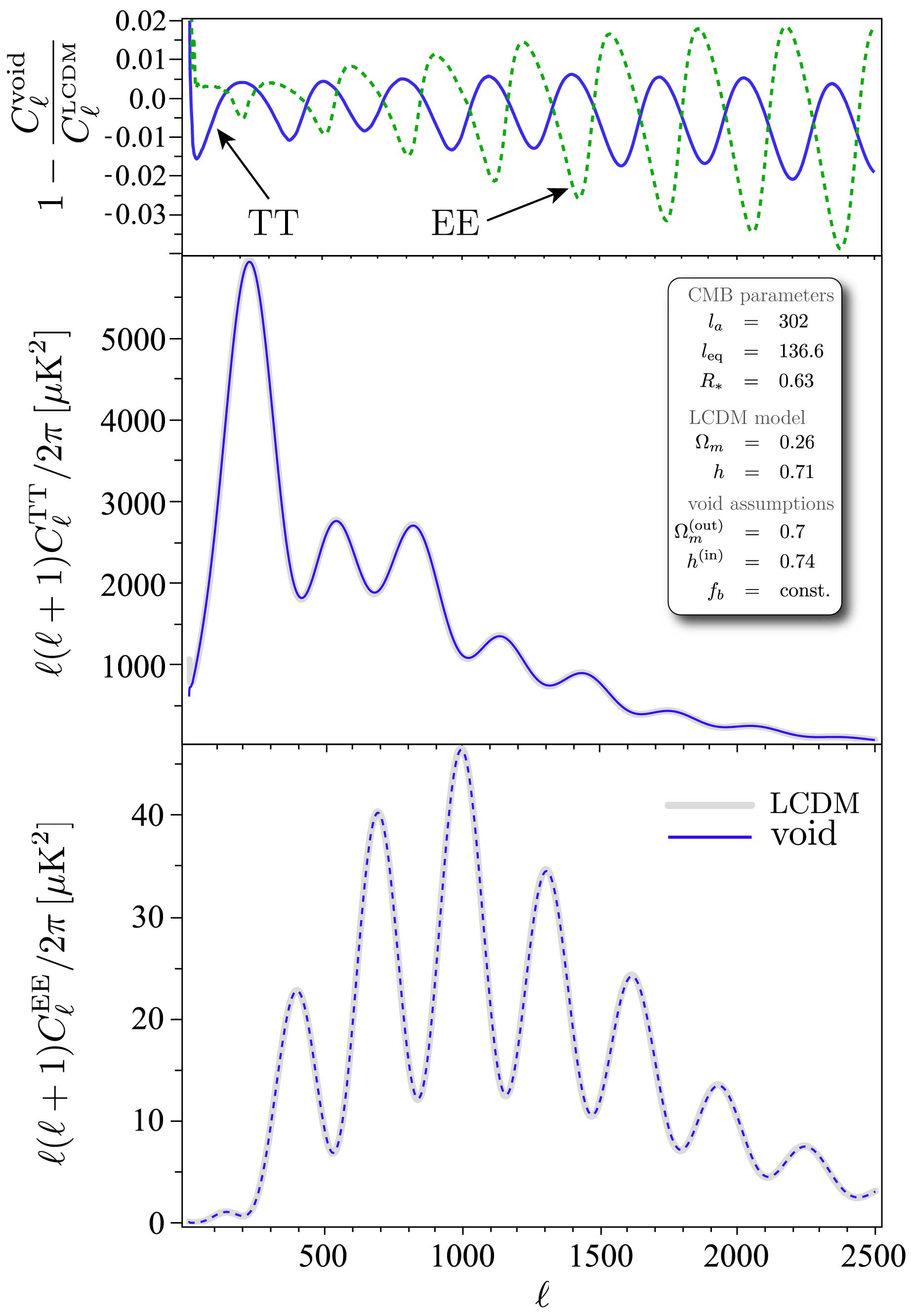}~~~~~~~
\caption{Left: The area distance to $z\sim1090$ in a Gaussian-profiled LTB void model with zero bang time. Adding bumps to the density profile changes this figure considerably. Whether the model can fit the CMB lies in the freedom of the value of $H_0$ for a measured $T_0$ and $z_*$. The simplest models require $h\sim0.5$.\\  Right: The normalised CMB angular power spectrum. The power spectrum is shown against a default flat concordance model with zero tilt. There is nothing between the two models for high $\ell$, with the maximum difference around 1\%. (From~\cite{Clarkson:2010ej})}
\label{cmb}
\end{figure}

The parameters given by Eqs.~(\ref{shift}) separate  the local physics of the CMB from the line-of-sight area distance. These parameters can be inverted to provide nearly uncorrelated constraints on $d_A(z_*)$, $f_b$ and $\eta$.
 Specifying asymptotic void model parameters to give the measured value of $R_*$ and $l_a/l_\eq$, leaves just the area distance of the CMB to be adjusted to fit the CMB shift parameters. This constrains a combination of the void profile and the curvature and Hubble rate at the centre. 

A final constraint arises when we integrate out along the past lightcone from the centre out to $z_\dec$. In terms of time it says that the local time at that $z_\dec$ must equal the time obtained by integrating up along the timelike worldline from the big bang up to decoupling. That is,
\be
t_0-t_\dec(z_\dec)=\int_0^{z_{\dec}}\frac{\d z}{1+z}\left.\left({\partial_t\ln\chi}\right)^{-1}\right|_{\text{nullcone}}\,,\label{glue}
\ee
where $\chi(t,r)=-\d t/\d r$ evaluated on the past nullcone, and $t_\dec(z_\dec)$ is the local time of decoupling at the redshift observed from the centre, which must be equal to 
\be
t_\dec=\int_{T_\dec}^\infty \frac{\d T}{T}\frac{1}{H(T)}\,,
\ee  
where the Hubble rate as a function of temperature, $H(T)$, is given locally at early times by 
\be\label{H(T)}
\frac{H(T)}{100\,\hu}=\sqrt{(\varpi_\gamma+\varpi_\nu)T^4+\varpi_b\frac{\eta}{f_b} T^3},
\ee
which also only has dependence on the local parameters $\eta$ and $f_b$ and with no reference to late times. We have defined the dimension-full constants 
\bea\label{varpis}
\varpi_\gamma=\frac{\Omega_\gamma h^2}{T_0^4}
\approx\left(\frac{0.02587}{1\,\text{K}}\right)^4,~~~
\varpi_\nu=\frac{\Omega_\nu h^2}{T_0^4}\approx0.227 N_\text{eff}\varpi_\gamma,~~~
\varpi_b=\frac{\Omega_b h^2}{\eta T_0^3}=\frac{30\zeta(3)}{\pi^4}m_p\varpi_\gamma \,. \label{varpi_b}
\eea
Note that these have {no dependence on any parameters of the model}. That is, we are not free to specify the $\varpi$'s, apart from $N_\text{eff}$. These are derived assuming that $f_b$ and $\eta$ are constant in time. An example from~\cite{Clarkson:2010ej} of how closely a void model can reproduce the CMB power spectrum found in a concordance model is shown in Fig.~\ref{cmb}.

In~\cite{ZMS,CFZ,BNV}, it was shown that the CMB can be very restrictive for adiabatic void models (i.e., those with $\eta=$\,const. spatially at early times) when the bang time is zero, the power spectrum has no features, and the universe is assumed to evolve from a homogeneous model. We can see this as follows. For a Gaussian profiled void with $\Omega_m\sim0.1$ at the centre, the area distance to the CMB favours a low $\Omega_m$ asymptotically, or else a low $H_0$ at the centre (see Fig.~\ref{cmb}). Thus, an asymptotically flat model needs $H_0\sim 50\,\hu$ to get the area distance right. Then, if the constraint, Eq.~(\ref{glue}), is evaluated either in an LTB model, or by matching on to an FLRW model, it is found that the asymptotic value of the density parameter must be high. Thus, in this approximation, we see that the CMB favours models with a very low $H_0$ at the centre to place the CMB far enough away. The difficulty fitting the CMB may therefore be considered one of fitting the local value $H_0$~\cite{BNV}, which is quite high.
However,~\cite{CFZ} showed that with a varying bang time, the data for $H_0$, SNIa and CMB can be simultaneously accommodated. This is because the constraint, Eq.~(\ref{glue}), must be modified by adding a factor of the difference between the bang time at the centre with the bang time asymptotically, so releasing the key constraint on $H_0$.

It was argued in~\cite{Clarkson:2010ej}  that Eq.~(\ref{glue}) can be accommodated by an $\mathcal{O}(1)$ inhomogeneity in the radiation profile, $\eta=\eta(r)$, at decoupling which varies over a similar scale as the matter inhomogeneity~-- giving an `isocurvature' void. The reason is because the constraint is sensitive to $t_*/t_0\sim10^{-5}$, which is mirrored by a sensitivity in $z_*$ at around the 10\% level when $z_*\sim1090$. Thus~\cite{Clarkson:2010ej} argue that a full two-fluid model is required in order to decisively evaluate Eq.~(\ref{glue}) in this case and so provide accurate constraints on isocurvature voids, though it remains unknown if this gives enough freedom to raise $H_0$ sufficiently. 

An important alternative solution, presented in~\cite{huntsarkar,Nadathur:2010zm}, is to add a bump to the primordial power spectrum around the equality scale. This then allows perfectly acceptable fits to the key set of observables $H_0$+SNIa+CMB. This is particularly important because to produce a void model in the first place, at least one extra scale must be present during inflation (or whatever model is used) so it is unreasonable to assume a featureless power spectrum which is the case for all other studies.  

Although the simplest models do not appear viable because their local $H_0$ is far too low~\cite{ZMS,CFZ,BNV}, it is still an open question exactly what constraints the small-scale CMB places on a generic void solution.

\subsection{Scattering of the CMB}\label{SZ}

The idea of using the CMB to probe radial inhomogeneity on large scales was initiated in a seminal paper by Goodman~\cite{Goodman:1995dt}. In essence, if we have some kind of cosmic mirror with which to view the CMB around distant observers we can measure its temperature there and constrain anisotropy of the CMB about them, and so constrain the type of radial inhomogeneity prevalent in void models. Observers in a large void typically have a large peculiar velocity with respect to the CMB frame, and so see a large dipole moment, as well as higher multipoles; that is, observers outside the void will see a large patch of their CMB sky distorted by the void. 

There are several mechanisms by which the temperature of the CMB can be measured at points away from us. The key probe is using the Sunyaev-Zel'dovich effect~\cite{1972CoASP...4..173S,1980MNRAS.190..413S}, where the CMB photons are inverse Compton scattered by hot electrons, typically in clusters. There are two effects from this: it heats up the CMB photons by increasing their frequency which distort the CMB spectrum; and it scatters photons into our line of sight which would not otherwise have been there. This changes the overall spectrum of the CMB in the direction of the scatterer because the primary CMB photons become mixed up with the photons from all over the scatterers' sky. If the scatterer measures an anisotropic CMB~-- so that their CMB temperature is different along different lines of sight~-- then this distorts the initially blackbody spectrum which is scattered into our line of sight. (The sum of two blackbodies at different temperature is not a blackbody.)

\begin{figure}[t]
\begin{center}
\begin{tabular}{lr}
\includegraphics[width=0.35\columnwidth]{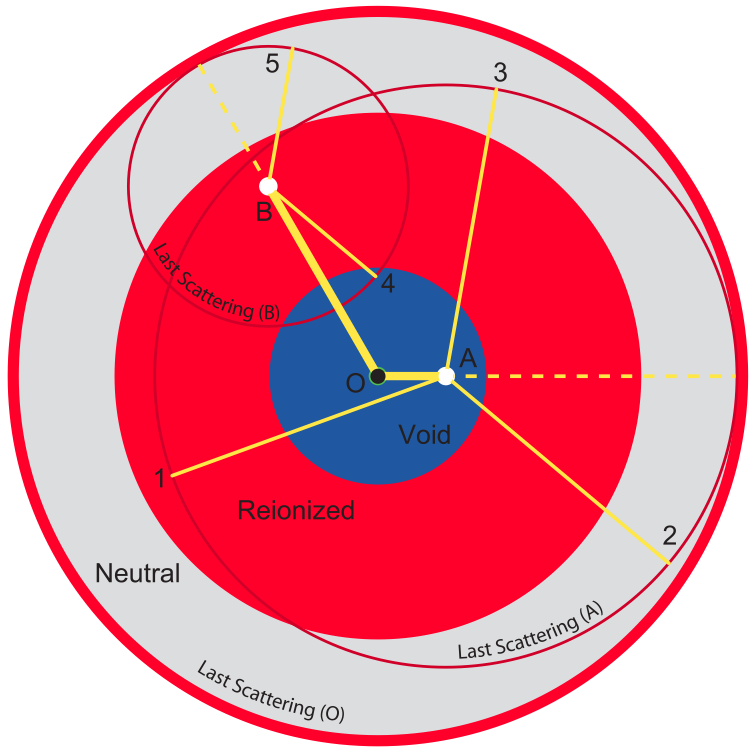}\hfill
&~~~~~~~~~\hfill\includegraphics[width=0.45\columnwidth]{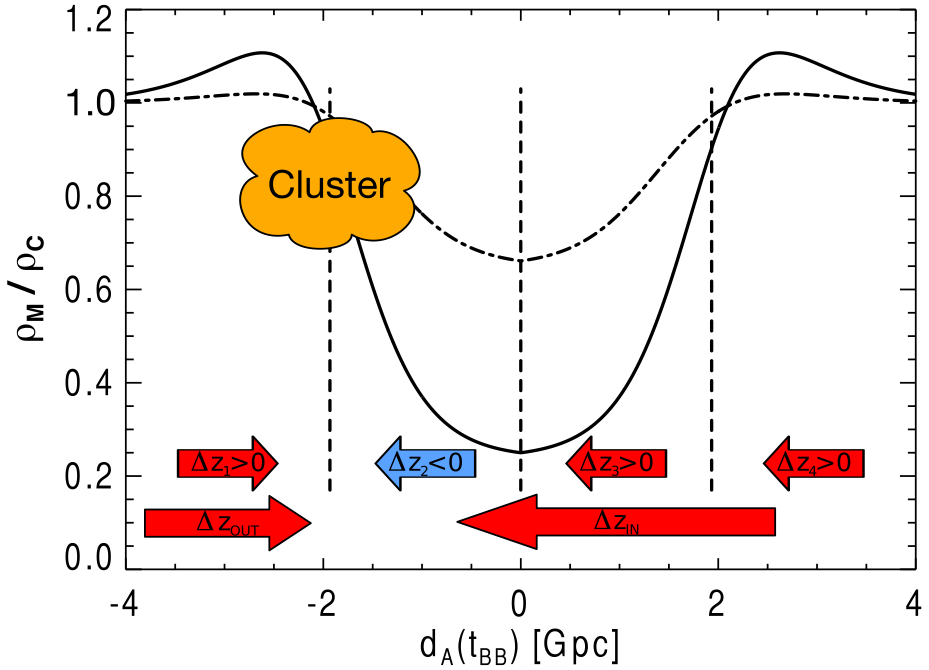}
\\
\includegraphics[width=0.42\columnwidth]{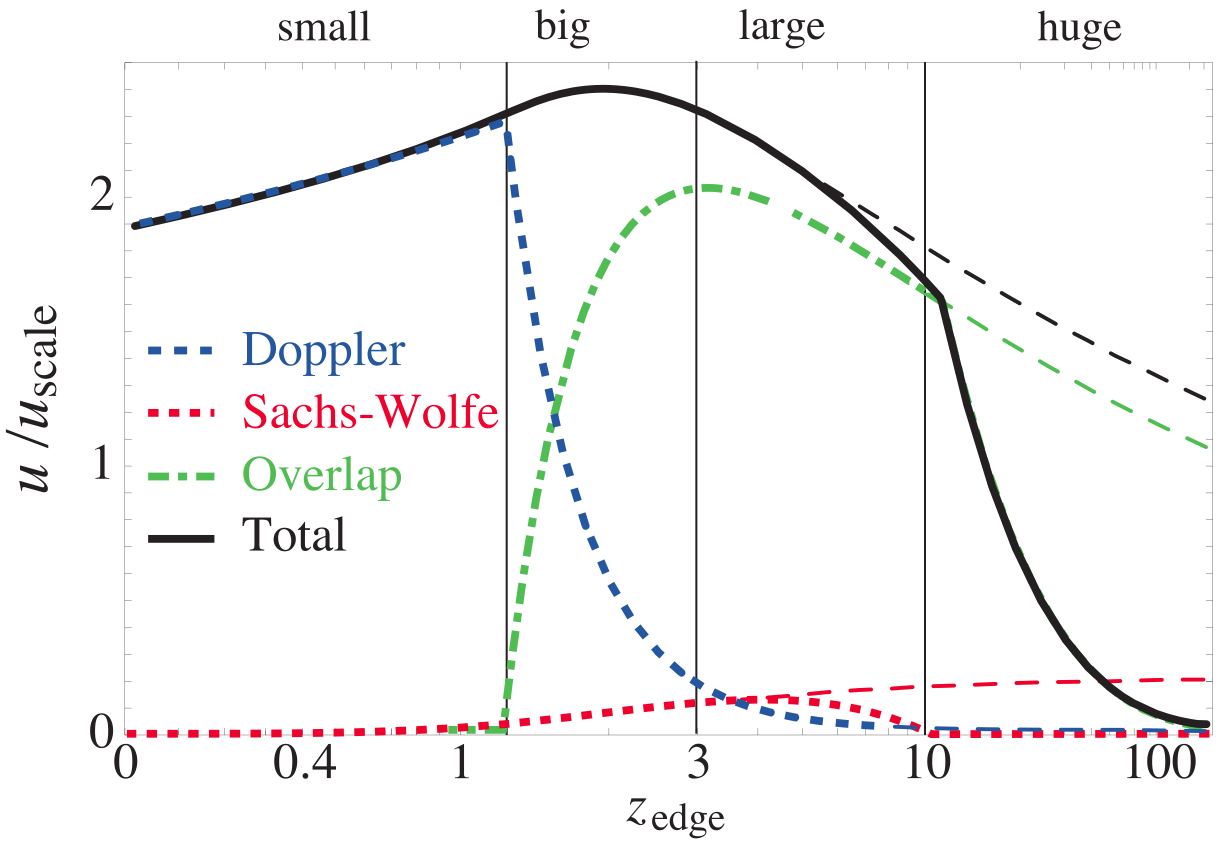}\hfill
&~~~~~~~~~\hfill\includegraphics[width=0.45\columnwidth]{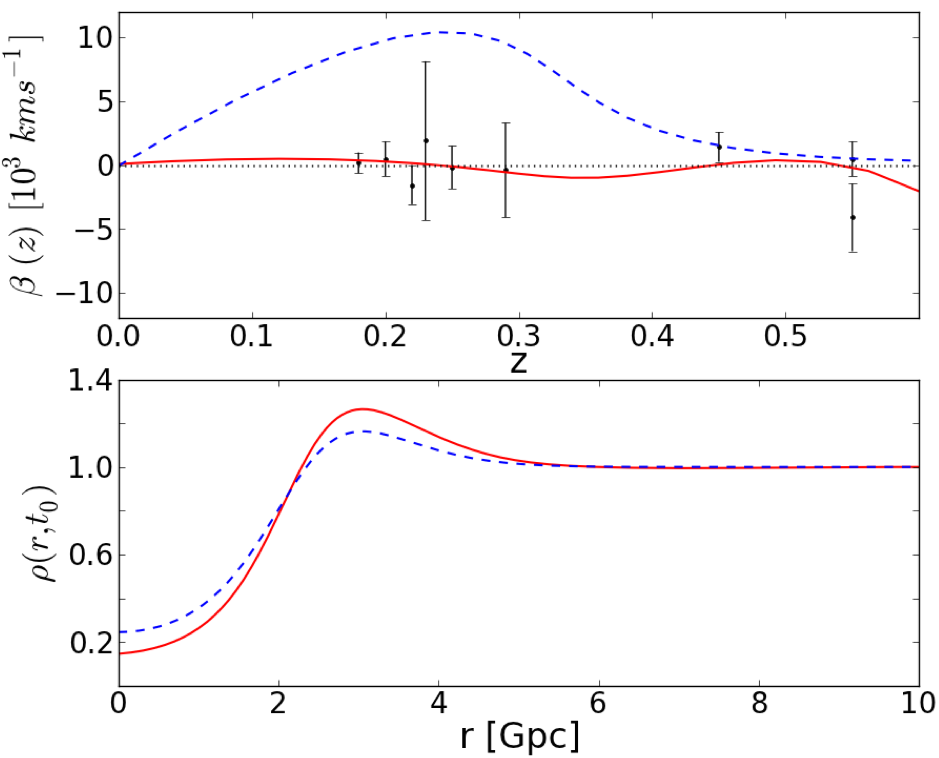}
\end{tabular}
\caption{Off-centre observers typically see an anisotropic CMB sky (top). This causes a large distortion in the CMB spectrum (bottom left). The kSZ effect allows us to infer the radial velocities of clusters which have a systematic drift in void models. In models with a homogeneous bang time, this is large (middle panel, right, dashed line) and already ruled out by present constraints~\cite{GarciaBellido:2008gd,Garfinkle:2009uf}; a small inhomogeneous bang time can drastically reduce this effect (middle panel, right, solid line) without changing the late time void (bottom right), and an isocurvature mode can do the same sort of thing~\cite{Yoo:2010ad}. (Figures from: left~\cite{Caldwell:2007yu}; top right~\cite{GarciaBellido:2008gd}; bottom right~\cite{Bull:2011wi}.)}
\label{ksz}
\end{center}
\end{figure}

The altered spectrum from scattering of CMB photons into our line of sight has two main contributions. The contribution from all the anisotropies at a cluster produces a so-called $y$-distortion  to the spectrum~\cite{Goodman:1995dt,Caldwell:2007yu,Stebbins:2007ve}. Adding up all sources, assuming single scattering results in a distortion proportional to
\be
y\propto\int_0^\infty\d z\frac{\d\tau}{\d z}\int\d^2\bm n'(1+\bm n\cdot\bm n')\left[\:{\Delta T}{T}(\bm n,\bm n, z)-\:{\Delta T}{T}(\bm n',\bm n, z)\right]^2
\ee
where $\:{\Delta T}{T}(\bm n',\bm n, z)$ is the CMB temperature anisotropy in the direction $\bm n'$ at the cluster located at $z$ in direction $\bm n$ according to the central observer. $\tau(z)$ is the optical depth. (Note that the thermal SZ effect also produces a $y$-distortion too, but through the monopole temperature rather than temperature anisotropies.)
The other contribution is from the kinetic SZ effect. This is an effect caused by the bulk motion of a cluster relative to the CMB frame~\cite{GarciaBellido:2008gd,Garfinkle:2009uf,Zhang:2010fa,Yoo:2010ad,Moss:2011ze,Bull:2011wi}. 
An observer looking in direction~$\bm n$ then sees a temperature fluctuation 
\be
\frac{\Delta T(\bm n)}{T_0}=\int_0^\infty\d z\frac{\d\tau}{\d z} v_r(z)\delta_e(\bm n,z)
\ee
where $\delta_e$ is the density contrast of electrons along the line of sight. In a void model clusters will have a systematic radial velocity $v_r(z)$ which can place constraints on the model.  In addition, given a primordial power spectrum and a way to evolve perturbations, the angular power spectrum associated with a continuum $\delta_e(\bm n,z)$ may be evaluated as a correction to the usual CMB $C_\ell$'s.

Constraints are impressive. All studies which consider a homogeneous early universe find them severely constrained using these effects~\cite{Goodman:1995dt,Caldwell:2007yu,GarciaBellido:2008gd,Garfinkle:2009uf,Moss:2010jx,Zhang:2010fa,Moss:2011ze}, and effectively rule out such models. 
However, it was shown in \cite{Bull:2011wi} that a bang time function with amplitude of order the decoupling time $t_*$ can be used to tune out the kSZ signal by fine tuning the peculiar velocity on the past lightcone. Removing the off-centre dipole in this way will weaken $y$-distortion constraints as well.   They found, however, that models which did this could not fit the CMB+$H_0$ constraints which requires a Gyr-amplitude bang time in their analysis. (Such a large amplitude bang time induces a huge $y$-distortion which is ruled out in FLRW~\cite{Zibin:2011ma}.) It was argued in~\cite{Yoo:2010ad,Clarkson:2010ej} that because the temperature of decoupling is not in general spatially constant that this should also be used to investigate these constraints, and will weaken them considerably; in this interpretation these constraints are really measurements of $f_b(r)$ and $\eta(r)$. Finally, the integrated kSZ constraints~\cite{Zhang:2010fa,Moss:2011ze} rely on structure formation and an unknown radial power spectrum, and so have these additional degrees of freedom and problems to consider.

Nevertheless, scattering of the CMB provides stringent constraints on the simplest voids, and show that for a void model to be a viable model of dark energy it will have to have inhomogeneity present at early times as well. Furthermore, if one is to tune a model to have vanishing dipole on the past lightcone of the central observer, this will typically be possible only for one  lightcone, and hence for one instant for the central observer. This will add further complications to the fine tuning problems of the models.

\subsection{Big Bang Nucleosynthesis and the Lithium problem}

Big-Bang nucleosynthesis (BBN) is the most robust probe of the first instants of the post-inflationary Universe. After three minutes, the lightest nuclei (mainly D, $^3$He, $^4$He, and $^7$Li) were synthesised in observationally significant abundances~\cite{Steigman:2007xt,Iocco:2008va}. Observations of these abundances provide powerful constraints on the primordial baryon-to-photon ratio $\eta=n_b/n_\gamma$, which is constant in time during adiabatic expansion. In the $\Lambda$CDM model, the CMB constrains $\eta_{CMB}=6.226\pm0.17\times10^{-10}$~\cite{Iocco:2008va} at a redshift $z\sim1100$. Observations of high redshift low metallicity quasar absorbers tells us D/H\,$=(2.8\pm0.2)\times 10^{-5}$~\cite{Pettini:2008mq} at $z\sim3$, which in standard BBN leads to $\eta_D=(5.8\pm0.3)\times10^{-10}$, in good agreement with the CMB constraint. 
In contrast to these distant measurements at $z\sim 10^3$ and $z\sim3$, primordial abundances at $z=0$ are either very uncertain (D and $^3$He), not a very sensitive baryometer ($^4$He), or, most importantly, in significant disagreement with these measurements~-- $^7$Li.
To probe the BBN yield of $^7$Li, observations have concentrated on old metal-poor stars in the Galactic halo or in Galactic globular clusters.
The ratio between $\eta_{Li}$ derived from $^7$Li at $z=0$ and $\eta_D$ derived from $D$ at $z\sim3$ is found to be $\eta_D/\eta_{Li}\sim 1.5$.  Within the standard model of cosmology, this anomalously low value for $\eta_{Li}$ disagrees with the CMB derived value by up to 5-$\sigma$~\cite{Cyburt:2008kw}.

\begin{SCfigure}
\includegraphics[width=0.7\textwidth]{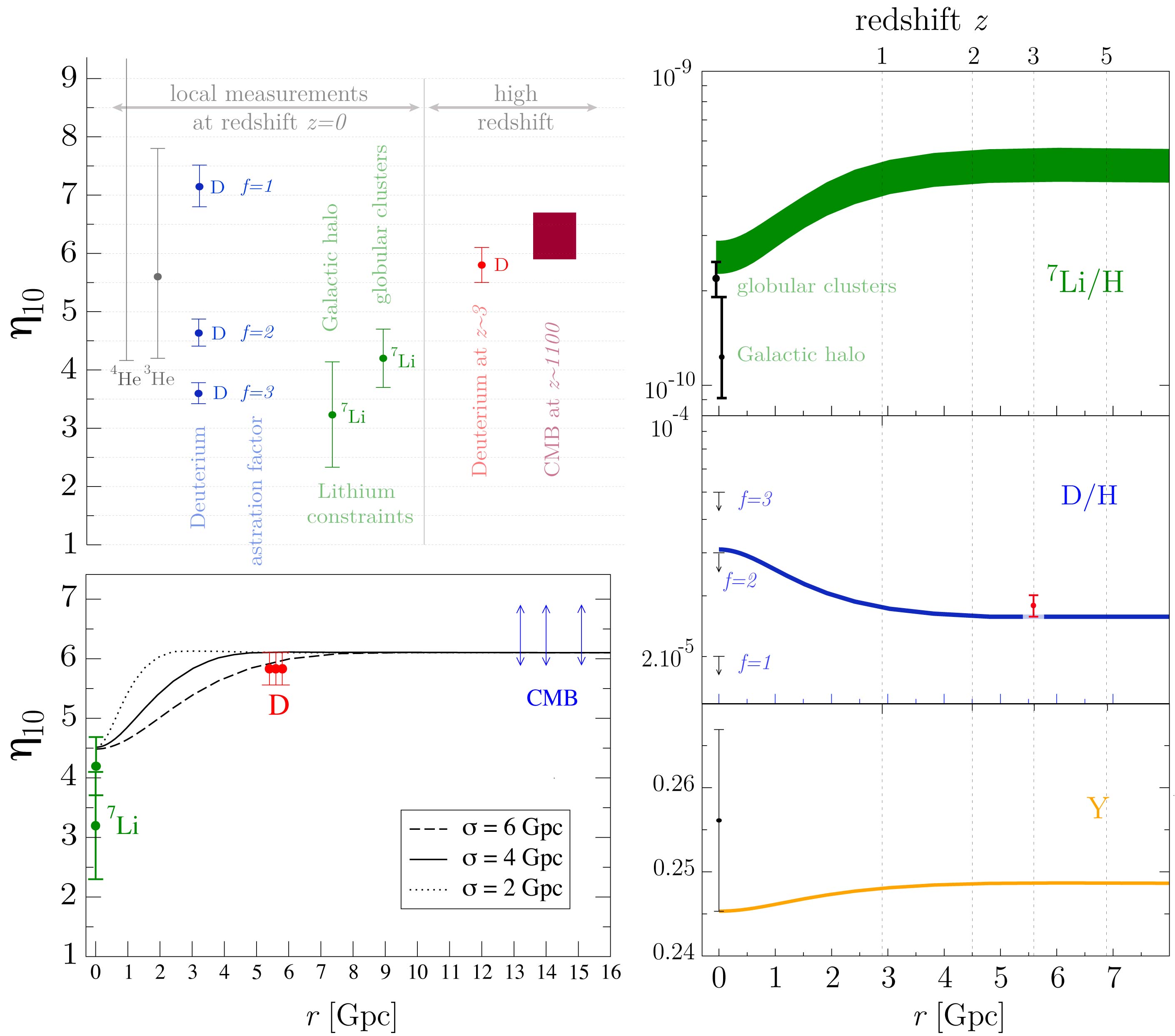}
\caption{Top left are recent constraints on $\eta_{10}=10^{10}\eta$ from different observations. Constraints from $^7$Li observations~\cite{Cyburt:2008kw} in Galactic globular clusters and Galactic halo are shown separately, alongside $^4$He~\cite{aver} and $^3$He~\cite{Steigman:2007xt}. These agree with each other if $\eta_{10}\sim 4.5$. 
On the other hand, D observations at high redshift (red)~\cite{Pettini:2008mq} and CMB require $\eta_{10}\simeq 6$. Bottom left we show how a varying radial profile for $\eta_{10}$ (from $\sim4.5$ at the centre to $\sim6$ asymptotically) can fit all the observational constraints, for differing inhomogeneity scales. On the right are the nuclei abundances as a function of $z$ in an example model. This model may be considered an `isocurvature void' model. (From~\cite{RC}.)}
\label{fig:eta}
\end{SCfigure}

A local value of $\eta\sim 4-5\times10^{-10}$ is consistent with all the measurements of primordial abundances at $z=0$, however (see top left panel in Fig.~\ref{fig:eta}). The disagreement with high-redshift CMB and D data (probing $\eta$ at large distances) shows up only when $\eta$ is assumed to be homogeneous on super-Hubble scales at BBN, as in standard cosmology. An inhomogeneous radial profile for $\eta$ can thus solve the $^7$Li problem, shown in Fig.~\ref{fig:eta}~\cite{RC}.

\subsection{The BAO}

During the process of recombination, when Compton scattering of the electrons and photons becomes low enough the baryons become freed from the photons. This `drag epoch', when the baryons are no longer dragged by the photons, happens at a temperature $T_d$. The size of the sound horizon at this time is consequently imprinted as a bump in the two-point correlation function of the matter at late times. Assuming an FLRW evolution over the scale of the horizon at this time, this the proper size of the sound horizon at the drag epoch is approximately given by (assuming $N_\text{eff}=3.04$)
\be
d_s=\frac{121.4\ln\left({2690f_b}/{\eta_{10}}\right)}{\sqrt{1+0.149{\eta_{10}}^{3/4}}}\left[\frac{1\,\text{K}}{T_d(f_b,\eta_{10})}\right]\,\text{Mpc}\,,
\ee
which is converted from~\cite{Eisenstein:1997ik} to make it purely local. In an FLRW model, this scale simply redshifts, and so can be used as a standard ruler at late times. In a void model, it shears into an axisymmetric ellipsoid through the differing expansion rates $H_\|$ and $H_\perp$. The proper lengths of the axes of this ellipse, when viewed from the centre, are given by 
\be
L_\|(z)=d_s\frac{a_\|(z)}{a_\|(t_d,r(z))}=\frac{\delta z(z)}{(1+z)H_\|(z)},~~~~L_\perp(z)=d_s\frac{a_\perp(z)}{a_\perp(t_d,r(z))}=d_A(z)\delta\theta(z)\,,
\ee
where the redshift increment $\delta z(z)$ and angular size $\delta\theta(z)$ are the corresponding observables. \\
\begin{wrapfigure}[19]{l}[0pt]{0.5\textwidth}
\includegraphics[width=0.48\columnwidth]{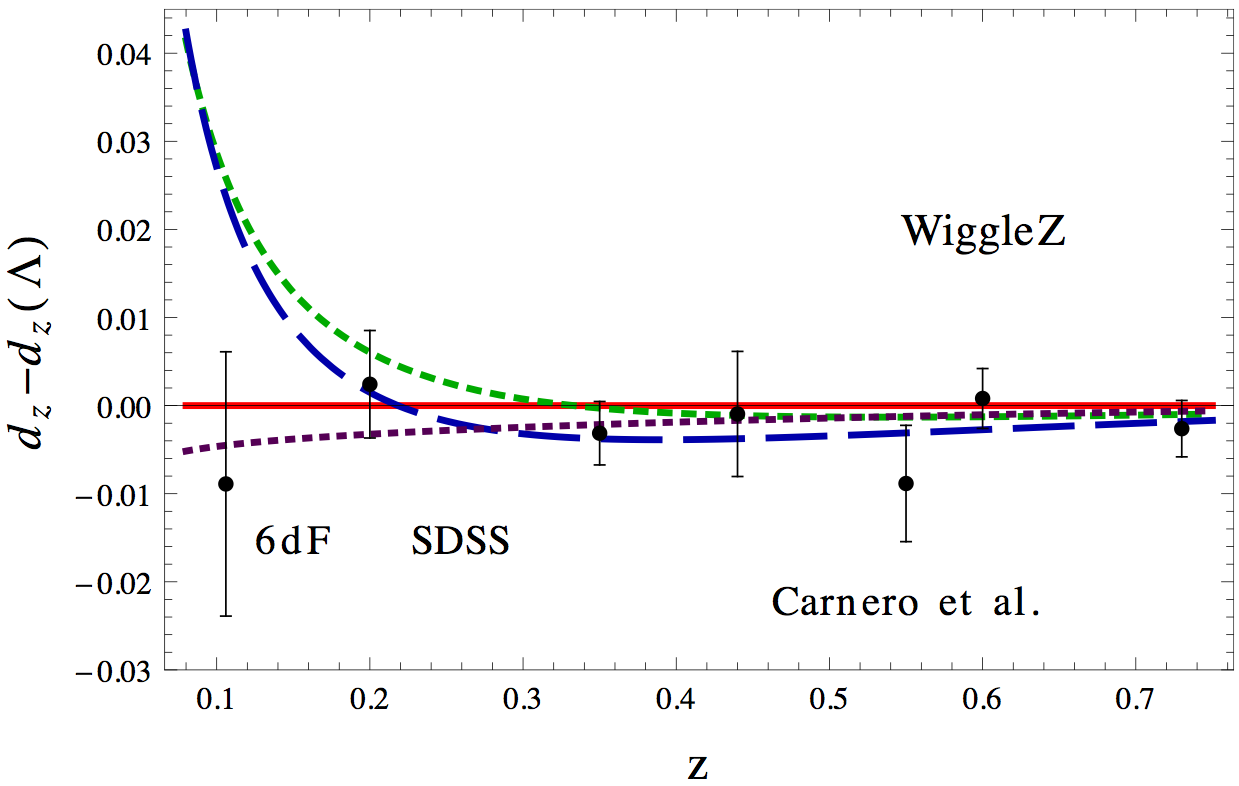}
\caption{The BAO distance measure $d_z=(\delta\theta^2\delta z/z)^{1/3}$ shown compared to a $\Lambda$CDM model. The green and blue dashed lines are void models, whereas the purple and red lines are FLRW models. Clearly, the low redshift data favours FLRW over these models. (From~\cite{Zumalacarregui:2012pq}.)}
\label{bao}
\end{wrapfigure} 
Even without this shearing effect, the BAO can be used to constrain void models because it can be used to measure $H(z)$ through the `volume distance':
\be
D_V=\left[\frac{zd_A^2}{H_\|(z)}\right]^{1/3}\,.
\ee
This quantity is given directly by current surveys. 

Thus, compared to $d_A(z)$ from SNIa, the BAO provide a complementary measurement of the geometry of the model, and in particular provide a probe of $H(z)$. For the simplest types of voids with zero bang time and no isocurvature modes, the BAO are strongly in tension with the SNIa data~\cite{ZMS,gbh1,BNV,Moss:2010jx,Zumalacarregui:2012pq}~-- see Fig.~\ref{bao}.
Note that this assumes there are no complications from the evolution of perturbations, and that scales evolve independently. While this is the case in FLRW, it is not in LTB where curvature gradients are present. Whether this is important to the analysis is yet to be shown (see below).  

While the BAO are indeed a restrictive test, it is clear that the constraints can easily be circumvented in the same way as the CMB. The bang time function can be used to free the constraint because it can be used to fix $H_\|(z)$ separately from $d_A(z)$ which is not the case if it is zero. Alternatively, we can use the freedom in $f_b=f_b(r)$ and $\eta=\eta(r)$ to change $d_s$ as a function of radial shell about the observer~\cite{Clarkson:2010ej,Zumalacarregui:2012pq}. The BAO can then be interpreted as a measurement of these parameters in different shells around us. Similarly, radial changes in the primordial power spectrum can significantly affect these results~\cite{Nadathur:2010zm}. While this might require some fine tuning to shift the BAO peak~\cite{Zumalacarregui:2012pq}, it is not yet clear if this is a significant issue.

\subsection{{Density perturbations}}

An important open problem in inhomogeneous models is the modelling of structure formation. This is important partly because it provides a means for distinguishing between FLRW and LTB. One example of where we might see an effect is in the peak in the two-point matter correlation function attributed to the Baryon Accoustic Oscillations (BAO). It has been shown that if LTB perturbations evolve as in FLRW, then BAO can be decisive in ruling out certain types of voids~\cite{ZMS,gbh1}. Whether this assumption is valid however requires a full analysis of perturbations.

There have been three approaches so far:
\begin{enumerate}
\item Using a covariant 1+1+2 formalism which was developed for gauge-invariant perturbations of spherically symmetric spacetimes~\cite{Clarkson:2002jz,Clarkson:2007yp}. The full master equations for LTB have not yet been derived, but some progress has been made in the `silent' approximation, neglecting the magnetic part of the Weyl tensor~\cite{zibin,Dunsby:2010ts}.
\item Using a 2+2 covariant formalism~\cite{GS,GMG}, developed for stellar and black hole physics. The full master equations for LTB perturbations were presented in~\cite{CCF} (see also~\cite{Tomita:1997pt}).
\item An N-body simulation has been used to study Newtonian perturbations of voids~\cite{Alonso:2010zv}.
\end{enumerate}

In FLRW cosmology, perturbations split into scalar, vector and
tensor modes that decouple from each other, and so evolve
independently (to first order). Such a split cannot usefully be performed in the same way in a spherically symmetric spacetime, as the
background is no longer spatially homogeneous, and modes written in this way couple together.  Instead, there exists a decoupling of the perturbations into two independent sectors, called `polar' (or even) and `axial' (or odd), which are analogous, but not equivalent, to scalar and vector modes in FLRW. These are based on how the perturbations transform on the sphere. Roughly speaking, polar modes are `curl' free on $S^2$ while axial modes are divergence free. Further decomposition may be made into spherical harmonics, so all variables are for a given spherical harmonic index $\ell$, and modes decouple for each $\ell$~-- analogously to $k$-modes evolving independently on an FLRW background. A full set of gauge-invariant variables were given by~\cite{GMG} who showed that there exists a natural gauge -- the Regge-Wheeler gauge -- in which all perturbation variables are gauge-invariant (rather like the longitudinal gauge in FLRW perturbation theory). Unfortunately, the interpretation of the gauge-invariant variables is not straightforward in a cosmological setting.

Most of the interesting physics happens in the polar sector, so we will discuss that case, following~\cite{CCF}. The general form of polar perturbations of the metric can be written, in Regge-Wheeler gauge, as
\ba
\mathrm{d} s^2 = -\left[1+(2 \gmgeta-\gmgchi-\gmgk) Y \right] \mathrm{d} t^2 -\displaystyle\frac{2
a_{\parallel} \gmgpsi Y}{\sqrt{1- \kappa r^2}} \mathrm{d} t \mathrm{d} r 
+\left[1+(\gmgchi+\gmgk) Y \right] \displaystyle\frac{a_{\parallel}^2
\mathrm{d} r^2}{(1-\kappa r^2)} +a_{\perp}^2 r^2 (1+ \gmgk Y) \mathrm{d}\Omega^2, \label{gpolar}
\ea
where $\gmgeta(t,r)$, $\gmgchi(t,r)$, $\gmgk(t,r)$ and $\gmgpsi(t,r)$ are gauge-invariant variables. The notation here is such that a variable times the spherical harmonic $Y$ has a sum over $\ell,m$, e.g., $\varphi Y=\sum_{\ell=0}^\infty\sum_{m=-\ell}^{m=+\ell}\varphi_{\ell m}(x^A) Y_{\ell m}(x^a)$, where $x^a$ are coordinates on $S^2$, and $x^A=(t,r)$.  The general form of polar matter perturbations in this gauge is given by
\ba
\label{upolar}
u_{\mu} = \left[\hat{u}_A+ \left( \gmggamma \hat{n}_A +\frac{1}{2} h_{AB}
\hat{u}^B \right) Y, \gmgalpha Y_{:a} \right],~~~~
\label{rhopolar}
\rho = \rho^{LTB} (1+\gmgomega Y),
\ea
where $\gmgalpha$, $\gmggamma$ and $\gmgomega$ are gauge-invariant velocity and density perturbations and $h_{AB}$ is the metric perturbation in the $x^A$ part of the metric; a colon denotes covariant differentiation on the 2-sphere.  The unit vectors in the time and radial directions are
\be
\hat{u}^A = (1,0)\,,~~~
\hat{n}^A = \left( 0, \frac{\sqrt{1- \kappa r^2}}{a_{\parallel}} \right).
\ee

The elegance of the Regge-Wheeler gauge is that the gauge-invariant metric perturbations are master variables for the problem, and obey a coupled system of PDEs which are decoupled from the matter perturbations. The matter perturbation variables are then determined by the solution to this system. We outline what this system looks like for $\ell\geq2$; in this case $\eta=0$. The generalized equation for the gravitational potential is~\cite{CCF}:
\be
\ddot\varphi+4H_\perp\dot\varphi-2\frac{\kappa} {a_\perp^2}\varphi=S_\varphi(\chi,\varsigma).
\ee
The left hand side of this equation has exactly the form of the usual equation for a curved FLRW model, except that here the curvature, scale factor and Hubble rate depend on $r$. On the right, $S_\varphi$ is a source term which couples this potential to gravitational waves, $\chi$, and generalized vector modes, $\varsigma$. These latter modes in turn are sourced by $\varphi$:
\ba
-\ddot{\gmgchi} + \gmgchi^{\prime \prime} -3 H_{\parallel}\dot{\gmgchi} -2 W \gmgchi^{\prime} + \Bigg[ 16 \pi G \rho -\frac{6M}{a_{\perp}^3}
-4 H_{\perp} (H_{\parallel}-H_{\perp})
-\frac{(\ell -1)(\ell +2)}{a_{\perp}^2 r^2}\Bigg] \gmgchi
&=&S_\chi(\gmgpsi,\gmgk),\\
\dot{\gmgpsi} + 2 H_{\parallel} \gmgpsi &=& -\gmgchi^{\prime}.
\ea
The prime is a radial derivative defined by $X'=n^A\del_A X$.

The gravitational field is inherently dynamic even at the linear level, which is not the case for a dust FLRW model with only scalar perturbations. Structure may grow more slowly due to the dissipation of potential energy into gravitational radiation and rotational degrees of freedom.
Since $H_\perp=H_\perp(t,r)$, $a_\perp=a_\perp(t,r)$ and $\kappa=\kappa(r)$, perturbations in each shell about the centre grow at different rates, and it is because of this that the perturbations generate gravitational waves and vector modes. This leads to a very complicated set of coupled PDEs to solve for each harmonic $\ell$.

In fact, things are even more complicated than they first seem. Since the scalar-vector-tensor decomposition does not exist in non-FLRW models, the interpretation of the gauge-invariant LTB perturbation variables is subtle. For example, when we take the FLRW limit we find that
\ba
\gmgk&=& -2\Psi-2\H V -2\frac{(1-\kappa r^2)}{r}h_r +\frac{1}{r^2}h^\T
+\left[-\H \p_\tau  +\frac{(1-\kappa r^2)}{r} \p_r
 +\frac{\ell(\ell+1)-4(1-\kappa r^2)}{2r^2}\right]h^\TF,
\ea
where $\Psi$ is the usual perturbation space potential, $V$ is the radial part of the vector perturbation, and the $h$'s are invariant parts of the tensor part of the metric perturbation. Thus $\varphi$ contains scalars, vectors and tensors. A similar expression for $\varsigma$ shows that it contains both vector and tensor degrees of freedom, while $\chi$ is a genuine gravitational wave mode, as may be seen from the characteristics of the equation it obeys. This mode mixing may be further seen in the gauge-invariant density perturbation which appears naturally in the formalism:
\ba
8 \pi G \rho \gmgomega&=& -\gmgk^{\prime \prime} - 2 W \gmgk^{\prime}+(H_{\parallel}+2 H_{\perp}) \dot{\gmgk}+W
\gmgchi^{\prime} + H_{\perp} \dot{\gmgchi} 
+ \left[ \frac{\ell (\ell +1)}{a_{\perp}^2r^2} +2 H_{\perp}^2 +4 H_{\parallel} H_{\perp} -8 \pi G \rho \right]
(\gmgchi +\gmgk)  \nonumber\\ && - \frac{(\ell -1)(\ell +2)}{2 a_{\perp}^2r^2} \gmgchi  +2 H_{\perp} \gmgpsi^{\prime}+2 (H_{\parallel}+H_{\perp}) W
\gmgpsi ,
\ea
where
\be
W \equiv \frac{\sqrt{1-\kappa r^2}}{a_{\perp} r}.
\ee
When evaluated in the FLRW limit the mode mixing becomes more obvious still: $\Delta$ contains both vector and tensor modes, while its scalar part is
\be
4\pi G a^2\rho\,\gmgomega = \sdel^2\Psi-3\H\p_\tau\Psi-3(\H^2-\kappa)\Psi,
\ee
which gives the usual gauge invariant density fluctuation
$\delta\rho^\GI\equiv\delta\rho+\p_\tau\rho (B-\p_\tau
E)$~\cite{Malik&Wands}. Here,
$\sdel^2$ refers to the Laplacian
acting on a 3-scalar.
 The fact that $\Delta$ is more complicated is because the gauge-invariant density perturbation includes metric degrees of freedom in its definition; gauge-invariant variables which are natural for spherical symmetry may not be natural for homogeneous backgrounds. A gauge-dependent $\Delta$ may be defined which reduces to $\delta\rho^\GI$ in the FLRW subcase, but its gauge-dependence will cause problems in the inhomogeneous case.

These equations have not yet been solved in full generality. We expect different structure growth in LTB models, but it is not clear what form the differences will take. It seems reasonable to expect that the coupling between scalars, vectors and tensors will lead to dissipation in the growth of large-scale structure where the curvature gradient is largest, as it is the curvature and density gradients that lead to mode coupling. In trying to use structure formation to compare FLRW to LTB models, some care must be taken over the primordial power spectrum and whatever early universe model is used to generate perturbations -- since there is a degeneracy with the primordial power spectrum and the features in the matter power spectrum.

\subsection{The Copernican problem: Constraints on the distance from the centre}

\begin{SCfigure}[1.0][t!]
\begin{minipage}{0.7\textwidth}
~~~~\includegraphics[width=0.95\textwidth]{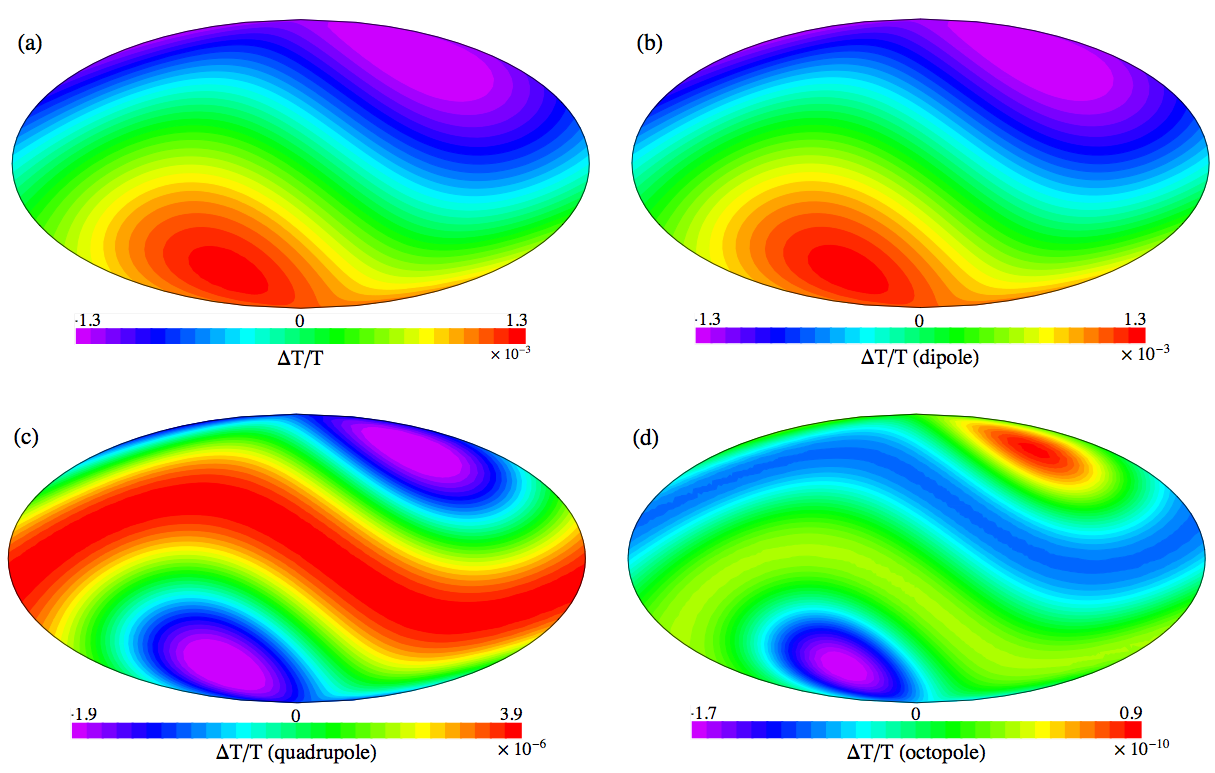}
\includegraphics[width=\textwidth]{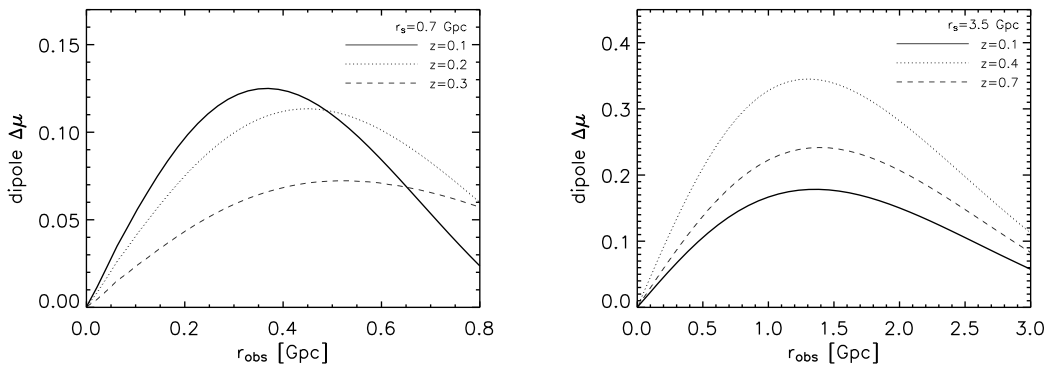}
\end{minipage}
\caption{Off-centre observers see anisotropy. \\
Top: The main contribution to the total CMB anisotropy (a) is in the form of a dipole (b) with higher-order moments suppressed (c), (d). (From~\cite{Grande:2011hm}, for an LTB model with dark energy.)\\
Bottom: There is also a dipole in the distance modulus, shown here for a large and small void at different redshifts. (From~\cite{mortsell}.)}
\label{default}
\end{SCfigure}

An off-centre observer will typically measure a large dipole in the CMB  due to their peculiar velocity with respect to the CMB frame, which is non-perturbative in this context~\cite{Humphreys:1996fd}. They will also measure a dipole in their local distance-redshift relation which is not due to a peculiar velocity effect nor a dipole in the Hubble law. Rather this first appears at $\mathcal{O}(z^2)$ in the distance-redshift relation through gradients in the expansion rate and the divergence of the shear~-- see Eq.~(\ref{jhsdbc}) below. Combined constraints on the dipole in the SNIa data and the CMB would require us to be within $\sim0.5\%$ of the scale radius of the void~\cite{Blomqvist:2009ps}, without fine-tuning our peculiar velocity to cancel out some of the dipole. Others reach similar conclusions~\cite{Humphreys:1996fd,Alnes:2006pf,alnes,Foreman:2010uj}. 

An intriguing alternative view was presented in~\cite{BNV}. Although they reach the same conclusion as to how close to the centre the observer should be, they argue that if we're slightly off centre, then one would expect to see a significant bulk flow in the direction of the CMB dipole. Such a `dark flow' has been tentatively measured~\cite{Kashlinsky:2008us,Kashlinsky:2012gy}, and has not been accounted for within $\Lambda$CDM at present.

\subsection{{ Summary and interpretation of inhomogeneous models}}

An inhomogeneous LTB void model, even if it over-simplifies nonlinear inhomogeneity at the background level, does produce some rather remarkable results. The apparent acceleration of the universe can be accounted for without dark energy, and the Lithium problem can be trivially solved. However, it seems that the simplest incarnation of such a model will not work: combining observables $H_0$, SNIa and the CMB reveals considerable tension with the data, with the main problem being a low $H_0$ in the models~\cite{BNV}; add in (even rudimentary) kSZ and $y$-distortion constraints, and the situation is conclusive. Results from the BAO, though only indicative and not yet decisive (as they do not take into account structure formation on an inhomogeneous background), also signal considerable tension. 

Each of these observables point to non-trivial inhomogeneity at early times (or $\Lambda$ of course). Most models which are ruled out have the assumption of evolving from a homogeneous FLRW model. Primary observables provide much weaker constraints if this restriction is removed, though it is still difficult to get a good fit using just the freedom of a bang time function~\cite{Bull:2011wi}.  But one can free up essentially any function which is assumed homogeneous in the standard model; in the context of inhomogeneous models, it doesn't make sense to keep them homogeneous unless we have a specific model in mind (perhaps derived from a model of inflation). Examples of such freedom include a radially varying bang time function, a radially varying primordial power spectrum (designed to have the required spectrum on 2-spheres perhaps), isocurvature degrees of freedom such as a varying baryon photon ratio or baryon fraction, and one can dream up more such as varying $N_\text{eff}$. 
Indeed, taken at face value the lithium problem~\cite{Cyburt:2008kw} can be interpreted as a direct measurement of an inhomogeneous isocurvature mode present at BBN in this context~\cite{RC}~-- this is actually the one observation which is potentially at odds with homogeneity. Most primordial numbers in the standard model are not understood well if at all, and if we remove slow roll inflation~-- as we must to make such a model in the first place~-- we remove significant motivation to keep them homogeneous. 

This suggests an important reverse-engineering way to handle such models. If we accept that presently any \emph{specific} inhomogeneous model is essentially pulled from thin air, then we have to conclude that what we are really trying to do is to invert observables to constrain different properties of the model in different shells around us.  Though it seems rather non-predictive, without an early universe model to create a void from it is really no different from making a map of the universe's history. This inverse-problem approach has been investigated in~\cite{7,7b,Araujo:2008rm,Hellaby:2008pp,Araujo:2009zh,Araujo:2010ag,vanderWalt:2010zd}. The idea is to specify observational data on our past lightcone, smooth it, and integrate into the interior. Whether this inverse problem is well-conditioned or not is crucial to the success of such an approach~\cite{Winfield:2011tj}. Nevertheless, this is a valuable strategy: If we specify data on our past lightcone and integrate into the interior, does it necessarily yield an FLRW model, or are there other solutions (perhaps without dark energy)? 

An important alternative view of these models is not to view them as an anti-Copernican alternative to dark energy, but rather to view them as the simplest probe of large scale inhomogeneity~\cite{Celerier:2011zh}. This is akin to considering Bianchi models as probes of large-scale anisotropy. 
We may therefore think of LTB models as models where we have smoothed all observables over the sky, thereby compressing all inhomogeneities into one or two radial degrees of freedom centred about us. In this interpretation, we avoid placing ourselves `at the centre of the universe' in the standard way. Furthermore, constraints which arise by considering anisotropy around distant observers~-- the Goodman constraints~-- are perhaps removed from the equation; distant observers would see an isotropic universe too.  

In this sense, these models are a natural first step in developing a fully inhomogeneous description of the universe. There is a vital caveat to this interpretation, however: we must include dark energy in the model for it to be meaningful, which is almost never done~(see~\cite{Marra:2010pg,Grande:2011hm,Marra:2012pj} for a first attempt). If we do not, then we are implicitly assuming that consequences of the averaging, backreaction and fitting problems really do lead to significant effects which solve the dark energy problem.
That is, by averaging observables at some redshift over the sky we are averaging the geometry out to that redshift, which can have a non-trivial `back-reaction' effect on our interpretation of the model~\cite{Clarkson:2011zq}. This could conceivably look like dark energy in our distance calculations (perhaps even dynamically too~\cite{Buchert:2011sx}). If that were indeed the case, we could have a significant effective energy-momentum tensor which would be very different from dust, and it would not be simple to calculate observables as they would not necessarily be derivable from the metric. Hence,  within this interpretation the dust LTB model would certainly be the wrong model to use. If one is to further peruse this idea, one might need to constrain deviations from homogeneity using the metric only, without resorting to the field equations at all (see~\cite{Clarkson:2011gm} for further discussion). 

This is nevertheless in some ways the most natural way to place constraints on inhomogeneity. Yet, if large-scale inhomogeneity were present, we shall see in the next section that within GR it is challenging~-- perhaps impossible~-- to reconcile it with the Copernican principle given the level of isotropy we observe.

\section{Routes to homogeneity}

Considering a specific inhomogeneous solution to the EFE which violates the CP helps us consider the types of observables which can be used to demonstrate homogeneity. Ruling out classes of solutions as viable models helps test the Copernican principle provided we understand where they sit in the space of solutions, so to speak. 
Under what circumstances does the Copernican principle, combined with some observable, actually imply an FLRW geometry? 

In the case of perfect observables and idealised conditions quite a lot is known, as we discuss below. These results are non-perturbative, and do not start from FLRW to show consistency with it. For the case of realistic observables in a lumpy universe, however, details are rather sketchy, with only one case properly considered.  

Many of these results rely on the following theorem~\cite{AndrzejKrasinski:1997zz}: 
\begin{description}
\item[\em The FLRW models\em :] For a perfect fluid solution to the Einstein field equations where the velocity field of the fluid is geodesic, then the spacetime is FLRW if either: 
\begin{itemize}
\item[--] the velocity field of the source is shear-free and irrotational; or,
\item[--] the spacetime is conformally flat (i.e., the Weyl tensor vanishes). 
\end{itemize}
\end{description}
The `perfect fluid source' here refers to the total matter content and not to the individual components, and is necessarily barotropic if either of the conditions are met. So, for example, the matter could be comoving dark matter and baryons, and dark energy in the form of a scalar field with its gradient parallel to the velocity of the matter.

\subsection{Isotropy of the CMB}

What can we say if the CMB is exactly isotropic for `fundamental observers'? This is the canonical expected observable which intuitively should imply FLRW with the CP. It does, usually, but requires assumptions about the theory of gravity and types of matter present.  The pioneering result is due to Ehlers, Geren and Sachs (1968)~\cite{EGS}. Without other assumptions we have:

\begin{description}
\item[{\bf \footnotesize [EGS]}~{\em Radiation isotropy} \withcp {\em conformally stationary}:]

In a region, if  observers on an expanding congruence $u^a$ measure a collisionless radiation field which is isotropic, then the congruence is shear-free, and the expansion is related to the acceleration via a potential $Q=-\frac{1}{4}\ln\rho_r$: $A_a=\D_aQ,~~\Theta=3\dot Q$; the spacetime must be conformal to a stationary spacetime in that region.

\end{description}

That doesn't tell us a great deal, but including geodesic observers changes things considerably. The original EGS work assumed that the only source of the gravitational field was the radiation, i.e., they neglected matter (and they had $\Lambda=0$). This has been generalised over the years to include self-gravitating matter and dark energy \cite{TreEll71,Ellis:1983,Stoeger:1994qs,FerMorPor99,CB,CC,CM}, as well as for scalar-tensor theories of gravity~\cite{Clarkson:2001qc}:

\begin{description}
\item[{\bf \footnotesize [EGS+]}~{\em Radiation isotropy with dust }\withcp FLRW:]

In a region, if dust observers on an expanding congruence $u^a$ measure a collisionless radiation field which has vanishing dipole, quadrupole and octopole, and non-interacting dark energy is the form of $\Lambda$, quintessence, a perfect fluid or is the result of a scalar-tensor extension of GR, then the spacetime is FLRW in that region. 

\end{description}
The dust observers are necessarily geodesic and expanding:
\begin{equation}\label{av}
A_\a=0\,,~~\Theta >0\,.
\end{equation}
Because the dust observers see the radiation energy flux to vanish (the dipole), $u^a$ is the frame of the radiation also. 
The photon distribution function $f(x,p,E)$ in momentum space depends  on components of the 4-momentum $p^\a$ along $u^\a$, i.e., on the photon energy $E=-u_\a p^\a$, and, in general, the direction $e^a$, and
 may be written in a spherical harmonic expansion as (see the appendix)
\begin{equation}
f=\sum_{\ell=0}^{\infty} F_{A_\ell}{e}^{A_\ell},
\end{equation}
where the spherical harmonic coefficients $F_{A_\ell}$ are symmetric, trace-free tensors orthogonal to $u^a$, and $A_\ell$ stands for the index string $a_1a_2\cdots a_\ell$. (In this notation, ${e}^{\<A_\ell\>}$ are a representation of the spherical harmonic functions.) The dust observers measure the first three moments of this to be zero which means
\be
 F_a=F_{ab}=F_{abc}=0. 
\ee
In particular, as follows from Eq.~(\ref{em3}), the momentum density (from the dipole), anisotropic stress (from the quadrupole), and the radiation brightness octopole vanish:
\begin{equation}
\label{qpv}
q_r^\a=\pi_r^{ab}=\Pi^{abc}=0\,.
\end{equation}
These are source terms in the anisotropic stress evolution equation, which is the $\ell=2$ case of Eq.~(\ref{r26}). In general fully nonlinear form, the $\pi_r^{\a\nu}$ evolution equation is
\begin{eqnarray}
&&\dot{\pi}_{r}^{\langle ab \rangle}+{{4\over3}}\Theta
\pi_{r}^{ab } +{{8\over15}}\rho_{r}\sigma^{ab }+
{{2\over5}}\D^{\langle \a}q_{r}^{b\rangle} +2 A^{\langle \a} q_{r}^{b\rangle} -2\omega^{c}\ep_{c d }{}{}^{(\a}
\pi_{r}^{b) d} ~~~~~
\nonumber\\
&& ~~~~~~~~  +{{2\over7}}\sigma_{c}{}^{\langle
\a}\pi_{r}^{b\rangle c} +{8\pi\over35}\D_{c}
\Pi^{abc}  -{32\pi\over315} \sigma_{c d } \Pi^{abcd}
= 0. \label{nl8}
\end{eqnarray}
Eq.~(\ref{qpv}) removes all terms on the left except the third and the last:
\be
\left(21\rho_{r}h_a^{~c}h_b^{~d} -4\pi{\Pi_{ab}}^{cd}\right)\sigma_{cd}=0\,,
\ee
which implies, since $\Pi_{abcd}$ is trace-free and the first term consists of traces,
 shear-free expansion of the fundamental congruence:
\begin{equation}\label{sv}
\sigma_{ab}=0\,.
\end{equation}

We can also show that $u^\a$ is irrotational as follows. Together with Eq.~(\ref{av}), momentum conservation for radiation, i.e., Eq.~(\ref{e3i}) with $I=r$, reduces to
\begin{equation}\label{mv}
\D_\a \rho_r=0\,.
\end{equation}
Thus the radiation density is homogeneous relative to fundamental observers.
Now we invoke the exact nonlinear identity for the covariant curl of the gradient, Eq.~(\ref{ri1}):
\begin{equation}\label{}
\mbox{curl}\, \D_\a \rho_r = - 2\dot \rho_r \omega_\a~ \Rightarrow~ \Theta \rho_r \omega_\a =0\,,
\end{equation}
where we have used the energy conservation equation~(\ref{e1i}) for radiation. By assumption $\Theta >0$, and hence we deduce that the vorticity must vanish:
\begin{equation}\label{vv}
\omega_a =0\,.
\end{equation}
Then we see from the curl shear constraint equation~(\ref{c3}) that the magnetic Weyl tensor must vanish:
\begin{equation}\label{hv}
H_{ab}=0\,.
\end{equation}

Furthermore, Eq.~(\ref{mv}) actually tells us that the expansion must also be homogeneous. From the radiation energy conservation equation~(\ref{e1i}), and using Eq.~(\ref{qpv}), we have $\Theta=-3{\dot\rho_r}/{4\rho_r}$.
On taking a covariant spatial gradient and using the commutation relation Eq.~(\ref{timespace}), we find
\be \label{tv}
\D_\a\Theta=0\,.
\ee
Then the shear divergence constraint, Eq.~(\ref{c2}), enforces the vanishing of the total momentum density in the matter frame,
\begin{equation}\label{qv}
q^\a \equiv \sum_I q_I^\a =0~ \Rightarrow ~\sum_I (1+\gamma_I^2v_I^2)(\rho^*_I + p^*_I)v_I^\a=0  \,.
\end{equation}
The second equality follows from Eq.~(A10) in~\cite{Maartens:1998xg}, using the fact that the baryons, CDM and dark energy (in the form of quintessence or a perfect fluid) have vanishing momentum density and anisotropic stress in their own frames, i.e.,
\begin{equation}\label{perf}
q_I^{*\a}=0= \pi_I^{*\a b}\,,
\end{equation}
where the asterisk denotes the intrinsic quantity (see Appendix A). If we include other species, such as neutrinos, then the same assumption applies to them. Except in artificial situations, it follows from Eq.~(\ref{qv}) that
\begin{equation}\label{pvan}
v_I^\a=0\,,
\end{equation}
i.e., the bulk peculiar velocities of matter and dark energy [and any other self-gravitating species satisfying Eq.~(\ref{perf})] are forced to vanish -- all species must be comoving with the radiation.

The comoving condition~(\ref{pvan}) then imposes the vanishing of the total anisotropic stress in the matter frame:
\begin{equation}\label{piv}
\pi^{a b}\equiv \sum_I \pi_I^{\a b} =\sum_I \gamma_I^2(\rho^*_I + p^*_I)v_I^{\langle \a}v_I^{b \rangle}=0  \,,
\end{equation}
where we used Eqs.~(A11) in~\cite{Maartens:1998xg}, (\ref{perf}) and (\ref{pvan}). Then the shear evolution equation~(\ref{e5}) leads to a vanishing electric Weyl tensor
\begin{equation}\label{evan}
E_{ab}=0\,.
\end{equation}
Equations~(\ref{qv}) and (\ref{piv}), now lead via the total momentum conservation equation~(\ref{e3}) and the $E$-divergence constraint~(\ref{c4}), to homogeneous total density and pressure:
\begin{equation}\label{rpv}
\D_\a\rho = 0 = \D_\a p\,.
\end{equation}

Equations~(\ref{av}), (\ref{sv}), (\ref{hv}), (\ref{tv}), (\ref{qv}), (\ref{piv}) and (\ref{rpv}) constitute a covariant characterisation of an FLRW spacetime. This establishes the EGS result, generalised from the original to include self-gravitating matter and dark energy. It is straightforward to include other species such as neutrinos. 

The critical assumption needed for all species is the vanishing of the intrinsic momentum density and anisotropic stress, i.e., Eq.~(\ref{perf}). Equivalently, the energy-momentum tensor for the $I$-component should have perfect fluid form in the $I$-frame (we rule out a special case that allows total anisotropic stress~\cite{CC}). The isotropy of the radiation and the geodesic nature of its 4-velocity~-- which follows from the assumption of geodesic observers~-- then enforce the vanishing of (bulk) peculiar velocities $v_I^\a$. Note that one does {\em not} need to assume that the other species are comoving with the radiation -- it follows from the assumptions on the radiation. A similar proof can be used for scalar-tensor theories of gravity, although it is somewhat more involved~\cite{Clarkson:2001qc}.

It is worth noting that we do have to assume the the dark matter and dark energy are non-interacting. If we do not, we cannot enforce the radiation congruence to be geodesic because the observers may not be, and one is actually left only with a very weak condition: only that the spacetime is conformally stationary~\cite{EGS,CB,Clarkson:2003ts}.  

In summary,  the EGS theorems, suitably generalised to include baryons and CDM and dark energy, are the most powerful basis that we have~-- within the framework of the Copernican Principle~-- for background spatial homogeneity and thus an FLRW background model.
Although this result applies only to the `background Universe', its proof nevertheless requires a fully nonperturbative analysis.

\subsection{Blackbody spectrum of the CMB}\label{BB}

The EGS results rely only on the isotropy of the radiation field, and do not utilise its spectrum. Of course, no geometry can affect the spectrum of the CMB because the gravitational influence on photons is frequency independent (except at very high frequencies). However, the fact that the CMB is a nearly perfect blackbody tells us much about the spacetime when there are scattering events present. The Sunyaev-Zel'dovich (SZ) effect is due to the scattering of CMB photons by charged  matter, and has already been shown to be a powerful tool for constraining radial inhomogeneity within the class of cosmological models constructed from the LTB solutions, as discussed in Sec.~\ref{SZ}. It has recently been shown~\cite{Clifton:2011sn} that under idealised circumstances similar to the EGS theorems above the SZ effect can actually be used as a proof of FLRW geometry \emph{for one observer} without requiring the Copernican principle at all, thus extending Goodman's tests to arbitrary spacetimes~\cite{Goodman:1995dt}.

\begin{description}
\item[{\bf \footnotesize [GCCB]}~{\em Isotropic blackbody} CMB \withoutcp FLRW:] An observer who sees an isotropic blackbody CMB in a universe with scattering events can deduce the universe is FLRW if either double scattering is present or they can observe the CMB for an extended period of time, assuming the CMB is emitted as a blackbody, and the conditions of the EGS theorem hold. 
\end{description}

\begin{SCfigure}
~~~~\includegraphics[width=0.7\columnwidth]{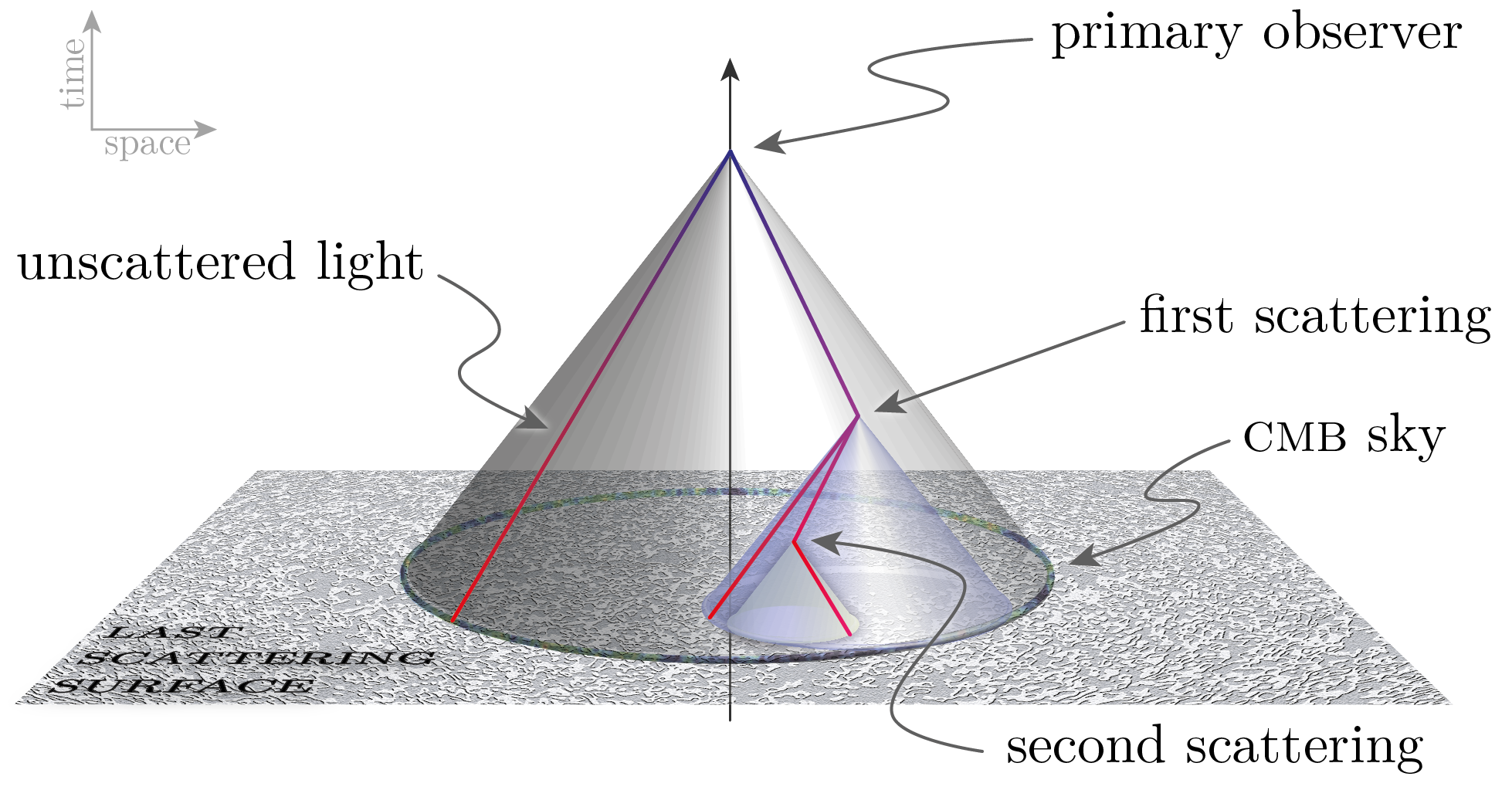}
\caption{The SZ effect provides information about the isotropy of
  the CMB sky at other points on our past lightcone.  It can also
  provide us with information about parts of the last scattering
  surface that would otherwise be inaccessible to us.  Multiple
  scattering events provide further information about
  the CMB sky at other points within our causal past. (From~\cite{Clifton:2011sn}.)}
\label{fig}
\end{SCfigure}

The observational effect of the scattering of CMB photons by baryonic matter 
is usually referred to in the literature as the SZ effect \cite{1972CoASP...4..173S,1980MNRAS.190..413S}, and 
is often divided into two different contributions; the thermal SZ effect (tSZ)
\cite{1972CoASP...4..173S} and the kinematic SZ effect (kSZ) \cite{1980MNRAS.190..413S}. The kSZ effect causes a distortion in the spectrum 
of the reflected light due to the anisotropy seen in the CMB sky 
of the scatterer, and maintains the same distribution function it had before the 
scattering event (all other changes being encapsulated in the tSZ). For the case of blackbody radiation this corresponds solely to a change in temperature of the scattered radiation. 

Thus, an observer who sees an exactly isotropic CMB in a universe where scattering takes place, can deduce that the scatterers themselves must also see an isotropic CMB, provided that decoupling emits the CMB radiation as an exact blackbody. The proof of this relies on the fact that blackbody spectra of differing temperatures cannot be added together to give another blackbody. 
This mechanism therefore provides, in effect, a set of mirrors that allows us to view the CMB from different locations~\cite{Goodman:1995dt}.

Is this enough to deduce FLRW geometry? Not on its own. Such an observation gives a single null surface on which observers see an isotropic CMB.  This allows us to use a much weaker version of the Copernican principle than used in the EGS theorems (which assume isotropy for all observers in a 4-dimensional patch of spacetime, and deduce homogeneity only in that patch) to deduce homogeneity. For example, if all observers in a spacelike region see an isotropic blackbody CMB when scattering is present, then the spacetime must be homogeneous in the causal past of that region.

The lone observer can say more, however~\cite{Clifton:2011sn}. If there are scattering events taking place throughout the universe, then each `primary scatterer' our observer sees must also be able to deduce that scatterers on their past nullcone see an isotropic CMB, or else their observations of a blackbody spectrum would be distorted. Consequently, the spacetime must be filled with observers seeing an isotropic CMB, and one can then use the EGS theorems to deduce FLRW geometry. Hence, under highly idealised conditions, a single observer at a single instant can deduce FLRW geometry within their past lightcone. Alternatively, the observer could wait for an extended period until their past nullcone sweeps out a 4-D region of spacetime. If no kSZ effect is measured at any time, then they can infer that their entire causal past must also be FLRW.

\subsection{Local observations}

Instead of using the CMB, local isotropy of observations can also provide a route to homogeneity.
Adopting the Copernican Principle, it follows from the `lightcone-isotropy implies spatial isotropy' theorem that if all observers see isotropy then the spacetime is isotropic about all galactic worldlines~-- and hence spacetime is FLRW.

\begin{description}
\item[{\em Matter lightcone-isotopy }\withcp FLRW:]

In an expanding dust region with $\Lambda$, if all fundamental observers measure isotropic area distances, number counts, bulk peculiar velocities, and lensing, then the spacetime is FLRW.
\end{description}

In essence, this is the Cosmological Principle, but derived from observed isotropy and not from assumed spatial isotropy. Note the significant number of observable quantities required.
Using the CP, we can actually give a much stronger statement than this, based only on distance data.
An important though under-recognised theorem due to Hasse and Perlick tells us that~\cite{HP}

\begin{description}
\item[{\bf \footnotesize [HP]}~{\em Isotropic distances }\withcp FLRW:]

If all fundamental observers in an expanding spacetime region measure isotropic area distances up to third-order in a redshift series expansion, then the spacetime is FLRW in that region.
\end{description}

The proof of this relies on performing a series expansion of the distance-redshift relation in a general spacetime, using the method of Kristian and Sachs~\cite{KS}, and looking at the spherical harmonic multipoles order by order. We illustrate the proof in the 1+3 covariant approach (in the case of zero vorticity). Performing a series expansion in redshift of the distance modulus, we have, in the notation of~\cite{MacEllis},
\ba\label{m-z-general}
m- M-25&=&5\log_{10}z-5\log_{10}
\left.K^\a K^b\del_\a u_b\right|_o+
\sfrac52\log_{10}e\Bigg\{
\Bigg[4-\frac{K^\a K^b K^c\del_\a\del_b u_c}{(K^d K^e\del_d u_e)^2}
\Bigg]_Oz
\nonumber\\
 &&-\Bigg[2+\frac{R_{a b}K^a K^b}{6(K^c K^d\del_c u_d)^2}-
\frac{3(K^\a K^b K^c\del_\a\del_b u_c)^2}{4(K^d K^e\del_d u_e)^4}
+
\frac{K^\a K^b K^c K^d\del_\a\del_b\del_c u_d}{3(K^e K^f\del_e u_f)^3}\Bigg]
_Oz^2
\Bigg\}+{\cal
O}(z^3),
\ea
where
\be
K^\a=\frac{k^\a}{\left.u^b k_b\right|_O}~~~~\hbox{and}~~~~
\left.K^\a\right|_O=\left.-u^\a+e^\a\right|_O,
\ee
denotes a past-pointing null vector at the observer $O$ in the direction $e^\a$. When fully decomposed into their projected, symmetric and trace-free parts, products of $e$'s represent a spherical harmonic expansion. Thus, this expression views the distance modulus as a function of redshift on the sky, with a particular spherical harmonic expansion on a sphere of constant redshift. (The inverse of this expression has coefficients which have a spherical harmonic interpretation on a sphere of constant magnitude.)
Comparing with the standard FLRW series expansion evaluated today, we define an observational Hubble rate and deceleration parameter as
\ba
\left.H^\mathrm{obs}\right|_{0}\li=\li\left.K^\a K^b\del_\a u_b\right|_0,
\label{hubble def from series}\\
\left.q^\mathrm{obs}\right|_0\li=\li\left.\;{K^\a K^b K^c\del_\a\del_b u_c}
{(K^d K^\e\del_d u_\e)^2}\right|_0-3.\label{q0 from series}
\ea
We can also give an effective observed cosmological constant parameter from the $O(z^2)$ term:
\ba
\Omega_\Lambda^\mathrm{obs}&=&\frac52\left(1-q_0^\mathrm{obs}\right)-5+
\left.\frac{R_{\a b}K^\a K^b}{12(H^\mathrm{obs})^2}\right|_0
+\left.\frac{K^\a K^b K^c K^d\del_\a\del_b\del_c u_d}{6(H^\mathrm{obs})^3}\right|_0.
\ea
The argument of~\cite{HP} relies on proving that if all observers measure these three quantities to be isotropic then the spacetime is necessarily FLRW.
In a general spacetime~\cite{MacEllis}
\be
H_0^\mathrm{obs}={\frac{1}{3}\Theta}-
{A_\a e^\a}+
{\sigma_{ab}e^\a e^b},
\label{hubble from series}
\ee
where ${A_\a e^\a}$ is a dipole and ${\sigma_{ab}e^\a e^b}$ is a quadrupole.
Hence, if all observers measure $H_0^\mathrm{obs}$ to be isotropic, then $\sigma_{ab}=0=A_\a$.
In a spacetime with isotropic $H_0^\mathrm{obs}$ the generalised deceleration parameter, defined on a sphere of constant redshift, is given by~\cite{Clarkson:1999zq,Clarkson:2011uk}
\ba
\left(H_0^\mathrm{obs}\right)^2\, q_0^\mathrm{obs}&=&\;16\rho+\;12p-\;13\Lambda
 -\;{1}{5}e^\a\left[
2q_\a-{3}\sdel_\a\Theta
\right]
 +e^{\<a} e^{b\>}\left[
 E_{ab}-\;12\pi_{ab}\right]
 .\label{q_0-observable}
\ea
If the dipole of this term vanishes then we see from Eq.~(\ref{c2}) that the energy flux must vanish as well as spatial gradients of the expansion. Excluding models with unphysical anisotropic pressure, Eq.~(\ref{e6}) then shows that the electric Weyl tensor must vanish, and it follows that the spacetime must be FLRW. The more general proof in~\cite{HP} uses $\Omega_\Lambda^\mathrm{obs}$ to show that the vorticity must necessarily vanish along with the anisotropic pressure.


\subsection{The Hubble rate on the past lightcone}

Measurements of the Hubble rate on our past lightcone can provide an important route to homogeneity assuming the Copernican principle. In a general spacetime, the spatial expansion tensor may be written as 
\be\label{expansion_tensor}
\Theta_{ab}=\Theta h_{ab}+\sigma_{ab}\,.
\ee
However, we do not measure exactly this as our observations are made on null cones. An observer can measure the expansion at a given redshift in three separate directions: radially, and in two orthogonal directions in the screen space. The \emph{observed} radial Hubble rate at any point is a generalisation of our observed Hubble constant above,
\be
H_\|(z;\bm e)=K^aK^b\del_a u_b={\frac{1}{3}\Theta}-
{A_\a e^\a}+
{\sigma_{ab}e^\a e^b}\,.
\ee
The Hubble rates orthogonal to this may be found by projecting into the screen space, $N_{ab}=g_{ab}-K_aK_b-K_au_b-u_aK_b$:
\be
{H_\perp}_{ab}(z;\bm e)=\frac{1}{2}{N_a}^c{N_b}^d\del_cu_d=\left(\frac{1}{6}\Theta-\frac{1}{4}\sigma_{cd}e^ce^d\right)N_{ab}+\frac{1}{2}\epsilon_{abc}e^c\,(e_d\omega^d)+\frac{1}{2}\left({N_a}^c{N_b}^d-\frac{1}{2}N_{ab}N^{cd}\right)\sigma_{ab}\,.
\ee
The trace of this gives the areal Hubble rate
\be
{H_\perp}(z;\bm e)={H_\perp}_{ab}N^{ab}=\frac{1}{3}\Theta-\frac{1}{2}{\sigma_{ab}e^\a e^b}\,,
\ee
which implies the spatial volume expansion rate is given via the observable expansion rates as:
\be
\Theta(z;\bm e)=H_\|(z;\bm e)+2H_\perp(z;\bm e)+A_ae^a.
\ee
If we measure ${H_\perp}_{ab}$ to be rotationally symmetric on the celestial sphere in each direction $e^a$ and at each $z$, and $H_\|(z;\bm e)=H_\perp(z;\bm e)$ then this, on its own, is not enough to set the shear, acceleration and rotation to zero, even on our past lightcone. However, we can see how measuring the Hubble rate can lead to homogeneity, when we apply the Copernican principle:
\begin{description}
\item[\emph{Isotropic Hubble rate} \withcp FLRW] If all observers in a perfect fluid spacetime measure: 
\begin{itemize}
\item[--] ${H_\perp}_{ab}$ to be rotationally symmetric on the screen space at each $z$\,; and,
\item[--] $H_\|(z;\bm e)=H_\perp(z;\bm e)$\,,
\end{itemize}
then the spacetime is FLRW.

\end{description}
The proof of this is straightforward: the first condition implies that $\omega^a=\sigma_{ab}=0$ and the second that $A^a=0$, which implies FLRW from the theorem above.

This forms the basis of the Alcock-Paczynski~\cite{Alcock:1979mp} test in a general spacetime. An object which is spherical at one instant will physically deform according to Eq.~(\ref{expansion_tensor}), if it is comoving (such as the BAO scale). A further effect is that it will be observed to be ellipsoidal with one of the principle axes scaled by $H_\|^{-1}$ and the other two by the inverse of the eigenvalues of ${H_\perp}_{ab}$.

\subsection{Ages: absolute and relative}

Measurement of the ages of objects~-- both absolute and relative~-- on our past lightcone provides important  information which can be used to deduce homogeneity. 

Neighbouring lines on the matter congruence $u_a=\partial_at$, measuring proper time $t$, may be thought of as connected by null curves $k^a$ (which are past pointing in our notation above). An increment of redshift on a null curve is related to an increment of proper time on the matter worldlines as
\be\label{dzdt}
\frac{\d t}{\d z}=-\frac{1}{(1+z)H_\|(z;\bm e)}\,.
\ee
This gives the age difference of objects observed on the past lightcone, in a redshift increment $\d z$. Over a finite redshift range, an observer at the origin may determine the age difference between objects $A$ and $B$ as
\be
t_A-t_B=\int_{z_{A}}^{z_B}\frac{\d z}{(1+z)H_\|(z;\bm e)}\,,
\ee
where the integral is along the null curve connecting $A$ and $B$. In a general spacetime, the absolute age of an object is given by the time interval from the big bang, which may not be homogeneous. Therefore, 
\be
\tau=\int_{t_\text{BB}}^t\d t=t-t_\text{BB}\,,
\ee
where the integral is along the worldline of the object. So, in terms of absolute ages we have
\be\label{bang}
\tau_A-\tau_B=\int_{z_{A}}^{z_B}\frac{\d z}{(1+z)H_\|(z;\bm e)}-t_\text{BB}(A)+t_\text{BB}(B)\,.
\ee

Measurements of ages, then, provide two important insights into inhomogeneity: firstly they probe the radial Hubble rate; secondly they give a direct measure of the bang time function, which arrises because surfaces of constant time may not be orthogonal to the matter. In a realistic model, this could represent the time at which a certain temperature was attained (so switching on a given type of cosmological clock). More generally, the `big bang' need not be homogenous, and ages provide a mechanism to probe this, given a separate measurement of~$H_\|$~\cite{Heavens:2011mr}.

\subsection{Does `almost' isotropy imply `almost' homogeneity?}

The results above are highly idealised because they assume perfect observations, and observables such as isotropy which are not actually representative of the real universe. In reality, the CMB temperature anisotropies, though small are not zero; local observations are nearly isotropic, but not to the same degree as the CMB. Are the above results stable to these perturbations?

The key argument is known as the almost-EGS theorem~\cite{Stoeger:1994qs,Maartens:1994qq,Maartens:1995hh,Maartens:1995zz,Stoeger:1999tb}:
\begin{description}
\item[{\bf \footnotesize [SME]}~{\em Almost isotropic} CMB \withcp {\em almost} FLRW:] 
In a region of an expanding Universe with dust and cosmological constant, if all dust observers measure an almost isotropic distribution of collisionless radiation, then the region is almost FLRW, provided certain constraints on the derivatives of the multipoles are satisfied.
\end{description}

The starting point for this proof is to assume that the multipoles of the radiation are much smaller than the monopole, which is exactly what we measure:
\be
\frac{|\Pi_{a_1a_2\cdots a_\ell}|}{\rho_r}={\cal O}(\epsilon_r)~~~\text{for}~~~ \ell=1,\ldots, 4
\ee
where ${\cal O}(\epsilon_r)$ is a smallness measure. However, we can see from Eq.~(\ref{nl8}) that we need further assumptions on the derivatives of the multipoles to prove almost homogeneity:
\be\label{gradaass}
\frac{|\D_b\Pi_{a_1a_2\cdots a_\ell}|}{\rho_r}=\frac{|\dot\Pi_{a_1a_2\cdots a_\ell}|}{\rho_r}={\cal O}(\epsilon_r)~~~\text{for}~~~ \ell=1,\ldots, 3
\ee
The proof proceeds as in the exact case above, but with $=0$ replaced by $={\cal O}(\epsilon_r)$ (except for $A_a=0$ exactly because the observers are dust). An `almost FLRW' condition is then arrived at:
\ba
\text{kinematics}:&&~~~
|\sigma_{ab}|=|\omega_a|=|\D_c\sigma_{ab}|=|\D_b\omega_a|=|\D_a\Theta|=\cdots={\cal O}(\epsilon_r)\\
\text{curvature}:&&~~~
|E_{ab}|=|H_{ab}|=|\D_cE_{ab}|=|\D_cH_{ab}|={\cal O}(\epsilon_r)\,,
\ea
in the region where the CMB is close to isotropic. Outside that region the spacetime need not be close to FLRW~\cite{Lim:1999uh}.

This proof is an important attempt to realistically arrive at a perturbed FLRW model using the CP and observables. It has been criticised as a `realistic' basis for near-homogeneity due to the reliance of the assumptions on the spatial gradients of the multipoles, Eq.~(\ref{gradaass})~\cite{Rasanen:2009mg}\footnote{Note that~\cite{Rasanen:2009mg} is discussing something slightly different to the almost EGS theorem. In almost EGS, the assumption is that all observers measure nearly isotropic radiation; in~\cite{Rasanen:2009mg} the assumption is that $p^a\del_a E$ is almost isotropic, but this is not the same thing. In the exact case this condition implies that the acceleration must be zero independent of the matter~-- this is not necessary for isotropic radiation, cf the radiation isotropy condition above.}. However, it seems reasonable that the gradients of the very low multipoles are small compared to their amplitude (i.e., it seems reasonable that the CMB power spectrum does not change a lot as we move from observer to observer). This is because the multipoles peak in power around the scale they probe, which for the low multipoles required for the theorem are of order the Hubble scale. So, while they may change rapidly on small scales, the power of such modes is significantly diminished.
Whether or not such criticisms  are justified, the fact we have to make such assumptions means that the observational case in the real universe is much harder than the exact results imply. How could we observe spatial gradients of the octopole of the CMB?

One route around this is to combine local distance information as well, and try for an almost-HP theorem. Consider the case of irrotational dust. Let us assume that all observers measure an approximately isotropic distance-redshift relation, and that the multipoles are bound by ${\cal O}(\epsilon_d)$. Clearly, we have, if this is the bound for all observers, $\sigma_{ab}={\cal O}(\epsilon_d)$.  Using the fully non-linear expression for the ${\cal O}(z^2)$ term requires near-isotropy of
\bea
\fl
K^aK^bK^c\nabla_a\nabla_bu_c = \sfrac{1}{6}(\rho+3p) + \sfrac{1}{3}\Theta^2 -\sfrac{1}{3}\Lambda
+ \sigma_{ab}\sigma^{ab}
+ e^a\left[\sfrac{1}{3}\D_a\Theta+\sfrac{2}{5}\div{\sigma}_a \right]\nonumber\\   
+e^{\<a}e^{b\>}\left[E_{ab}
+ \sfrac{5}{3}\Theta\sigma_{ab} + 2 {\sigma^c}_{a}\sigma_{bc}\right]
+ e^{\<a}e^be^{c\>} \D_{a}\sigma_{bc}\label{jhsdbc}
\end{eqnarray}
which, together with Eq.~(\ref{c1}), and assuming that time derivatives of ${\cal O}(\epsilon_d)$ quantities are ${\cal O}(\epsilon_d)$, give the conditions 
\be
|\div\sigma_{a}|=|\D_{\<a}\sigma_{bc\>}|=|\D_a\Theta|=|E_{ab}|=|\div E_a|=|\div H_a|={\cal O}(\epsilon_d)\,.
\ee
Yet, we still cannot arrive at the almost-FLRW condition as there is not enough information to switch off `curl' degrees of freedom: $H_{ab}$, $\curl E_{ab}$, and $\curl\sigma_{ab}$ are unconstrained. Polarisation of the CMB could be used to constrain these, but it would be interesting to see how everything fits together.

A critical issue with these arguments lies in the question of what we mean by `almost-FLRW', and what `$=\mathcal{O}(\epsilon)$' means. A sensible definition might be a dust solution which is `almost` conformally flat. (Exact conformal flatness implies exactly FLRW.) Or, as used in the almost-EGS case, a set of conditions on all 1+3 irreducible vectors and PSTF tensors (which are necessarily gauge-invariant in the covariant approach) having small magnitudes: 
\be
|X_{ab\cdots c}|=\sqrt{X_{ab\cdots c}X^{ab\cdots c}}={\mathcal{O}(\epsilon)}.
\ee
Most quantities at first order average to zero in a standard formulation, so taking magnitudes is important. 
One of the conditions which does not rely on coordinates might therefore be $E_{ab}E^{ab}={\cal O}(\epsilon^2)$ in the notation above. Unfortunately, this is non-trivial. Consider a linearly perturbed flat $\Lambda$CDM model in the Poisson gauge. Then,
\be
E_{ab}\propto\partial_{\<a}\partial_{b\>}\Phi~~~\Rightarrow~~~E_{ab}E^{ab}\propto (\partial_{\<a}\partial_{b\>}\Phi)\,(\partial^{\<a}\partial^{b\>}\Phi)\,.
\ee
Evaluating the ensemble average of this gives its expectation value, which, assuming scale-invariant initial conditions gives~\cite{Clarkson:2011uk}
\be\label{EE}
\frac{\sqrt{\overline{E_{ab}E^{ab}}}}{H_0^2}\sim\Delta_\mathcal{R}\left(\frac{k_\text{eq}}{k_H}\right)^2\ln^{3/2} \frac{k_\text{uv}}{k_\text{eq}}\sim 2.4\, \Omega_m^2 h^2\,\,\ln^{3/2} \frac{k_\text{uv}}{k_\text{eq}}\,.
\ee
Here, $k_\text{uv}$ is the wavenumber of some UV cutoff, and $k_\text{eq}$ is the wavenumber of the equality scale. The ensemble average has been evaluated assuming ergodicity via a spatial integral over a super-Hubble domain.
We have a scalar which is $\mathcal{O}(1)$ times a term which diverges in the UV. The divergence here represents modes which are smoothed over, and seems to be a necessary condition for writing down an approximately FLRW model. Eq.~(\ref{EE}) is certainly not ${\cal O}(\epsilon)$ in the normal sense of the meaning.  This is because, roughly speaking, the Weyl tensor has a large variance even though we would like it to be `small' for our covariant characterisation of FLRW. This is true also for most covariant objects with indices~-- their magnitudes are actually quite large in a perturbed FLRW model! Products of  quantities such as $\sigma_{ab}\sigma^{ab}$ and $E_{c\<a}{\sigma_{b\>}}^c$ act as sources in the 1+3 equations. 

Instead we might try to define almost FLRW in the conventional perturbative sense. That is, we write the metric in the Newtonian gauge and require that the potential $\Phi$, its first derivatives, and the peculiar velocity between the matter frame and the coordinate frame is small. This is often claimed to be a sufficient condition for a spacetime to be close to FLRW~\cite{Ishibashi:2005sj}. While this is fine in certain contexts, it is not necessarily so for large-scale inhomogeneities we are concerned with here. The LTB models provide a counter-example, as they can also be written as perturbed FLRW~\cite{VanAcoleyen:2008cy}, yet are clearly not `almost FLRW' in the sense we are interested in.


Consequently, a robust, non-perturbative, covariant definition of almost-FLRW is lacking and we are forced into the realm of the averaging problem: to define almost-FLRW, we must smooth away power on scales smaller than a few Mpc. How should this be done covariantly and what are the implications~\cite{Clarkson:2011zq,Buchert:2011sx}?

\section{Null hypotheses for FLRW and tests for the Copernican principle}

As we have seen, it is rather difficult to conclude homogeneity even given ideal observations. Indeed, it is subtle to robustly deduce (approximate) spherical symmetry about us given (near) isotropy of observations. It is rather surprising how sparse our current observations are in comparison to what is required from the theorems above, one of which includes observing transverse proper motions for an extend period! 

Nevertheless, the results above allow us to formulate some generic tests of the Copernican and cosmological principles, and the underling assumption of an FLRW geometry and its accompanying observational relationships~-- i.e., the usual observational relationships which are derived assuming there are no issues from averaging. We refer to this as the `FLRW assumption' below. (If there are non-trivial effects associated with averaging, then some of the tests below can also be used to signal it~-- see e.g.,~\cite{Wiltshire:2009db} for some specific examples.)

The types of tests we consider range in power from focussed tests for the concordance model (flat $\Lambda$CDM), to generic null tests which can signal if something is wrong with the FLRW assumption under a wide variety of circumstances irrespective of dark energy, initial conditions or theory of gravity. In this sense, they formulate our understanding of the Copernican principle as a null hypothesis which can be refuted but not proven. One of the conceptually important issues with some of these tests is that they can, in principle, be utilised by simply smoothing observed data, and so do not require an underlying model to be specified at all. These tests should be considered as additional to checking various observables for isotropy, which we assume below.

\subsection{Tests for the concordance model}

\begin{SCfigure}
~~~~\includegraphics[width=0.27\textwidth]{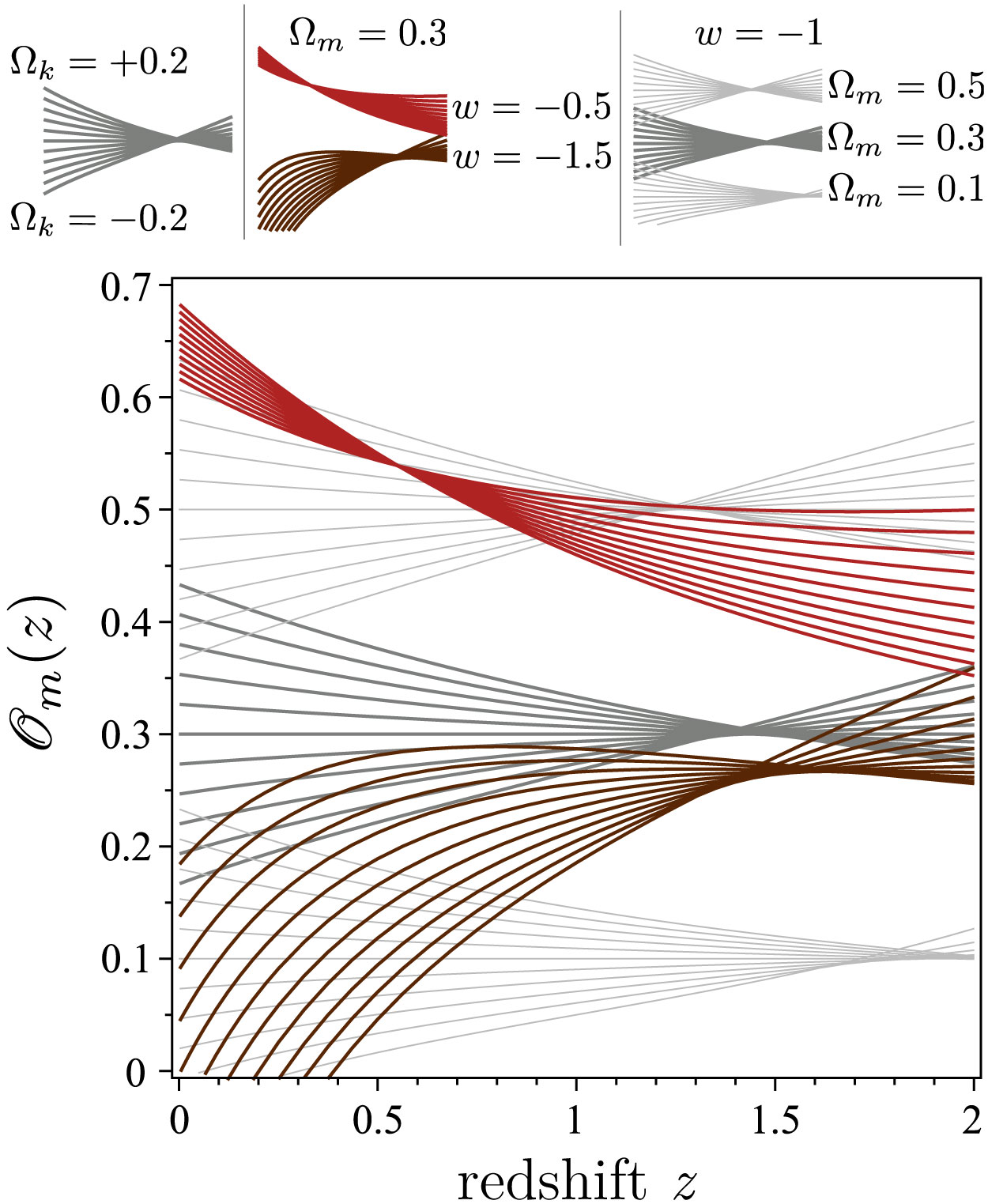}~~
\includegraphics[width=0.4\textwidth]{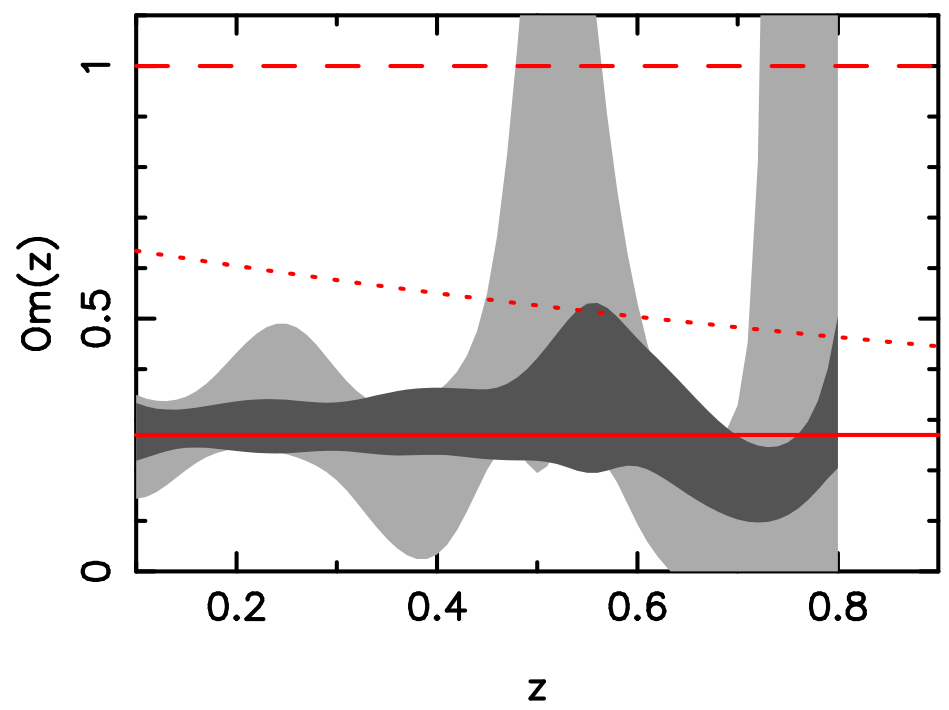}
\caption{Left: $\mathscr{O}_m(z)$ obtained using distances for different FLRW models. The figure shows the range of behaviour from curvature in each fan of curves (key, top left). The three grey fans show how differing $\Omega_m$ values interact with curvature for $\Lambda$CDM (key, top right). The effect of changing w by a constant is illustrated in the red and brown fans. (From~\cite{Shafieloo:2009hi}.) \\
Right: Constraints on $\mathscr{O}_m(z)$ (called $Om(z)$ in~\cite{Sahni:2008xx}) from the latest Wigglez BAO data. (From~\cite{Blake:2011ep}.)}
\label{om-wigglez}
\end{SCfigure}

A simple consistency test for flat $\Lambda$CDM may be formulated easily. In a flat $\Lambda$CDM model we can rearrange the Friedmann equation to read~\cite{Zunckel:2008ti,Sahni:2008xx}
\be\label{om}
\Omega_m=\frac{h^2(z)-1}{(1+z)^3-1}=\frac{1-D'(z)^2}{[(1+z)^3-1]D'(z)^2}\equiv\mathscr{O}_m(z)\inflcdm\text{const.},
\ee
where $h(z)=H(z)/H_0$ is the dimensionless Hubble rate. Viewed in terms of the observable functions on the rhs, these equations tells us how to measure $\Omega_m$ from $H(z)$ or $D'(z)$.  Furthermore, if flat $\Lambda$CDM is correct the answer should come out to be the same irrespective of the redshift of measurement. This provides a simple consistency test of the standard paradigm~-- deviations from a constant of $\mathscr{O}_m(z)$ at any $z$ indicates either deviations from flatness, or from $\Lambda$, or from the FLRW models themselves~\cite{Zunckel:2008ti,Sahni:2008xx}. 
 See Fig.~\ref{om-wigglez}.

To implement it as a `consistency test' one firsts smooths some data in an appropriate model-independent way, and then constructs the function $\mathscr{O}_m(z)$ to determine if it is consistent with the null hypothesis that it should be constant. See~\cite{Shafieloo:2009hi} for a discussion of how to do this with SNIa data, and~\cite{Blake:2011ep} for an example using the BAO to measure $H(z)$.

More generally, if we don't restrict ourselves to $\Omega_k=0$, we have that~\cite{Clarkson:2009jq}\footnote{Thanks to Marina Seikel and Sahba Yahya for correcting an error in these formula in~\cite{Clarkson:2009jq}.} 
\ba
\Omega_{{k}}&=&\Upsilon(z)\left\{ 2\left( 1-(1+z) ^{3}
 \right)D''+3D'(D'^2-1)( 1+z)^{2}
\right\}\equiv\mathscr{O}^{(2)}_k(z)\label{O2k}\inlcdm\text{const.}\\
\Omega_m&=&2\Upsilon(z)\left\{ \left[(1+z
 )^{2} -D^{2}-1 \right]D'' -
 \left( D'^2 -1 \right)  \left[( 1+z)D' -D 
 \right] 
\right\}\equiv\mathscr{O}^{(2)}_m(z)\label{O2m}\inlcdm\text{const.}
\ea
where 
\ba
\Upsilon(z)^{-1}= 2\left[ 1-(1+z) ^{3} \right]  D^2 D'' 
  -
 \left\{  (1+z)\left[  ( 1+z)^{3}-3( 1+z)+2 \right]  D'^{2}-
 2\left[ 1-( 1+z)^{3} \right] DD' -3(1+z)^{2}D^{2} \right\}D' .\nonumber
\ea
The numerator of the formula for $\Omega_k$ forms the basis of the test presented in~\cite{Zunckel:2008ti}. Again, these formulae for $\Omega_m$ and $\Omega_k$ can be used to test consistency of the $\Lambda$CDM model. Note that each of these tests depend only on $D(z)$, and not on any parameters of the model. 

\subsection{Tests for FLRW geometry}

\subsubsection{Hubble rate(s) and ages}

In an FLRW model, the two Hubble parameters we have discussed, $H_\|(z;\bm e)$ and $H_\perp(z;\bm e)$ measured in any direction must be the same at the same redshift. That is,
\be
H_\|(z;\bm e)=H_\perp(z;\bm e)=H(z)\,.
\ee
In a typical LTB model, for example, this simple relation is violated (with the exception of a fine-tuned sub-class of models of course), and so this forms a basic check of the FLRW assumption, and Copernican principle. Thus we have an isotropic expansion test:
\be
\mathscr{H}(z)=H_\|(z;\bm e)-H_\perp(z;\bm e)\inflrw0\,.
\ee
An important question here is how $H_\perp$ can be measured. In~\cite{UCE,Clarkson:2009jq,Quartin:2009xr,Yoo:2010hi,Dunsby:2010ts} it was shown that, in LTB models for a central observer, the `redshift drift'~-- the change in redshift of a source measured over an extended time~-- is determined by the angular Hubble rate:
\be 
\dot z(z)=(1+z)H_0-H_\perp(z)\,. 
\ee 
Although it is a rather sensitive observable~\cite{ELT}, it gives an important consistency check for the standard model. Another possibility to measure $H_\perp(z)$ is the polarised SZ effect (see below).

Measuring ages of objects is notoriously difficult, yet can provide important insights as it probes the geometry inside our past null cone. At the simplest level, the \emph{relative} ages of objects can be measured between redshift bins giving the Hubble rate. At a redshift $z$, the Hubble rate is may be found from Eq.~(\ref{dzdt}) if we know the age difference between two nearby objects, $\delta t$ separated by $\delta z$:
\be
H_\|(z)=-\frac{1}{1+z}\frac{\delta z }{\delta t}\,.
\ee
The fossil record of galaxies provides, in principle, the absolute ages of galaxies as a function of $z$ assuming a model of galaxy evolution.  In an FLRW model, this constrains the line of sight integral of the Hubble rate as~\cite{Heavens:2011mr}
\be
\mathscr{S}(z)=\tau_\text{us}-\tau_\text{galaxy}-\int_0^{z_\text{galaxy}}\frac{1}{(1+z)H_\|(z)}\inflrw0\,.
\ee
This has yet to be implemented.

The dipole in the distance data can measure $H(z)$ in a perturbed FLRW context~\cite{Bonvin:2006en}. It would be interesting to see in a more general context whether this is $H_\|(z)$, $H_\perp(z)$ or $\Theta(z)$, thereby determining how it can be used for~$\mathscr{H}(z)$.

\subsubsection{Curvature test}

A rather general test of FLRW geometry may be found by considering the distance-redshift relation in an open or closed FLRW model. Consider the dimensionless comoving distance:
\begin{equation}\label{d_L}
D(z)=(1+z)H_0d_A(z)=\frac{1}{\sqrt{-\Omega_k}}\sin{\left( 
\sqrt{-\Omega_k}\int_0^z{\frac{\mathrm{d}z'}{h(z')}}\right)}\,,
\end{equation}
where $h(z)=H(z)/H_0$ and $d_A$ is the area or angular diameter distance. Rearranging Eq.~(\ref{d_L}), we have the curvature parameter today given by, for any curvature~\cite{Clarkson:2007bc,CBL}
\begin{equation}\label{OK}
\Omega_k=\frac{\left[h(z)D'(z)\right]^2-1}{[D(z)]^2}\equiv\mathcal{O}_k(z)\inflrw\text{const.}
\end{equation}
On the face of it, this gives a way to measure the curvature parameter today by combining distance data with Hubble rate data, irrespective of the redshift of measurement. In FLRW this will be constant as a function of $z$, independently of the dark energy model, or theory of gravity. Alternatively, we may re-write this as the condition that~\cite{CBL}
\begin{equation}\label{C(z)}
\mathscr{C}(z)=1+h^2\left(DD''-D'^2\right)+hh'DD'\inflrw0\ ,
\end{equation}
by virtue of Eq.~(\ref{OK}). In more general spacetimes this will not be the case. In particular, in LTB models, even for a  for a central observer, we have $\mathscr{C}(z)\neq0$, or $\mathscr{O}_k(z)\neq$\,const. 

\begin{figure}[htbp]
\includegraphics[height=0.32\textwidth]{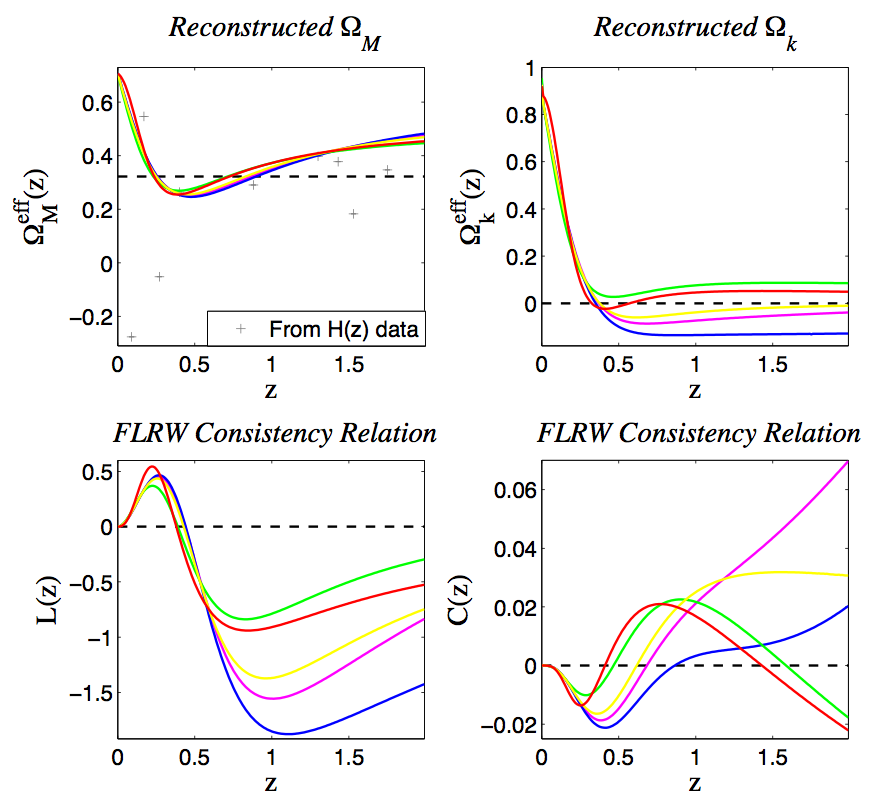}\hfill
\includegraphics[height=0.32\textwidth]{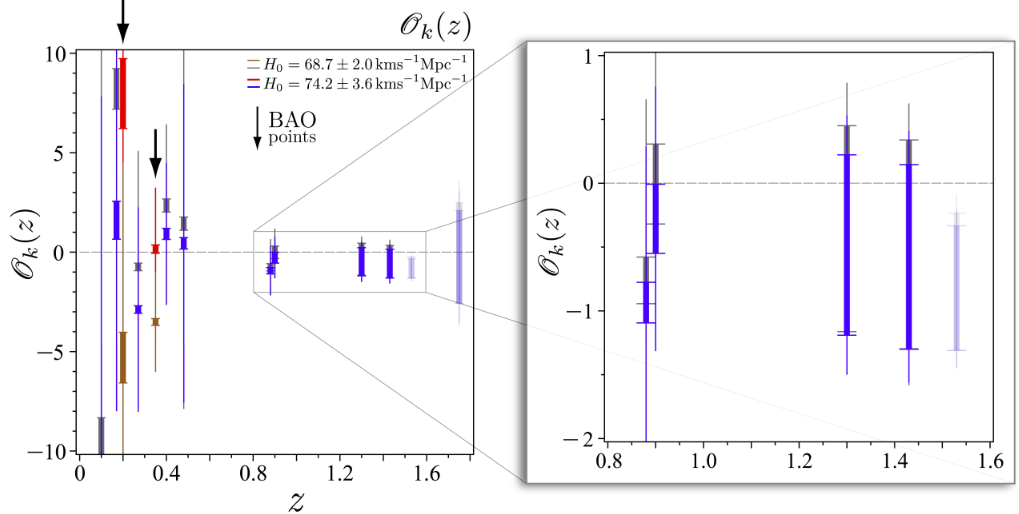}
\caption{Left: Consistency functions, shown for anti-Copernican models. The functions $\mathscr{O}_m(z)$, $\mathscr{O}_k(z)$, $\mathscr{L}(z)=[\text{numerator of}~\mathscr{O}^{(2)}_k(z)]$~\cite{Zunckel:2008ti} and $\mathscr{C}(z)$ are shown for a variety of void models. Deviations from constant indicate inconsistency with the standard paradigm, but only $\mathscr{O}_k(z)\neq\,$const. and  $\mathscr{C}(z)\neq0$ definitely indicates the FLRW assumption is wrong. (From~\cite{FLSC}.)\\
Right: Constraints on $\mathscr{O}_k(z)$ from a combination of SNIa and $H(z)$ data from the relative ages of LRGs. This function is most usefully constrained at high redshift (blow-up). (From~\cite{Shafieloo:2009hi}.)
} 
\label{consistency}
\end{figure}

This tells us that in all FLRW models there exists a precise relationship between the Hubble rate and distance measurements as we look down our past null cone. This relationship can be tested experimentally \emph{without specifying a model at all}, if we reconstruct the functions $H(z)$ and $D(z)$ in a model independent way and independently of each other. This provides a model-independent method by which to experimentally verify the Copernican assumption, and so verify the basis of the FLRW models themselves (see Fig.~\ref{consistency}). 

The curvature test Eq.~(\ref{OK}) has been implemented in~\cite{Shafieloo:2009hi} using a mixture of SNIa, ages [to give $H(z)$], and BAO data; constraints are weak at present (Fig.~\ref{consistency}). This is because the parameter is unstable at low redshift, while at high-$z$ the SNIa data is very sparse. A variant of this was presented in~\cite{Mortsell:2011yk} where it is reformulated as an integral test to measure curvature with much better results. Consistency between $D(z)$ and $H(z)$ measurements was explored and found in~\cite{Avgoustidis:2009ai}. The dipole in the distance data can measure $H(z)$~\cite{Bonvin:2006en} so can be used directly in the curvature tests $\mathscr{C}(z)$ and $\mathscr{O}_k(z)$. This means that these curvature tests can be formulated from distance data alone if we interpret $D(z)$ as the monopole of the distance-redshift relation in Eq~(\ref{C(z)}), though this has not yet been investigated.

\subsubsection{CMB}

Goodman's tests allow us to directly `observe' the CMB from other locations. Backed up with the EGS theorems, and the scattering theorem discussed in Sec.~\ref{BB}, they help us to deduce homogeneity.  Different multipoles can be probed in different ways, and so provide independent consistency conditions on the geometry at late times. If we consider fundamental observers comoving with the matter flow (i.e., without boosting to the CMB frame) we have:

\begin{description}

\item[Monopole] In FLRW, the monopole temperature around distant observers redshifts as $(1+z)$, which is not the case in an inhomogeneous geometry. (The direct temperature along a line of sight redshifts like this of course, but the monopole does not.) Thus we have, assuming standard physics, 
\be
\mathscr{T}(z)=\frac{\<T_\text{CMB}\>(z)}{T_0(1+z)}-1\inflrw0\,,
\ee
where $\<T_\text{CMB}\>(z)$ is the monopole temperature observed at redshift $z$ in some frame $u^a$, and $T_0$ is our CMB temperature. This can be measured, for example, via the thermal SZ effect~\cite{Goodman:1995dt} and via atomic fine-structure transitions in cool absorption-line systems along the line of sight to high-redshift quasars~\cite{LoSecco:2001zz}. Constraints are reasonably tight already: \cite{Avgoustidis:2011aa} find that possible deviations of the form $\<T_\text{CMB}\>(z)= T_0(1 + z)^{1-\beta}$ to be $\beta = 0.004 \pm 0.016$ up to  $z \sim 3$ from SZ measurements.

\item[Dipole] Measuring the radial dipole around other observers using the kSZ effect constrains a mixture of their local peculiar velocity (which has a component to the mean matter frame as well as to the CMB frame) as well as spatial temperature gradients in the monopole (and higher-order effects) along the line of sight (cf Eq.~\ref{e3i} with $I=r$). Obviously this must be small in the standard model, so a strong kSZ signal would violate homogeneity. The integrated kSZ signal is consistent with the standard model~\cite{Zhang:2010fa} (see Sec.~\ref{SZ}), but there have been measurements of very large dipoles measured around X-ray clusters~\cite{Kashlinsky:2012gy}, which has been interpreted as a `dark flow', and is not consistent with $\Lambda$CDM. 

The $y$-distortion also depends on the total dipole which, if other contributions to the $y$-distortion can be accounted for, offers the possibility of inferring transverse velocities by comparing it with the kSZ signal (which depends only on the radial dipole). Transverse velocities are a key ingredient for determining isotropy of the lightcone, as well as for determining $H_\perp(z)$, so constraining these are a key test for homogeneity.

\item[Higher Multipoles] Higher multipoles generate $y$-distortion of the CMB, as we have seen in Sec.~\ref{SZ}. A large quadrupole principally indicates large shear, which in turn generates an octopole and higher moments. Using the EGS theorem, and the scattering theorem in Sec.~\ref{BB} we can see that a small $y$-distortion indicates strong evidence for FLRW geometry under quite general conditions. Scattering of the quadrupole seen by clusters induces a polarisation signal in our CMB sky and correlations between clusters can be used to place constraints on homogeneity~\cite{Bunn:2006mp}.


\item[Polarisation] The SZ effect also affects the intrinsic polarisation of the CMB~\cite{1980MNRAS.190..413S}. The transverse velocity of the cluster relative to the CMB frame can be extracted and used to constrain homogeneity~\cite{Maartens:2011yx} (see also~\cite{Goto:2011ru}).

\end{description}

In addition to these features, the CMB can be used with the integrated Sachs-Wolfe effect to constrain the geometry~\cite{Tomita:2009wz,Tomita:2010zz}, but it is challenging to do so in a model independent manner.


\begin{SCfigure}
~~~~~~\includegraphics[width=0.6\textwidth]{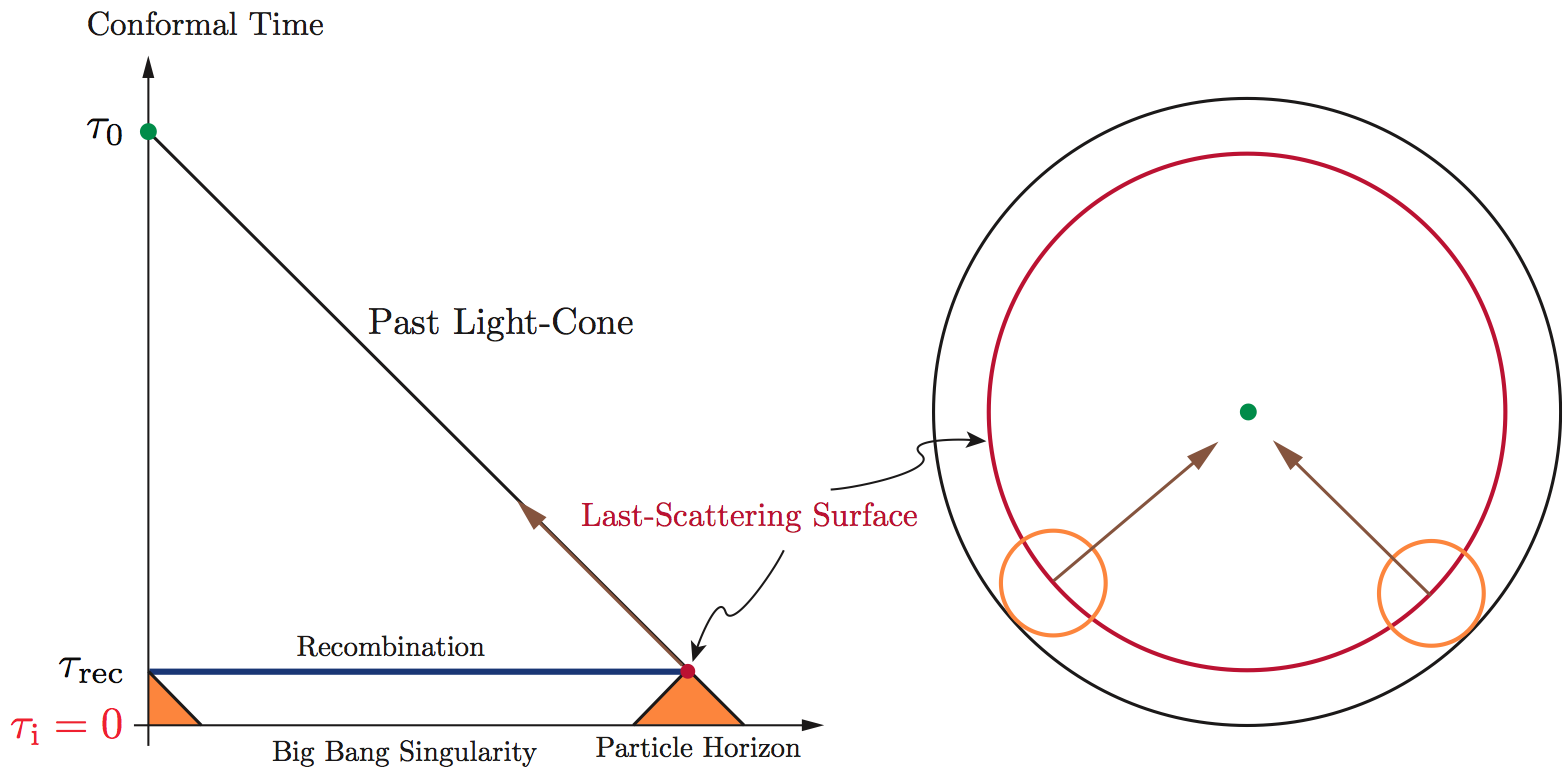}
\caption{Without inflation the last-scattering surface consists of $10^5$ causally disconnected regions. We only directly observe those on our primary CMB sky and we assume the others inside our past lightcone share precisely the same history. (From~\cite{Baumann:2009ds}.)}
\label{default}
\end{SCfigure}

\subsection{Probing homogeneity in the early universe.}

Homogeneity in the early universe may be constrained in different ways. In particular, the early universe provides the primary place we can place constraints on various parameters such as the baryon fraction whose physical origins are poorly understood, yet are assumed to be homogeneous. This follows from inflation of course, so probing FLRW geometry at early times, and in particular whether widely separated regions share a common thermal history~\cite{1986MNRAS.218..605B}, is an important consistency check for the standard model. As we have seen, inhomogeneous models require a significant reworking of the early universe in order to have a chance of working as dark energy alternatives. Each era provides opportunities for confirming homogeneity:

\begin{description}

\item[Last scattering] To a first approximation, this happens at fixed temperature $T_*$ on a surface of constant proper time in an FLRW model. This temperature depends principally on the baryon fraction and the baryon-photon ratio, as well as $N_\text{eff}$, all of which should be constrained to be homogeneous. While we have tight constraints on the variability of these parameters across our CMB sky (e.g., the constraints on isocurvature modes are quite tight~-- see e.g.,~\cite{Valiviita:2009bp}), there are no constraints from within our past lightcone. The temperature of the last scattering surface may be probed using the SZ effect, though this must be disentangled from inhomogeneity in the region of the scatterer.

\item[Drag epoch] The sound horizon size at the drag epoch also depends on the the (effective) number of relativistic degrees of freedom in addition to the baryon-photon ratio and the baryon fraction. This scale is probed inside our past lightcone via the BAO. Assuming that any large scale inhomogeneity we are probing has a length scale much larger than the horizon size at this time, the proper size of the sound horizon at the drag epoch may be written locally as 
\bea
d_s=\frac{h}{\sqrt{3}H_0T_{d}}\int_{T_{d}}^\infty dT\,\, T^{-3/2}
\left[(\varpi_\gamma+\varpi_\nu)T + \varpi_b\frac{\eta}{f_b}\right]^{-1/2}\left(1+\frac{3}{4}\frac{\varpi_b\eta}{\varpi_\gamma T}\right)^{-1/2}\label{rs},
\eea
where $h/H_0\approx2998\,\text{Mpc}$  The temperature of the drag epoch $T_d$ is also a function of $f_b$ and $\eta$. This scale is imprinted in the galaxy correlation function. Assuming we could measure the Hubble rates $H_\|$ and $H_\perp$ through other means, the BAO peak is then a direct measurement of $d_s(r;\bm e)$, and therefore constrains $f_b(\bm x)$ and $\eta(\bm x)$. Note that this highlights the problem of using the BAO to infer the Hubble rate at late times: we typically have to assume homogeneity at early times to do so in order to find $d_s$.

\begin{SCfigure}
~~~~\includegraphics[width=0.6\textwidth]{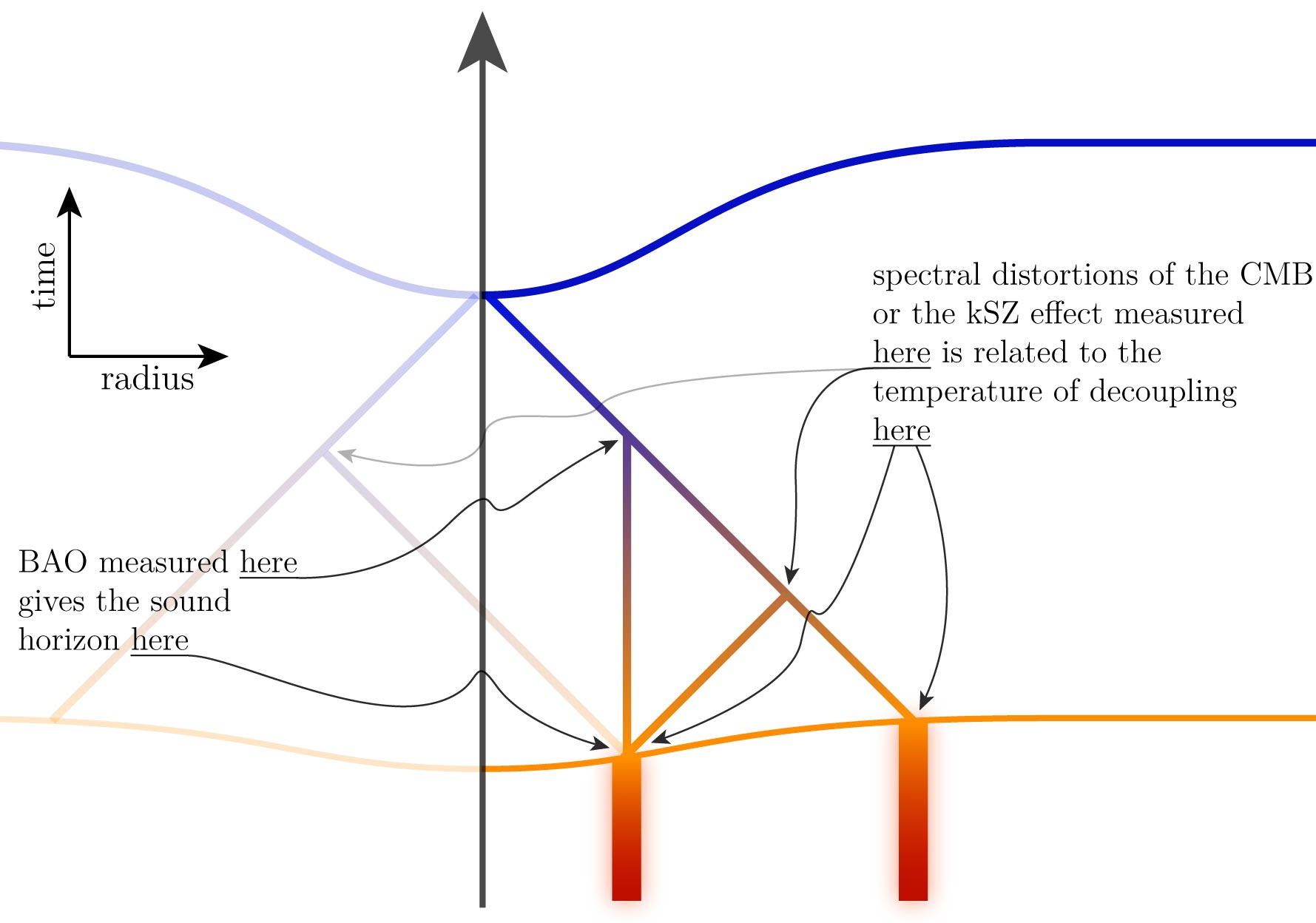}
\caption{BAO and measurements of the CMB anisotropies around distant observers tell us conditions around the time of decoupling inside our past lightcone, and so allow us to constrain homogeneity at that time. To use these probes in a model-independent way they need to be cross-correlated with other observations in different directions such that they both measure conditions at the same radius. In principle a spatial map of the interior of our past lightcone can be made of this era. (From~\cite{Clarkson:2010ej}.)}
\label{ksz-fig}
\end{SCfigure}

\item[BBN] Primordial nucleosynthesis is a powerful test of early time homogeneity, because we can measure local light element abundances which were synthesised on our past worldline at early times. This can be compared with element the element abundance at high redshift and at the CMB~-- i.e., in the past of worldlines very far from us. The ratios of light element abundances depend strongly on the baryon-photon ratio $\eta$ which can be probed spatially by measuring it in the CMB, and through local abundances of hydrogen, helium and lithium~\cite{Goodman:1995dt,RC}.  It is intriguing that abundances of $^7$Li implies a value for the local baryon-photon ratio at odds with the CMB observed one by about 50\%, and the disagreement is around 5-$\sigma$. It is easy to interpret this as a spatially varying $\eta$~\cite{RC}. This is one of the few observations potentially at odds with the homogeneity assumption. 

\item[Neutrino decoupling] The neutrino background contains information from inside our past lightcone if neutrinos are massive. In that case, they do not travel on null geodesics, and therefore contain information about inhomogeneity from the time of neutrino decoupling~\cite{Jia:2008ti}.


\item[Primordial power spectrum] The amplitude of the power spectrum measured via the CMB compared to low redshift tracers may be different for several reasons. In the standard model, a tilt to the spectrum implies that different scales have different power. In a more general model with large-scale inhomogeneity the spectrum in one region may be effectively disconnected from the spectrum in another. In the case of a spherically symmetric model this is easy to envisage because the spectrum in each shell may have different amplitude, while maintaining scale-invariance (say) within each shell. Thus, the primary CMB delivers the amplitude of perturbations in a comoving sphere $\sim13$\,Gpc away from us (as well as the tilt, over that sphere), while local measurements of $\sigma_8$ tell us the amplitude of primordial perturbations smoothed over a comoving region close to our past worldline. Tracers at different redshifts probe power in different shells around us, and thus can be used to constrain homogeneity at very early times. Inhomogeneity could also be signalled by deviations from scale invariance which is different in the radial and angular directions (e.g., a different tilt may be found in the CMB versus LSS), as well as from shell to shell.


\item[The end of inflation/`bang time'] Many inhomogeneous models contain a `bang time' functional degree of freedom in them. In a comoving frame such a function has the interpretation of a big bang singularity along each worldline, except it happens at different times in the past according to observers on a hypersurface orthogonal to the fluid flow. What such a degree of freedom might actually represent in a realistic model remains to be seen, but it can be thought of as a `decaying mode' degree of freedom which can in principle be constrained observationally. Recall Eq.~(\ref{bang}):
Given an observational probe of $H_\|(z)$, and a model for local structure evolution (galaxy and star formation), this allows us to constrain any bang time inhomogeneity between worldlines $A$ and $B$~\cite{Heavens:2011mr}. Much tighter constraints come from Goodman's tests~-- it must be $\mathcal{O}(t_*)$~\cite{Bull:2011wi,Zibin:2011ma}.

\end{description}

\section{Conclusions}

It is the stunning perfection of the CMB that persuades most of us that the FLRW models are the correct basis for our cosmological model. Geometrical alternatives which can reproduce such a perfect CMB require fine-tuning in space (e.g., LTB models) or time (e.g., Bianchi models~\cite{Lim:1999uh}), or require ill-defined matter content (e.g., Stephani models~\cite{CB,Barrett:1999fd,Clarkson:1999zq}). Bayesian arguments suffice to persuade us our working model is most likely correct for now.  But until we understand $\Lambda$~-- or whatever dark energy is~-- with what level of confidence should we ascribe it? Rigorously testing the Copernican assumption, and with it the FLRW framework, is therefore a vital aspect of dark energy studies. 

We have seen three distinct approaches which help us establish homogeneity:
\begin{description}
\item[Model Building:] This has been the most prolific method as it is relatively straightforward to make a model such as LTB and compare it to observations. One can then compare an alternative directly to FLRW and decide one way or the other. However, this approach has focussed almost exclusively on providing an alternative to dark energy rather than testing the Copernican principle per se, so it is not yet clear exactly what conclusions we may draw directly. Constraining inhomogeneity when dark energy is present would help test the Copernican principle generically (e.g. \cite{Grande:2011hm,Valkenburg:2011ty}).   

\item[Consistency Tests:] The FLRW models can be formulated as a series of observational null hypotheses which can be used to test the Copernican principle (among other things). The idea is to combine observables in such a way that they can reveal unambiguously if the FLRW assumption is wrong. Ideally these are independent of dark energy or theory of gravity. Where possible, `observables' such as the distance redshift relation are constructed by smoothing data in as model-independent way as possible to remove any residual bias which might exist. Robust application of these requires removing as many model dependent assumptions as possible: for example, in the case of SNIa this is relatively straightforward; in the case of the BAO this is much harder. 

\item[Inverse Approach:] The idea of using data directly to construct the metric on the past lightcone, and then integrating into the interior, goes back to a pioneering paper by Ellis et. al.~\cite{7b} based on a very old thesis by Maartens~\cite{7}. This is an approach to cosmology which circumvents the Copernican assumption entirely: in theory, with perfect observations, we have just enough information from our past lightcone to reconstruct the correct model of the universe within it~-- whatever it is. One of the key requirements is for transverse velocities (or ${H_\perp}_{ab}$), so in practise this is a \emph{very} challenging method, but important nevertheless. 

\end{description}

Each of these approaches reveals significant insights into the geometry of the universe, and the restrictions that different observables place upon it. However, it is not yet clear from a theoretical point of view when we are able to say enough is enough. That is, a principle question of establishing confidence levels remains unanswered. If we constrain, say, $\mathscr{C}(z)<\varepsilon_\mathscr{C}$ and $\mathscr{T}(z)<\varepsilon_\mathscr{T}$ over some range of redshift, then under what circumstances is that sufficient to conclude approximate FLRW geometry? A recent study suggests this might be critical to their efficacy~\cite{Busti:2012yd}. Similarly, if we can constrain a variety of radially inhomogeneous models with dark energy to be homogeneous, is this sufficient? 

Some key questions remain:
\begin{description}
\item[{\em Stability}:] Which of the consistency tests are perturbatively stable? That is, for which of the tests does `$=\mathcal{O}(\epsilon)$' imply the geometry is similarly close to FLRW? What does close to FLRW mean? 
\item[{\em Sufficiency}:] What are the sets of consistency tests which are necessary and  sufficient to determine FLRW geometry? 
\item[{\em Confidence}:] How do we use these to place a measure on model-independent limits of deviations from FLRW, given realistic data?
\end{description}
In the case of exact spherical symmetry and a dust equation of state, say, sufficiency is relatively straightforward~-- the LTB model has two free functional degrees of freedom, so if these are chosen to have the same observables as an FLRW model with the same matter content, then the models must be the same~\cite{MHE}. This could be phrased in terms of the consistency relations as: $\mathscr{C}(z)=0=\mathscr{H}(z)$ iff FLRW. Similarly, one can probably say that a model which passes Goodmans CMB tests and has $\mathscr{C}(z)=0$ is necessarily FLRW~-- but there is no proof of this as yet. Furthermore, in a realistic setting with near isotropy, approximate consistency and unknown dark energy, are these enough? How do these tests manage with realistic data? 

It is actually surprising that it is possible to test the Copernican principle at all, and it has only lately been seen as a testable assumption. As recently as 2006, George Ellis asserts in relation to the Copernican principle~\cite{Ellis:2006fy}: ``Establishing a Robertson-Walker geometry for the universe relies on plausible philosophical assumptions. The deduction of spatial homogeneity follows not directly from astronomical data, but because we add to the observations a philosophical principle that is plausible but untestable.'' It seems that this is rather too pessimistic. We have discussed quite general ideas for testing the Copernican principle which can observationally establish homogeneity within our horizon which are achievable over the coming years.

\acknowledgments 
I would like to thank Tim Clifton, George Ellis, Juan Garc\'\i a-Bellido, Roy Maartens, Valerio Marra, Marco Regis, Wessel Valkenburg and  Miguel Zumalac\'arregui P\'erez for comments on the draft, and my collaborators for many discussions. This work is funded in part by the NRF (South Africa).

\appendix

\section*{Covariant formulation of the field equations}

The 1+3 covariant Ehlers-Ellis formalism provides a physically transparent formulation of the field equations in fully nonlinear generality (see~\cite{Ellis:1998ct,Tsagas:2007yx} for reviews). The Ehlers-Ellis formalism is a covariant Lagrangian approach to gravitational dynamics, based on a decomposition relative to a chosen 4-velocity field $u^\a$.
The fundamental tensors are
\begin{equation}\label{hep}
h_{\a\b}=g_{\a\b}+u_\a u_\b,~ ~ \ep_{\a\b c}=\eta_{\a\b c d}u^ d,
\end{equation}
where $h_{\a\b}$ projects into the instantaneous rest space of comoving
observers, and $\ep_{\a\b c}$ is the projection of the spacetime alternating tensor $\eta_{\a\b c  d}=-\sqrt{-g}
\delta^0{}_{[\a}\delta^1{}_\b\delta^2{}_ c\delta^3{}_{ d]}$, and so
\begin{equation}
\eta_{abcd} = 2u_{[a}\ep_{b]cd}-2\ep_{ab[c}u_{d]}\,,~
\ep_{abc}\ep^{def}=3!h_{[a}{}^dh_b{}^eh_{c]}{}^f\,.
\end{equation}
The projected symmetric tracefree (PSTF) parts of vectors and
rank-2 tensors are
\begin{eqnarray}
V_{\langle \a\rangle}=h_\a{}^\b V_\b\,,~ S_{\langle \a\b \rangle }=
\Big\{h_{(a}{}^ c h_{\b)}{}^ d-
{{1\over3}}h^{ c  d }h_{\a\b }\Big\}S_{ c  d }\,. \label{pstf}
\end{eqnarray}
The skew part of a
projected rank-2 tensor is spatially dual to the projected vector,
$S_{\a}={1\over2}\ep_{\a\b  c}S^{[\b  c]}$, and then any projected rank-2
tensor has the decomposition
$S_{\a\b }={1\over 3}Sh_{\a\b }+\ep_{\a\b  c}S^{ c}+S_{\langle \a\b \rangle}$, where $S=S_{ c d} h^{ c  d}$. Scalars, projected vectors and PSTF tensors are the fundamental objects which represent the gravitational field. 

The covariant derivative $\nabla_{\a}$ defines 1+3 covariant time and spatial derivatives:
\begin{eqnarray}
\dot{J}^{\a\cdots}{}{}_{\cdots \b}= u^{ c} \nabla_{ c}
J^{\a\cdots}{}{}_{\cdots \b},~~~~~ \D_{ c} J^{\a\cdots}{}{}_{\cdots \b} =    h_{ c}{}^ d h^{\a}{}_ e\cdots h_{\b}{}^f
\nabla_ d J^{ e\cdots}{}{}_{\cdots f}. \label{dd}
\end{eqnarray}
The projected derivative $\D_{\a}$
defines a covariant PSTF divergence, $\div V=\D^\a V_\a\,, ~ \div S_a=\D^\b S_{\a\b}$,
and a covariant PSTF curl,
\begin{eqnarray}
\c V_{\a}=\ep_{\a\b c}\D^{\b}V^{ c}\,,~~~~ \c
S_{\a\b }=\ep_{ c  d (\a}\D^{ c}S_{\b)}{}^ d\,. \label{curl}
\end{eqnarray}

The relative motion of comoving observers
is encoded in the PSTF kinematical quantities: the volume expansion rate,  4-acceleration, vorticity
and shear, given respectively by
\begin{eqnarray}
\Theta=\D^{\a}u_{\a},~~~ A_{\a}=\dot{u}_{\a},~~~ {\omega}_\a=\c u_\a ,~~~ \sigma_{\a\b }=\D_{\langle \a}u_{\b\rangle }. \label{kin}
\end{eqnarray}
Thus
\begin{eqnarray}
\nabla_{\b}u_{\a}={{1\over3}}\Theta h_{\a\b }+\ep_{\a\b  c}\omega^{ c}
+\sigma_{\a\b }-A_{\a}u_{\b}\,. \label{du}
\end{eqnarray}

A key identity (valid in the fully nonlinear case) is
\begin{eqnarray}
\c\D_{\a}\psi \equiv \ep_{\a\b  c}\D^{\b}\D^{ c}\psi=
-2\dot{\psi}\omega_{\a} \,, \label{ri1}
\end{eqnarray}
which shows that curl grad is nonzero in the presence of vorticity (a purely relativistic feature, with no Newtonian analogue). A crucial nonlinear commutation relation for scalars is
\be
h_\a^{~\b}(\D_{\b}\psi\dot)-\D_{\a}\dot\psi = \dot\psi A_\a-\left(\frac{1}{3}\Theta h_{\a\b}+\sigma_{\a\b}+\varepsilon_{\a\b  e}\omega^ e\right)\D^\b\psi\, .\label{timespace}
\ee

The PSTF dynamical quantities which describe the
sources of the gravitational field are:
the (total) energy density $\rho=T_{\a\b }u^{\a}u^{\b}$, isotropic pressure
$p={1\over3}h_{\a\b }T^{\a\b }$, momentum density $q_{\a}=-T_{\langle \a\rangle \b}u^{\b}$,
and anisotropic stress $\pi_{\a\b }=T_{\langle \a\b \rangle}$, where $T_{\a\b }$ is
the total energy-momentum tensor. The locally free gravitational
field, i.e. the part of the spacetime curvature not directly
determined locally by dynamical sources, is given by the Weyl tensor
$C_{\a\b  c  d }$. This splits into the PSTF gravito-electric and gravito-magnetic fields
\begin{eqnarray}
E_{\a\b }=C_{\a  c \b  d}u^{ c}u^ d\,,~~
H_{\a\b }={{1\over2}}\ep_{\a c  d }C^{ c  d }{}{}_{\b  e}u^ e  \,,
\end{eqnarray}
which provide a covariant description of tidal forces
and gravitational radiation.

The Ricci and Bianchi identities,
\begin{eqnarray}
\nabla_{[\a} \nabla_{\b]} u_ c= R_{\a\b  c   d}u^{ d}, ~~~ \nabla^ d
C_{\a\b  c  d } =- \nabla_{[\a}\Big\{ R_{\b] c} - {1\over6}Rg_{\b] c} \Big\}, \label{rbi}
\end{eqnarray}
produce the
fundamental evolution and constraint equations governing the
covariant quantities. Einstein's equations are
incorporated via the algebraic replacement of the Ricci tensor
\begin{equation}\label{efe}
R^{\a\b }=T^{\a\b }-{1\over2}T_{ c}{}^{ c}g^{\a\b }+\Lambda g^{\a\b},
\end{equation}
where $T^{\a\b}$ is the total energy-momentum tensor.

The resulting equations, in fully nonlinear form and for a general
source of the gravitational field, are:\\
\noindent{\em Evolution:}
\begin{eqnarray}\fl
 \dot{\rho} +(\rho+p)\Theta+\div q &=& -2A^{\a} q_{\a}
 -\sigma^{\a\b }\pi_{\a\b }\,, 
 \label{e1}\\ \fl
 \dot{\Theta} +{{1\over3}}\Theta^2 +{{1\over2}}(\rho+3p)-\Lambda-\div
A 
&=& -\sigma_{\a\b }\sigma^{\a\b }
+2\omega_{\a}\omega^{\a}+A_{\a}A^{\a} \,,
\label{e2}\\ \fl
 \dot{q}_{\langle \a\rangle }
+{{4\over3}}\Theta q_{\a}+(\rho+p)A_{\a} +\D_{\a} p +\div\pi_{\a} 
&=& -\sigma_{\a\b }q^{\b}
+\ep_{\a\b c}\omega^\b q^{ c} -A^{\b}\pi_{\a\b } \,,
\label{e3} \\ \fl
 \dot{\omega}_{\langle \a\rangle } +{{2\over3}}\Theta\omega_{\a}
+{{1\over2}}\c A_{\a} &=& \sigma_{\a\b }\omega^{\b} \,,\label{e4}
\\ \fl
 \dot{\sigma}_{\langle \a\b \rangle } +{{2\over3}}\Theta\sigma_{\a\b }
+E_{\a\b }-{{1\over2}}\pi_{\a\b } -\D_{\langle \a}A_{\b\rangle } 
&=& -\sigma_{ c\langle \a}\sigma_{\b\rangle }{}^{ c} - \omega_{\langle \a}\omega_{\b\rangle }
+A_{\langle \a}A_{\b\rangle }\,,
\label{e5}\\ \fl
 \dot{E}_{\langle \a\b \rangle } +\Theta E_{\a\b }
-\c H_{\a\b } +{{1\over2}}(\rho+p)\sigma_{\a\b }
+{{1\over2}}
\dot{\pi}_{\langle \a\b \rangle } +{{1\over6}}
\Theta\pi_{\a\b } +{{1\over2}}\D_{\langle \a}q_{\b\rangle } 
&=&-A_{\langle \a}q_{\b\rangle } +2A^{ c}\ep_{ c  d (\a}H_{\b)}{}^ d
\nonumber\\
+3\sigma_{ c\langle \a}E_{\b\rangle }{}^{ c} 
-\omega^{ c} \ep_{ c  d (\a}E_{\b)}{}^ d &&\!\!\!\!\!\!\!\!\!\!\!-{{1\over2}}\sigma^{ c}{}_{\langle
\a}\pi_{\b\rangle  c} -{{1\over2}}\omega^{ c}\ep_{ c  d (\a}\pi_{\b)}{}^ d \,,
\label{e6}\\ \fl
 \dot{H}_{\langle \a\b \rangle } +\Theta H_{\a\b } +\c E_{\a\b }
-{{1\over2}}\c\pi_{\a\b } 
&=& 3\sigma_{ c\langle \a}H_{\b\rangle }{}^{ c}
-\omega^{ c} \ep_{ c  d (\a}H_{\b)}{}^ d \nonumber\\
-2A^{ c}\ep_{ c  d (\a}E_{\b)}{}^ c &&\!\!\!\!\!\!\!\!\!\!\!-{{3\over2}}\omega_{\langle \a}q_{\b\rangle
}+ {{1\over2}}\sigma^{ c}{}_{(\a}\ep_{\b) c  d }q^ d \,. \label{e7}
\end{eqnarray}

\noindent{\em Constraint:}
\begin{eqnarray}
\fl
 \div\omega &=& A^{\a}\omega_{\a} \,,  \label{c1}\\ \fl
 \div\sigma_{\a}-\c\omega_{\a} -{{2\over3}}\D_{\a}\Theta +q_{\a} &=&-
2\ep_{\a\b c}\omega^\b A^{ c}  \,,\label{c2}\\ \fl
  \c\sigma_{\a\b }+\D_{\langle \a}\omega_{\b\rangle }
 -H_{\a\b }&=& -2A_{\langle \a}
\omega_{\b\rangle } \,,\label{c3}\\\fl
\div E_{\a}
+{{1\over2}}\div\pi_{\a}
 -{{1\over3}}\D_{\a}\rho
+{{1\over3}}\Theta q_{\a} 
~&=& \ep_{\a\b c}\sigma^\b{}_ d H^{ c  d} -3H_{\a\b}
\omega^{\b} +{{1\over2}}\sigma_{\a\b }q^{\b}-{{3\over2}}
\ep_{\a\b c}\omega^\b q^{ c}  \,,\label{c4}\\\fl
 \div H_{\a}
+{{1\over2}}\c q_{\a}
 -(\rho+p)\omega_{\a}
 &=&
-\ep_{\a\b c}\sigma^\b{}_ d E^{ c  d}-{{1\over2}}\ep_{\a\b c}\sigma^\b{}_ d \pi^{ c  d}   +3E_{\a\b }\omega^{\b}
-{{1\over2}}\pi_{\a\b } \omega^{\b}  \,.\label{c5}
\end{eqnarray}
The energy and momentum conservation equations are the evolution equations~(\ref{e1}) and (\ref{e3}). The dynamical quantities $\rho, p, q_\a, \pi_{\a\nu}$ in the evolution and constraint equations
(\ref{e1})--(\ref{c5}) are the total quantities, with
contributions from all dynamically significant particle species.
That is,
\begin{eqnarray}
T^{ab } &=& \sum_I T_{I}^{ab } = \rho
u^{\a}u^{b}+ph^{ab }+2q^{(\a}u^{b)}
+\pi^{ab } \,, \label{t1}\\
T_{I}^{ab }&=&
\rho^*_{I}u_{I}^{\a}u_{I}^{\a}+p^*_{I}h_{I}^{\a b }
+2q_{I}^{*(\a}u_{I}^{ b)}+\pi_{I}^{*ab}\,, \label{t2}
\end{eqnarray}
where $I=r,n,b,c,\Lambda$ labels the species. The asterisk on the
dynamical quantities $\rho^*_{I},\cdots$ is intended to emphasize that these quantities are measured, not in the $u^\a$-frame, but in the $I$-frame, whose 4-velocity is given by
\begin{eqnarray}
u_{I}^{\a}=\gamma_{I}\left(u^{\a}+v_{I}^{\a}\right)\,, ~v_{I}^{\a}u_{\a}=0\,, ~\gamma_I=\left( 1-v_I^2 \right)^{-1/2}.\label{t3}
\end{eqnarray}
The total dynamical quantities in Eqs.~(\ref{e1})--(\ref{c5}), are given by
\begin{eqnarray}
\rho=\sum_I\rho_{I}\,,~p=\sum_I p_{I}\,,~ q^{\a}=\sum_I
q_{I}^{\a}\,,~\pi^{\a\nu }=\sum_I\pi_{I}^{\a\nu }\,.
\end{eqnarray}
Assuming that the species are non-interacting, they each separately obey the energy and momentum conservation equations~(\ref{e1}) and (\ref{e3}):
\begin{eqnarray}
\dot{\rho}_I +(\rho_I+p_I)\Theta+\D_\a q_I^\a &=& -2A_{\a} q_I^{\a}
 -\sigma_{\a b }\pi_I^{\a b }\,, 
 \label{e1i}\\
 \dot{q}_I^{\langle \a\rangle }
+{{4\over3}}\Theta q_I^{\a}+(\rho_I+p_I)A^{\a} +\D^{\a} p_I +\D_\nu\pi_I^{\a\nu} 
& =& -\sigma^\a{}_b q_I^{b}
+\ep^\a{}_{bc}\omega^b q_I^{c} -A_{b}\pi_I^{\a b } \,.
\label{e3i}
\end{eqnarray}

The Ehlers-Ellis covariant kinetic theory description starts by splitting the photon 4-momentum as
\begin{equation}
p^{\a}=E(u^{\a}+e^{\a})\,,~~e^{\a} e_{\a}=1\,,~e^{\a} u_{\a}=0\,. \label{E}
\end{equation}
Here $E=-u_{\a}p^{\a}$ is the energy and $e^{\a}=p^{\langle \a\rangle}/E$ is the
direction, as measured by a comoving fundamental observer. Then
the photon distribution function is decomposed into covariant
harmonics via the expansion
\begin{eqnarray}
f(x,p)=f(x,E,e) &=& F+F_{\a}e^{\a}+F_{\a b }e^{\a}e^{b}+\cdots \nonumber \\ &=& \sum_{\ell\geq0}
F_{A_\ell}(x,E) e^{\langle A_\ell\rangle}, \label{r3}
\end{eqnarray}
where
${A_\ell}\equiv {\a_1}{\a_2}\cdots {\a_\ell}$ and $e^{A_\ell} \equiv e^{\a_1}\cdots e^{\a_\ell}$. The multipoles $F_{A_\ell}$ are a covariant alternative to the usual expansion in spherical harmonics. They are PSTF:
\begin{eqnarray}
F_{\a\cdots b}=F_{\langle \a\cdots b\rangle}\,\Leftrightarrow \, F_{\a\cdots b}=F_{(\a\cdots b)},~~~F_{\a\cdots b}u^{b}=0= F_{\a\cdots bc}h^{bc}.~~~~~~\label{r3'}
\end{eqnarray}

The first 3 multipoles determine the radiation energy-momentum
tensor,
\begin{eqnarray}
T_{r}^{\a b }(x)& \equiv &\int p^{\a}p^{b}f(x,p)\mathrm{d}^3p \nonumber\\ &=&
\rho_{r}u^{\a}u^{b}+{{1\over3}}\rho_{r}h^{\a  b }
+2q_{r}^{(\a}u^{b)}+\pi_{r}^{\a b }\,,
\end{eqnarray}
where $\mathrm{d}^3p=E\mathrm{d} E\mathrm{d}\Omega$ is the covariant volume element on the
future light cone at event $x$. It follows that the dynamical quantities of the radiation (in the $u^{\a}$-frame) are: \begin{eqnarray}
\rho_{r} &=& 4\pi\int_0^\infty E^3F\,\mathrm{d} E\,,~~~~ ~q_{r}^{\a} =
{4\pi\over 3}\int_0^\infty E^3F^{\a}\,\mathrm{d} E \,,
~~~~\pi_{r}^{\a b } =
{8\pi\over 15}\int_0^\infty E^3F^{\a b }\,\mathrm{d} E\,. \label{em3} \end{eqnarray}
We extend these dynamical quantities to all multipole orders by
defining the brightness multipoles
\begin{eqnarray}
\Pi_{\a_1\cdots \a_\ell} = \int E^3
F_{\a_1\cdots \a_\ell}\mathrm{d} E\,, \label{r10}
\end{eqnarray}
so that
\begin{equation}\label{}
\Pi={1\over 4\pi}\rho_{r},~ \Pi^{\a}={3\over 4\pi}q_{r}^{\a}, ~\Pi^{\a b }={15\over 8\pi}\pi_{r}^{\a b }.
\end{equation}
From these the non-linear temperature fluctuations may be defined as, for $\ell\geq1$,
\be
\mathcal{T}_{A_\ell}=\left(\frac{\pi}{\sigma T^4}\right)\Pi_{A_\ell},
\ee
where the monopole temperature is defined from $\rho_r=\sigma T^4$.

The collisionless Boltzmann (or Liouville) equation is
\begin{equation}
{\mathrm{d} f\over \mathrm{d} v}\equiv p^{\a}{\p f\over \p x^{\a}}-\Gamma^{\a}{}_{bc}
p^{b}p^{c}{\p f\over \p p^{\a}}=0 \,, \label{boltz}
\end{equation}
where $p^{\a}=\mathrm{d} x^{\a}/\mathrm{d} v$, and we neglect polarization.
The covariant multipoles of $\mathrm{d} f/\mathrm{d} v$ are given by
\begin{eqnarray}
&&{1\over E}\left({\mathrm{d} f \over \mathrm{d} v}\right)_{A_\ell} =\dot{F}_{\langle A_\ell \rangle}-{{1\over3}}\Theta
EF'_{A_\ell} +\D_{\langle \a_\ell} F_{A_{\ell-1}\rangle}
+{(\ell+1)\over(2\ell+3)} \D^{\a}F_{\a A_\ell} ~~~~
\nonumber\\
&&{}
-{(\ell+1)\over(2\ell+3)}E^{-(\ell+1)}\left[E^{\ell+2}F_{\a A_\ell}
\right]'A^{\a}-E^\ell\left[E^{1-\ell}F_{\langle A_{\ell-1}}\right]'
A_{\a_\ell\rangle}
\nonumber\\
&&{} -\ell\omega^{\b}\ep_{\b c( \a_\ell} F_{A_{\ell-1})}{}^{c}
-{(\ell+1)(\ell+2)\over(2\ell+3)(2\ell+5)}E^{-(\ell+2)}
\left[E^{\ell+3}F_{\a\b A_\ell}\right]'\sigma^{\a\b }
\nonumber\\
&&{} -{2\ell\over (2\ell+3)}E^{-1/2}\left[E^{3/2}F_{\b\langle
A_{\ell-1}}
\right]'\sigma_{\a_\ell\rangle}{}^{\b} \nonumber\\
&&{}
 -E^{\ell-1}\left[E^{2-\ell} F_{\langle
A_{\ell-2}}\right]'\sigma_{\a_{\ell-1}\a_\ell\rangle}\,,
\label{r25}\end{eqnarray}
where a prime denotes $\p/\p E$. This is a fully nonlinear expression.

Multiplying Eq. (\ref{r25}) by $E^3$ and integrating over all
energies leads to the brightness multipole evolution equations:
\begin{eqnarray}
0 &=& \dot{\Pi}_{\langle A_\ell\rangle}+{{4\over3}}\Theta
\Pi_{A_\ell}+ \D_{\langle \a_\ell}\Pi_{A_{\ell-1}\rangle}
+{(\ell+1)\over(2\ell+3)}\D^{\b} \Pi_{\b A_\ell}
\nonumber\\
&&{} -{(\ell+1)(\ell-2)\over(2\ell+3)} A^{\b} \Pi_{\b A_\ell} +(\ell+3)
A_{\langle \a_\ell} \Pi_{A_{\ell-1}\rangle}\nonumber\\
&&{} -\ell\omega^{\b}\ep_{\b c( \a_\ell}
\Pi_{A_{\ell-1})}{}^{ c}
 -{(\ell-1)(\ell+1)(\ell+2)\over(2\ell+3)(2\ell+5)}
\sigma^{\b c}\Pi_{\b c A_\ell} \nonumber\\
&&{} +{5\ell\over(2\ell+3)} \sigma^{\b}{}_{\langle
\a_\ell} \Pi_{A_{\ell-1}\rangle \b} -(\ell+2) \sigma_{\langle
\a_{\ell}\a_{\ell-1}} \Pi_{A_{\ell-2}\rangle}\,.
\label{r26}\end{eqnarray}
Once again, this is a fully nonlinear result.
The monopole evolution equation is just the energy conservation equation, i.e., Eq.~(\ref{e1i}) with $I=r$, and the dipole evolution equation is the momentum conservation equation~(\ref{e3i}), with $I=r$. The quadrupole evolution is given by Eq.~(\ref{nl8}).

\end{document}